# Multiway Turing Machines

Stephen Wolfram*

*Multiway Turing machines (also known as nondeterministic Turing machines or NDTMs) with explicit, simple rules are studied. Even very simple rules are found to generate complex behavior, characterized by complex multiway graphs, that can be visualized in multispace that combines "tape" and branchial space. The threshold for complex behavior appears to be machines with just $s = 1$ head states, $k = 2$ tape colors and $p = 3$ possible cases, and such machines may potentially be universal. Other characteristics of multiway Turing machines are also studied, including causal invariance, cyclic tapes and generalized busy beaver problems. Multiway Turing machines provide minimal examples of a variety of issues encountered in both concurrent computing and the theory of observers in quantum mechanics, especially in our recent models of physics.*

---

Over the years I've studied the simplest ordinary Turing machines quite a bit, but I've barely looked at multiway Turing machines (also known as nondeterministic Turing machines or NDTMs). Recently, though, I realized that multiway Turing machines can be thought of as "maximally minimal" models both of concurrent computing and of the way we think about quantum mechanics in our Physics Project. So now this piece is my attempt to "do the obvious explorations" of multiway Turing machines . And as I've found so often in the computational universe, even cases with some of the very simplest possible rules yield some significant surprises....

## Ordinary vs. Multiway Turing Machines

An ordinary Turing machine has a rule such as

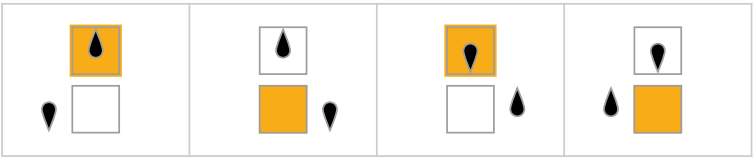

---

A version of this document with immediately executable code is available at
https://www.wolframphysics.org/bulletins/2021/02/multiway-turing-machines/

*Email: s.wolfram@wolfram.com

that specifies a unique successor for each configuration of the system (here shown going down the page starting from an initial condition consisting of a blank tape):

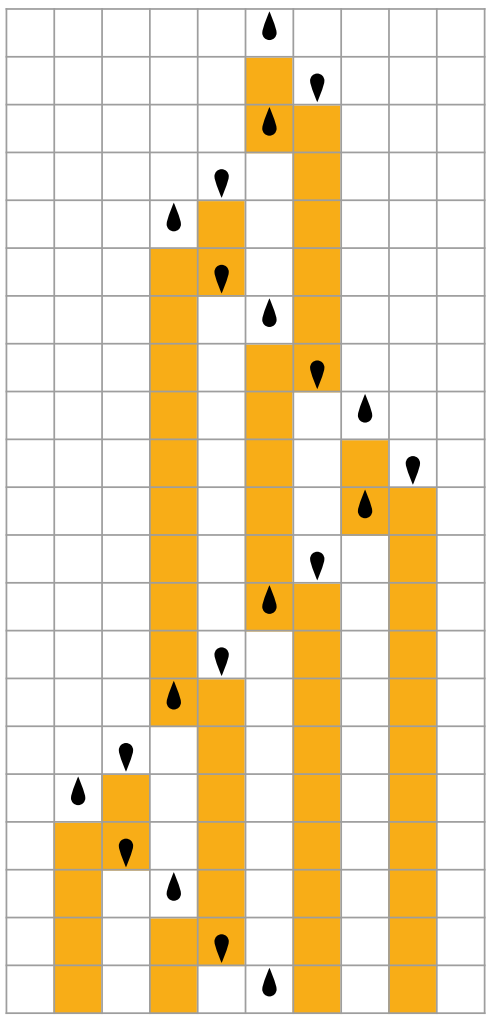

In a multiway Turing machine more than one possible outcome can be specified for a particular case in the rule:

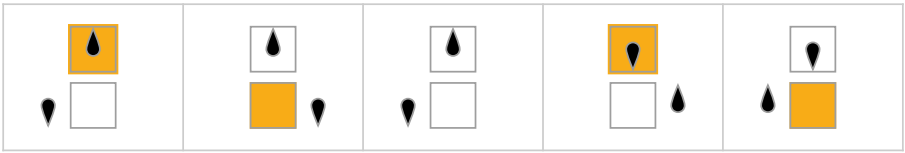

The result is that there can be many possible paths of evolution for the system—as represented by a multiway graph that connects successive possible configurations:

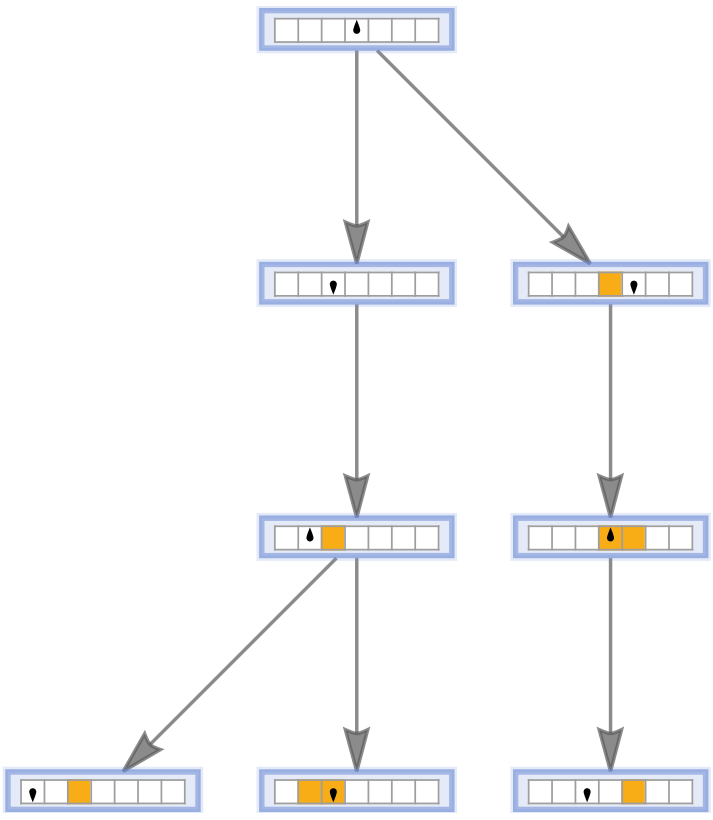

The evolution according to the ordinary Turing machine rule corresponds to a particular path in the multiway graph:

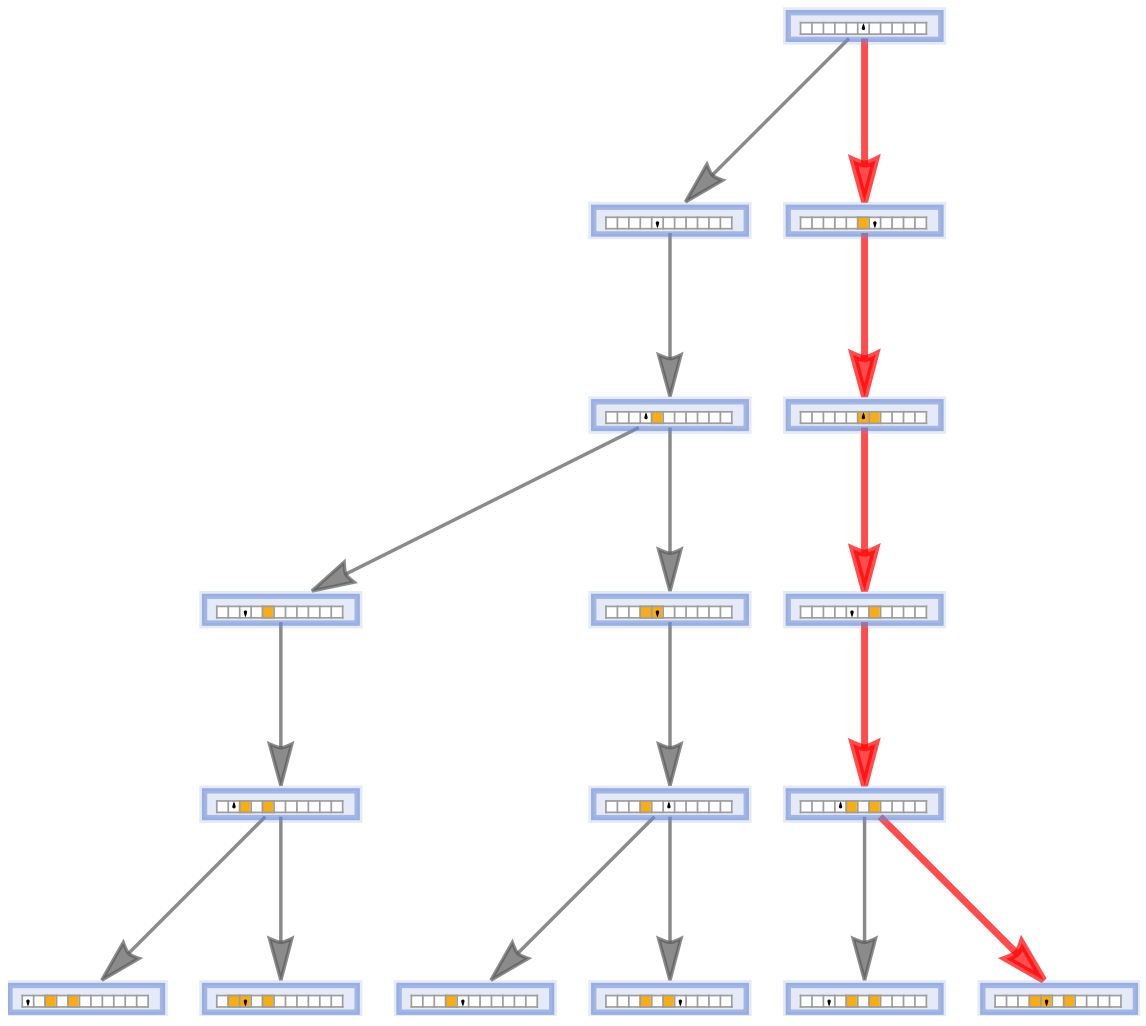

Continuing for a few more steps, one begins to see a fairly complex pattern of branching and merging in the multiway graph:

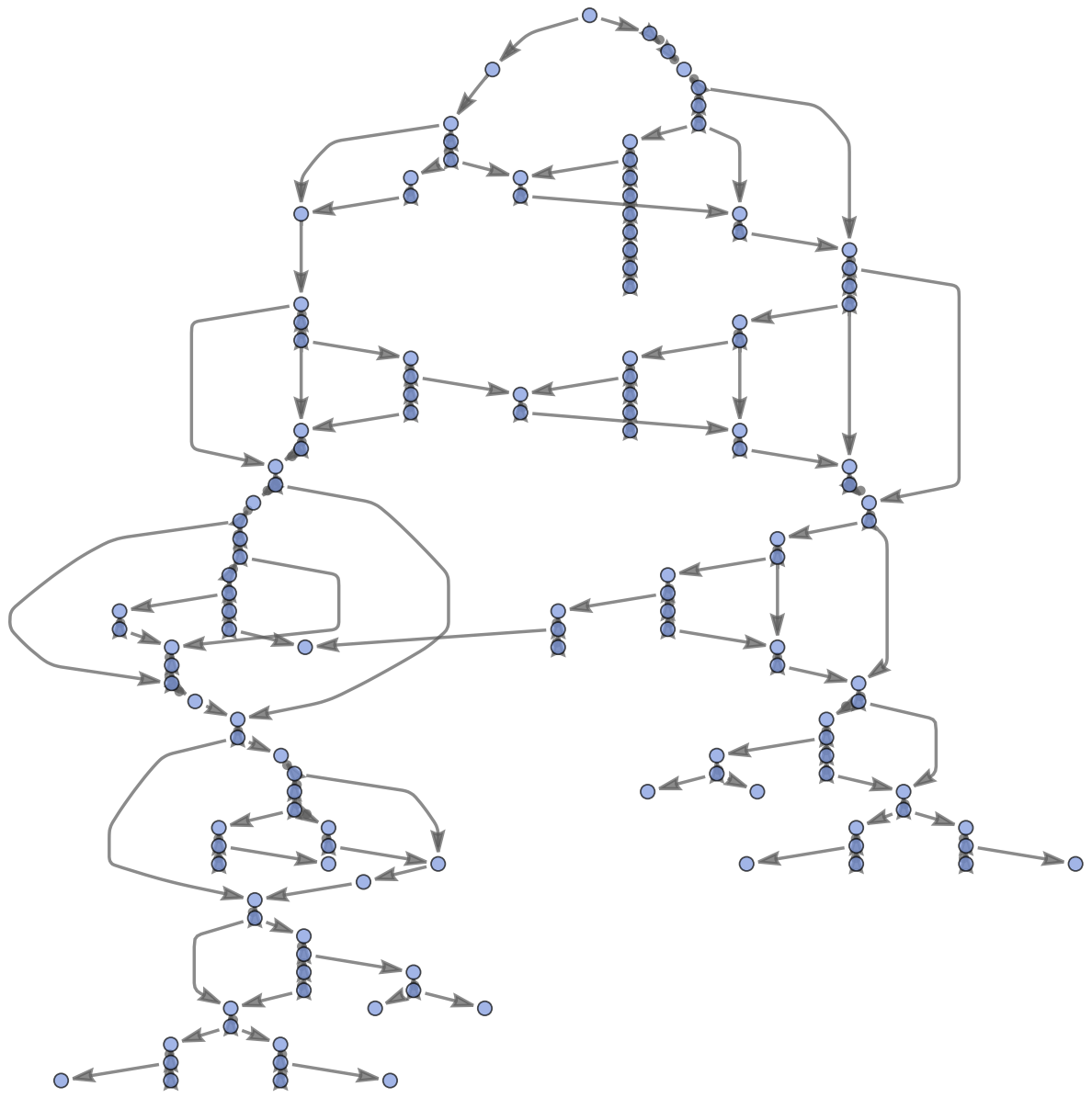

My purpose here is to explore what multiway Turing machines with simple rules can do.

The basic setup for my multiway Turing machines is the same as for what are usually called “nondeterministic Turing machines” (NDTMs), but in NDTMs one is usually interested in whether single paths have particular properties, while we will be interested in the complete multiway structure of all possible paths. (And by using "multiway" rather than "nondeterministic" we avoid the confusion that we might be thinking about probabilistic or random paths—and emphasize that we're studying the structure of all possible paths.)

In our Physics Project, multiway systems are what lead to quantum mechanics, and our multiway Turing machines here correspond quite directly to quantum Turing machines, with the magnitudes of quantum amplitudes being determined by path weighting, and their phases being determined by locations in the branchial space that emerges from transverse slices of the multiway graph.

## Turing Machines with Simple Rules

There are 4096 possible ordinary Turing machines with s = 2 head states and k = 2 colors. All these machines ultimately behave in simple ways, with examples of the most complex behavior being (the last case is a binary counter with logarithmic growth):

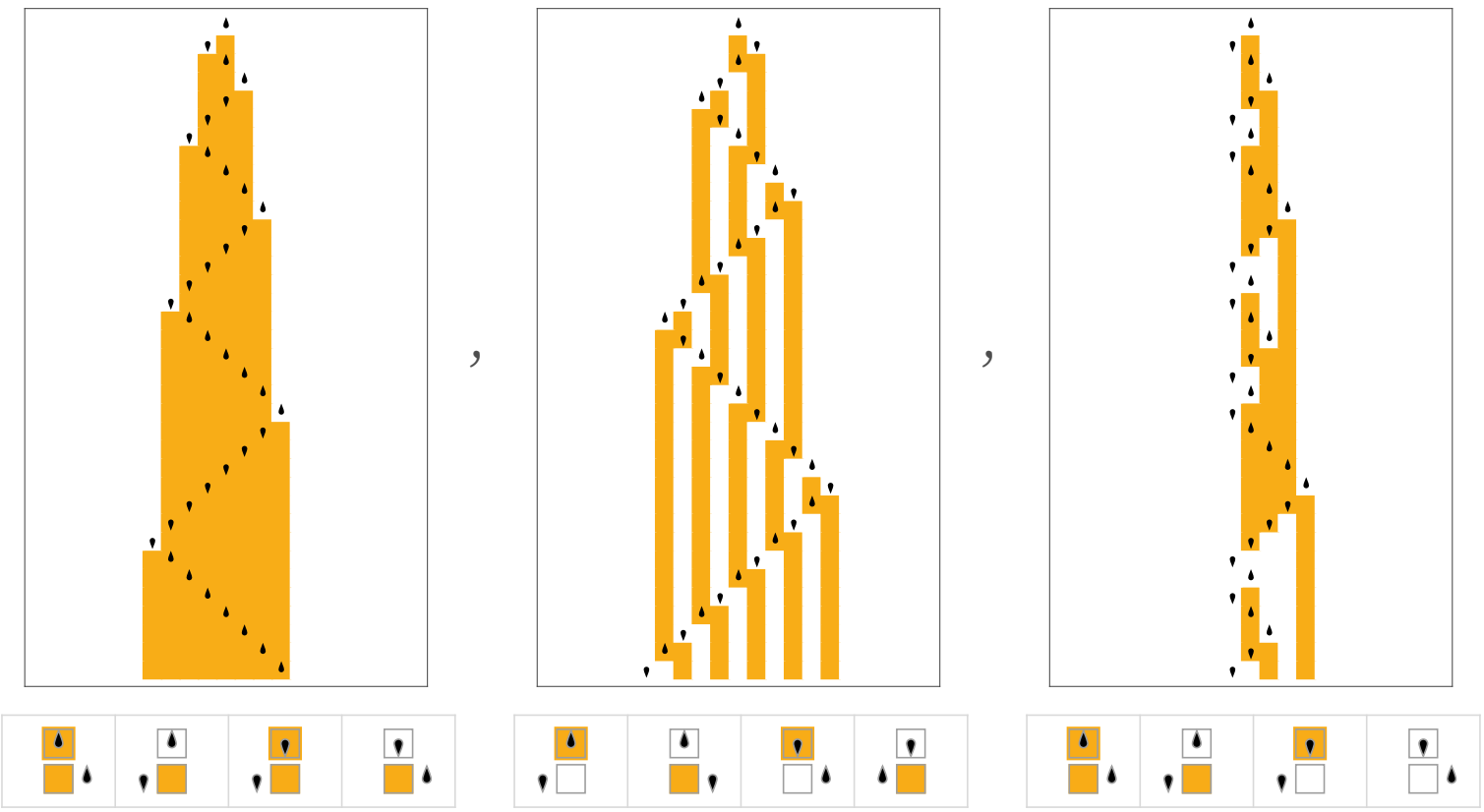

Things get only slightly more complex with s = 3, k = 2, but with s = 2, k = 3 one finds the simplest Turing machine with complex behavior (and this behavior occurs even with a blank initial tape):

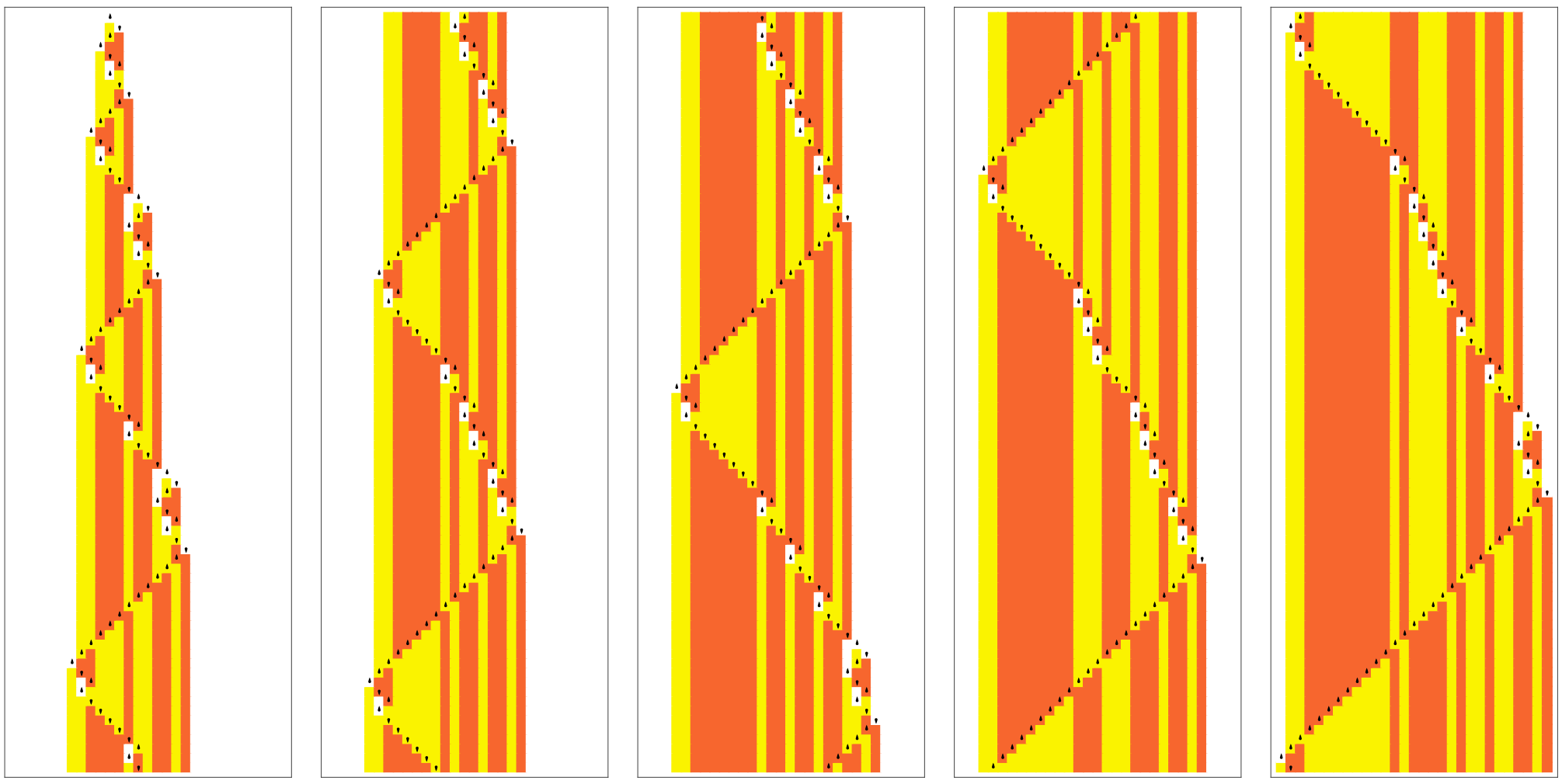

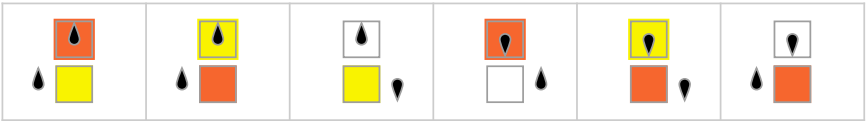

As suggested by the Principle of Computational Equivalence, this Turing machine turns out be computation universal (making it the very simplest possible universal Turing machine).

So what this means is that the threshold for universality in ordinary Turing machines is s = 2, k = 3: i.e. rules in which there are 6 cases specified. But one question we can then ask is: what is the threshold for universality in multiway Turing machines?

To say that our s = 2, k = 3 ordinary Turing machine is universal is to say that with appropriate initial conditions the evolution of this Turing machine can emulate the evolution of any other Turing machine (at least with a suitable encoding, implementable by a bounded computation).

The closest analogous definition of computation universality for multiway Turing machines is to say that with appropriate initial conditions the multiway evolution of a Turing machine can emulate (up to suitable encoding) the multiway evolution of any other Turing machine. (As we will discuss later, however, the "nondeterministic interpretation" suggests a nominally different definition based on whether initial conditions can be set up to make corresponding individual paths with particular identifying features exist.)

To specify the rule for an ordinary Turing machine, one normally just defines the outcome for each of the s k possible "inputs" such as

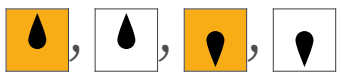

or:

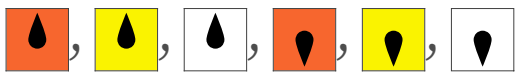

But what if one omits some of these inputs? In analogy to our usual treatment of string or hypergraph rewriting systems it is natural to just say that if one of the omitted inputs is reached in the evolution of the system, then no rewrite is performed, or—in standard Turing machine terms—the system "halts".

With $s = 2$, $k = 2$, the longest that machines that will eventually halt survive (starting from a blank tape) is just 6 steps—and there are 8 distinct machines (up to reflection) which do this, 3 with only 2 cases out of 4 in their rules, and 5 with 3 cases:

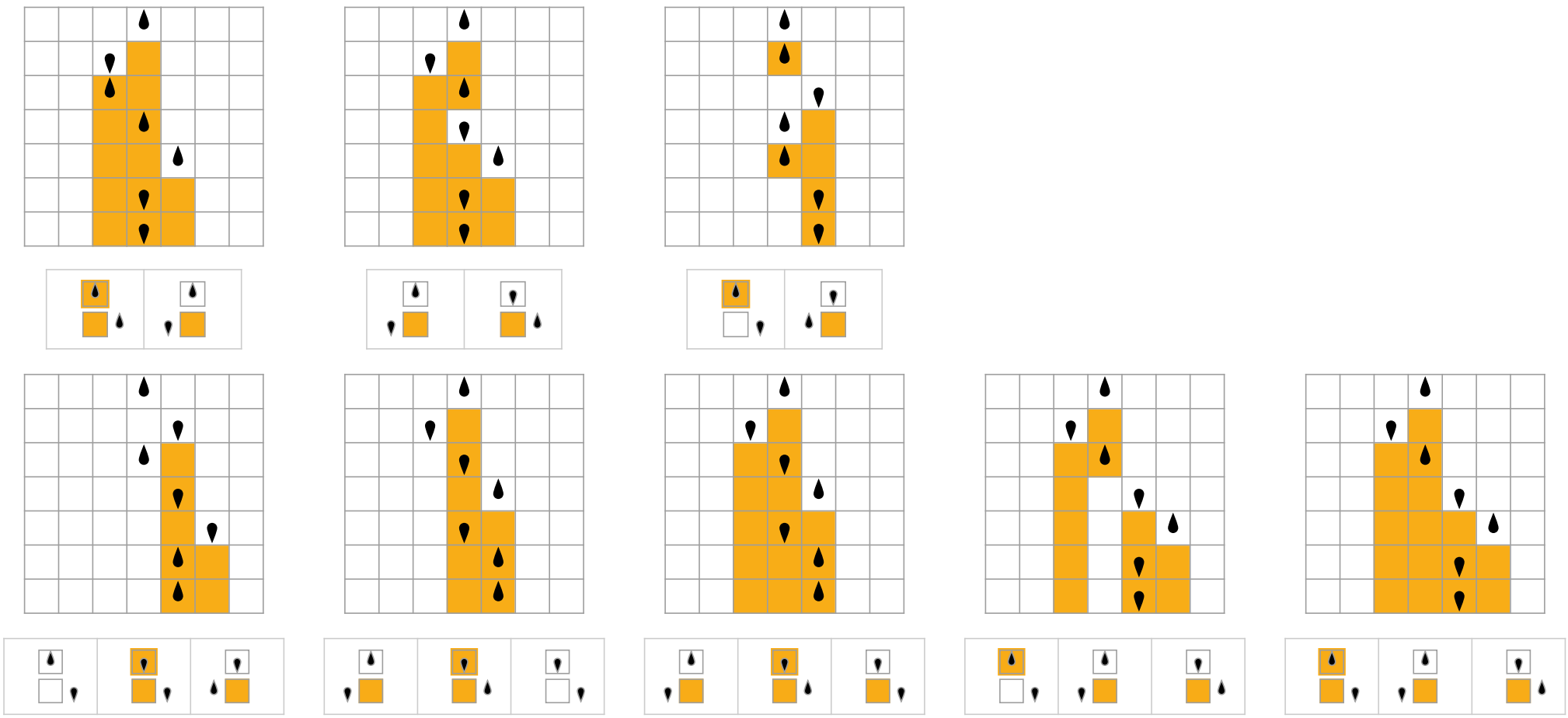

In a sense each of these rules can be thought of as using some subset of all $2\, s^2\, k^2$ possible rule cases, which for $k = 2$, $s = 2$ is:

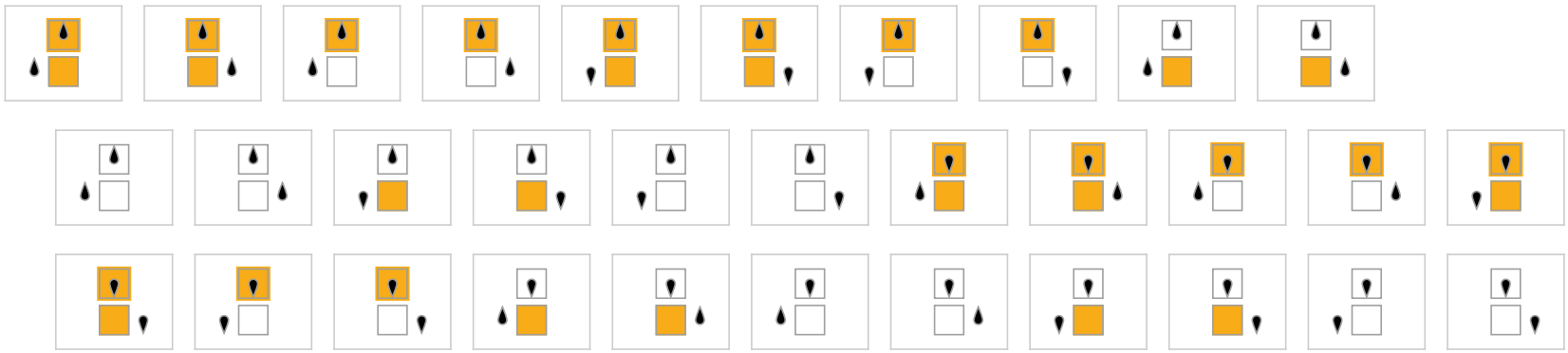

And in general any multiway Turing machine rule corresponds to some such subset. If all possible inputs appear exactly once, one gets an ordinary Turing machine that does not halt. If inputs appear at most once, but some do not appear at all, the Turing machine may halt (though if it never reaches that input it won't halt).

But if any input appears more than once, the Turing machine may exhibit nontrivial multiway evolution. If all inputs appear at least once, no branch of the multiway evolution can halt. But if some inputs appear more than once, while some do not appear at all, some branches of the multiway evolution may halt, while others may continue forever.

In general, there are $2^{2 s^2 k^2}$ possible multiway Turing machine rules. In talking about rulial space and rulial multiway systems I previously discussed the particular example of the ("rulial") multiway Turing machine rule in which every single case is included (i.e. all 32 cases for s = 2, k = 2), so that the Turing machine is in a sense "as multiway as possible". But here we are concerned with multiway Turing machines whose rules are instead as simple as possible—and "far from the rulial limit".

The simplest nontrivial multiway Turing machines have just two cases in their rules. In general, there are Binomial[2 $s^2 k^2$, 2] such rules, or for s = 2, k = 2, 496---of which 112 have actual multiway behavior.

A very simple example is the rule

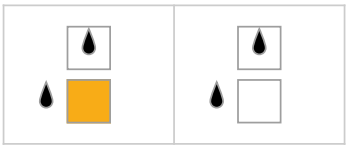

which generates tape configurations containing all binary numbers

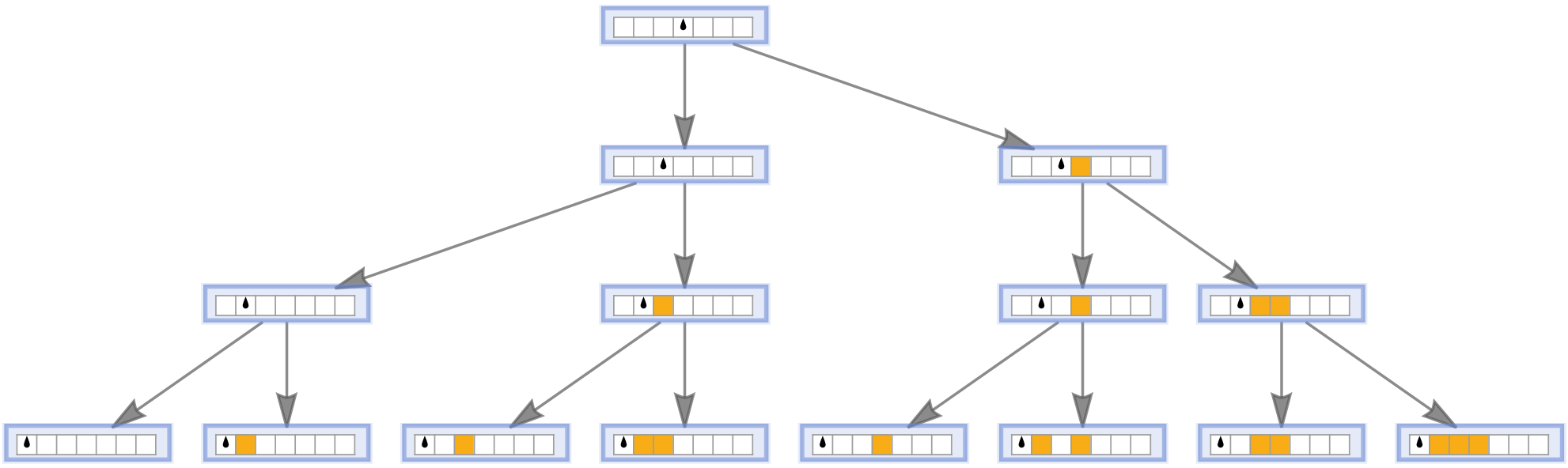

and gives a multiway system which forms an infinite binary tree:

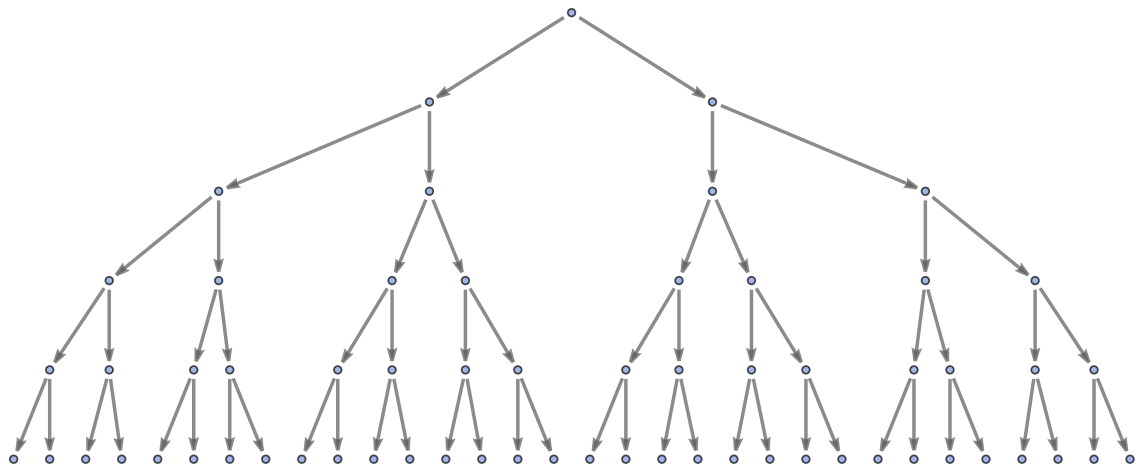

A slightly less trivial example is the rule

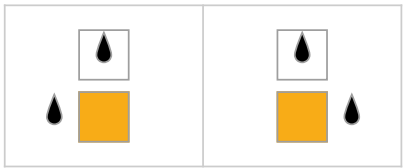

which produces a multiway graph in which branches stop as soon as the head is on a non-blank cell:

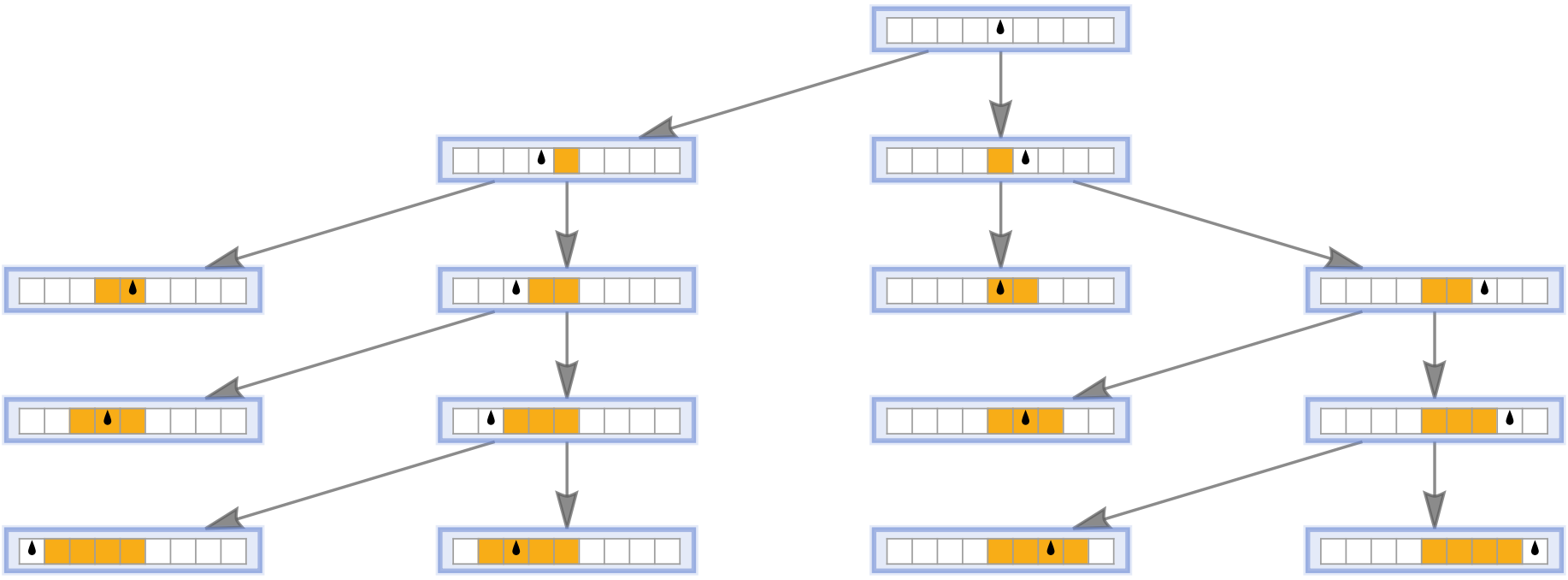

Continuing this, the overall structure of the multiway graph is (where halting states are indicated by red dots):

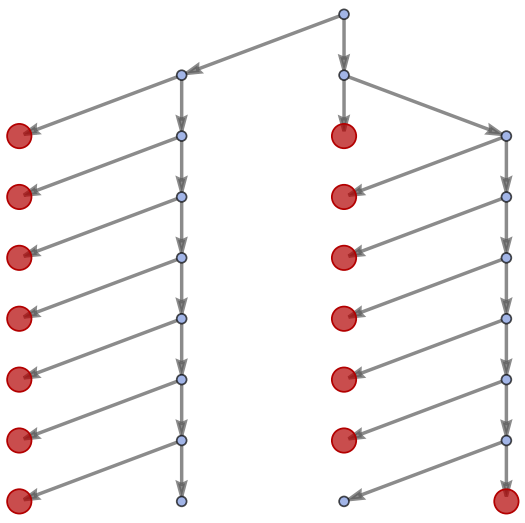

As another example, consider the rule:

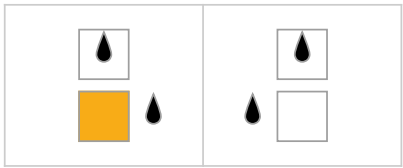

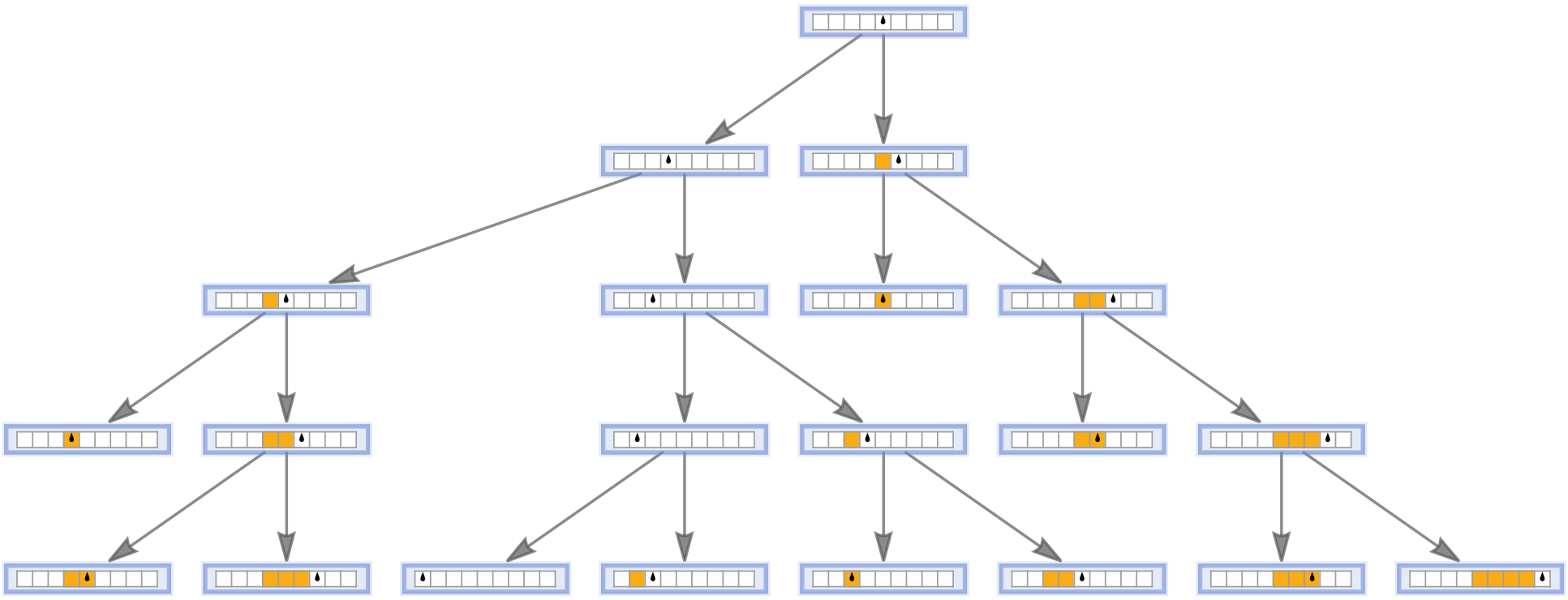

Once again, the system halts if the head gets onto a non-blank cell, but this happens in a slightly more elaborate way, giving a slightly more complicated multiway graph:

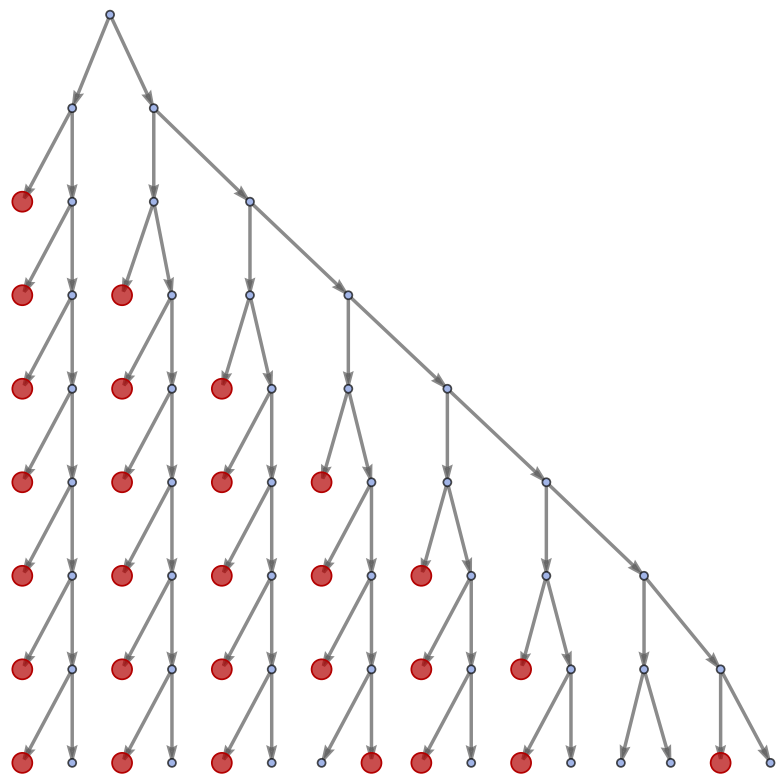

In the example we've just seen, there is explicit halting, in the sense that the Turing machine can reach a state in which none of its rules apply. Another thing that can happen is that Turing machines can get into infinite loops—and since in our multiway graphs identical states are merged, such loops show up as actual loops in the multiway graph, as in this example:

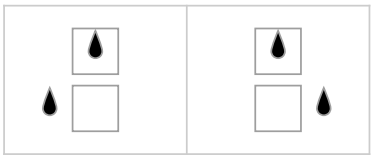

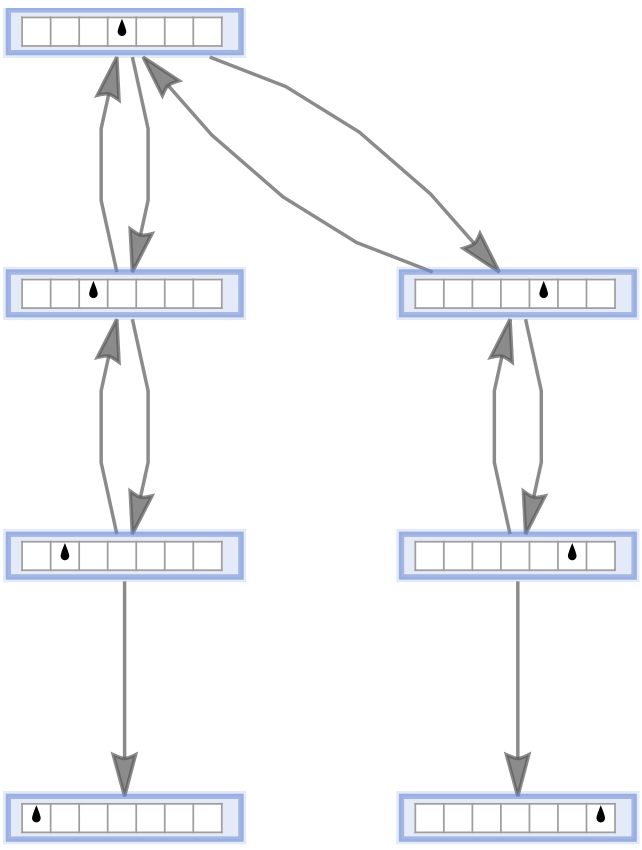

As soon as one allows 3 cases in the rules, things rapidly get more complicated. Consider for example the rule (which can never halt):

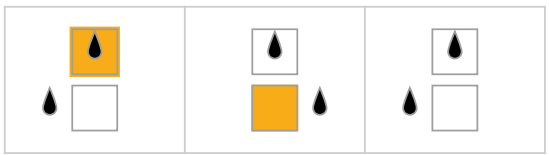

The first few steps in its multiway graph are:

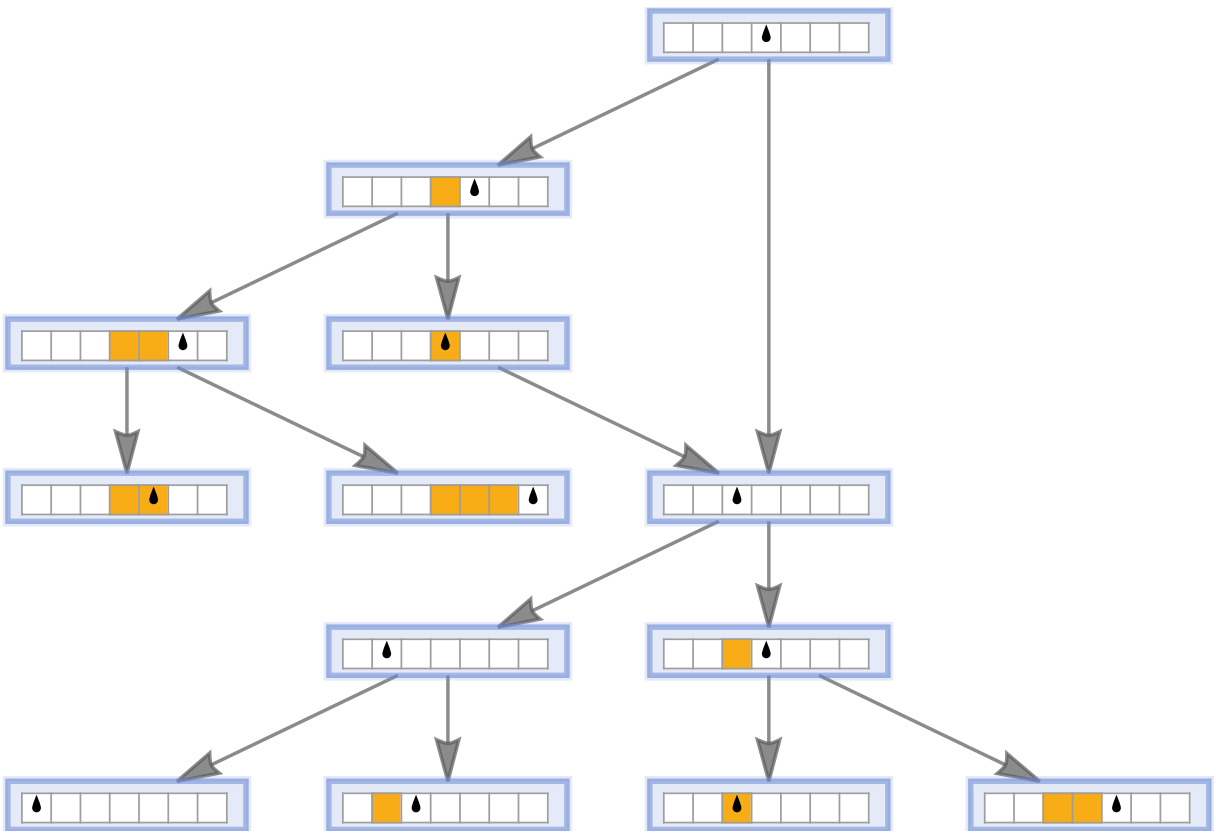

This example exposes a somewhat subtle issue. In our earlier examples, all the states generated on a particular step could consistently be arranged in a single layer in the multiway graph. But here the state 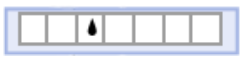 is generated both at step 2 and at step 4—so no such "global layering" is possible (at least, assuming, as we do, that we combine identical copies of this state).

And a consequence of this lack of global layering is that if we compute for a given number of steps we can end up with "dangling states"—like 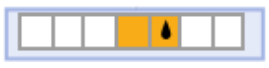—that appear in our rendering well above the "final layer" shown.

After another step, the previous "dangling states" now have successors, but there are new dangling states:

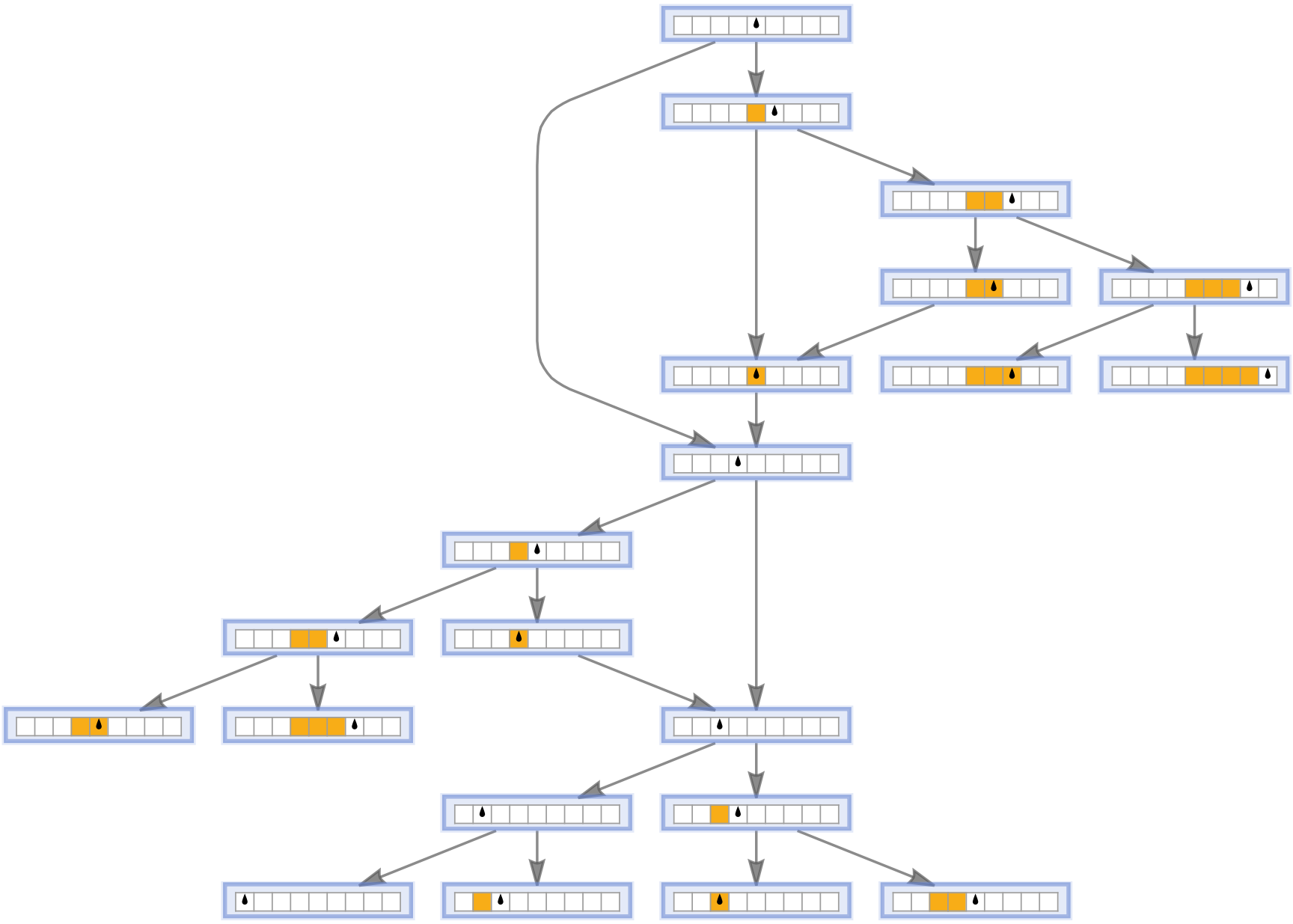

Continuing for more steps, a somewhat elaborate but nevertheless fairly regular multiway begins to develop:

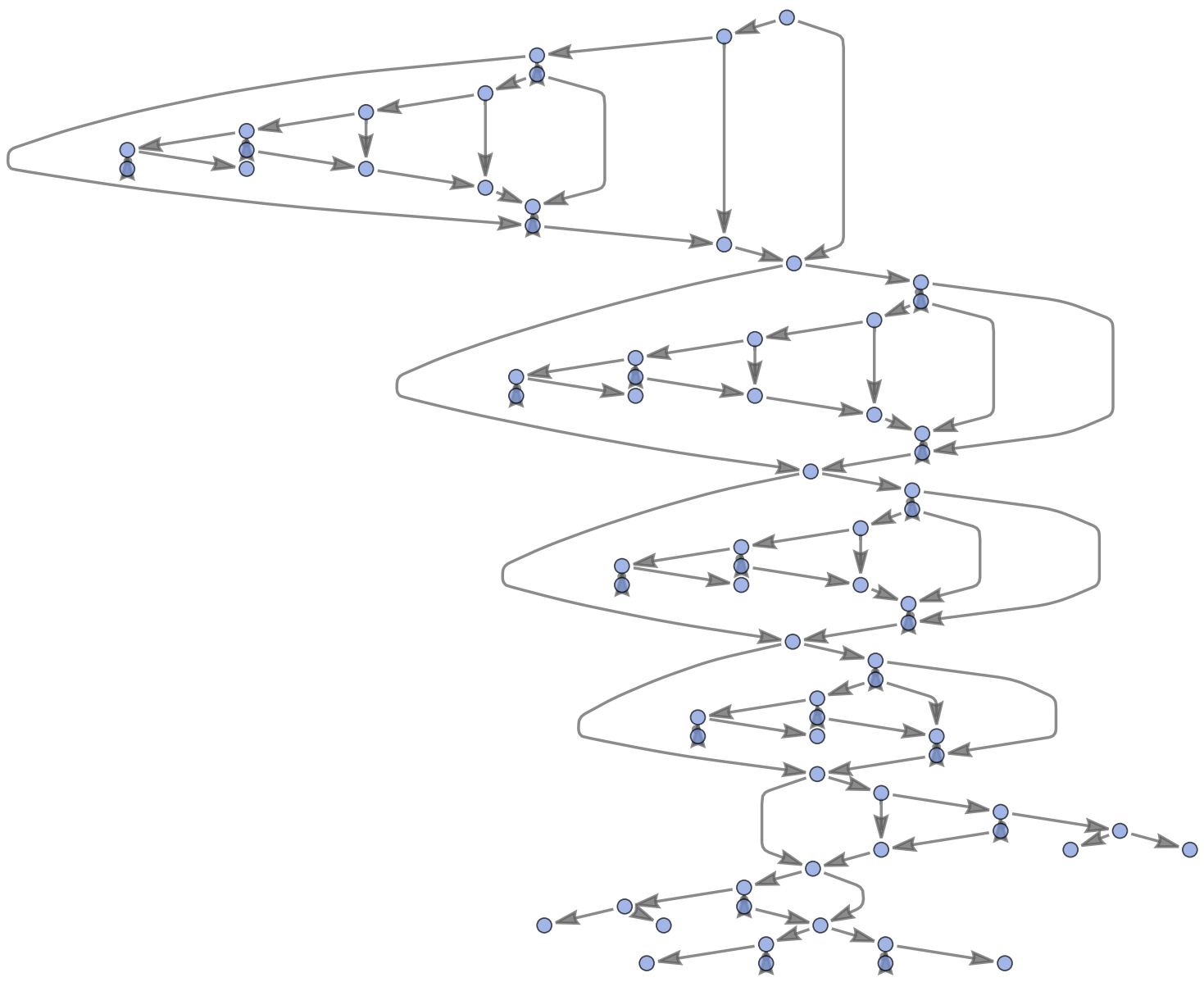

Here's another example of a multiway Turing machine with 3 cases in its rule:

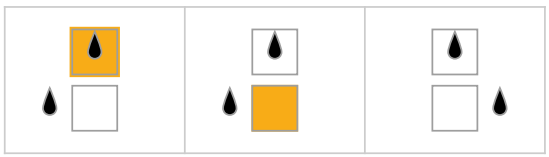

The first few steps in its multiway graph are:

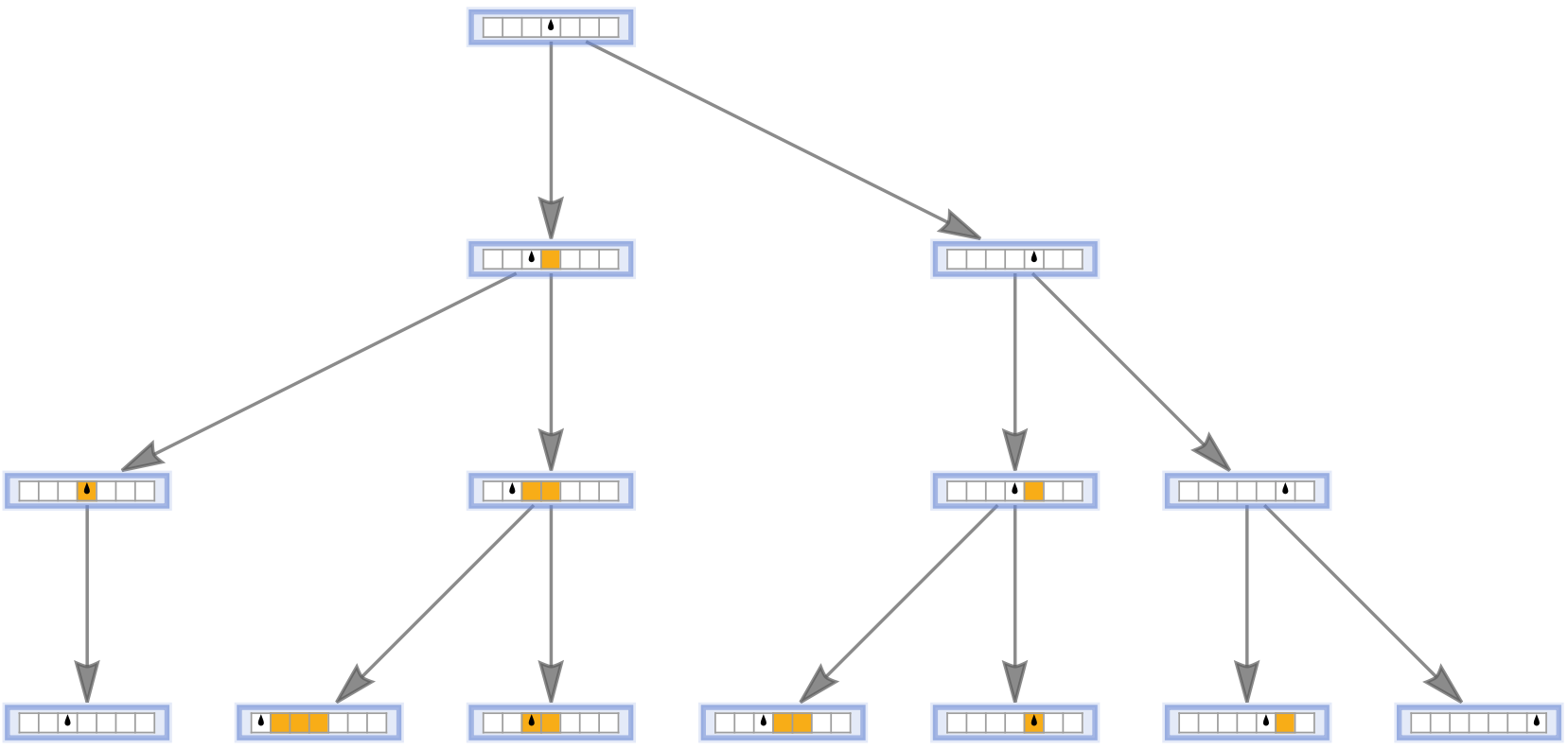

Continuing for another step, things get more complicated, and, notably, we see a loop:

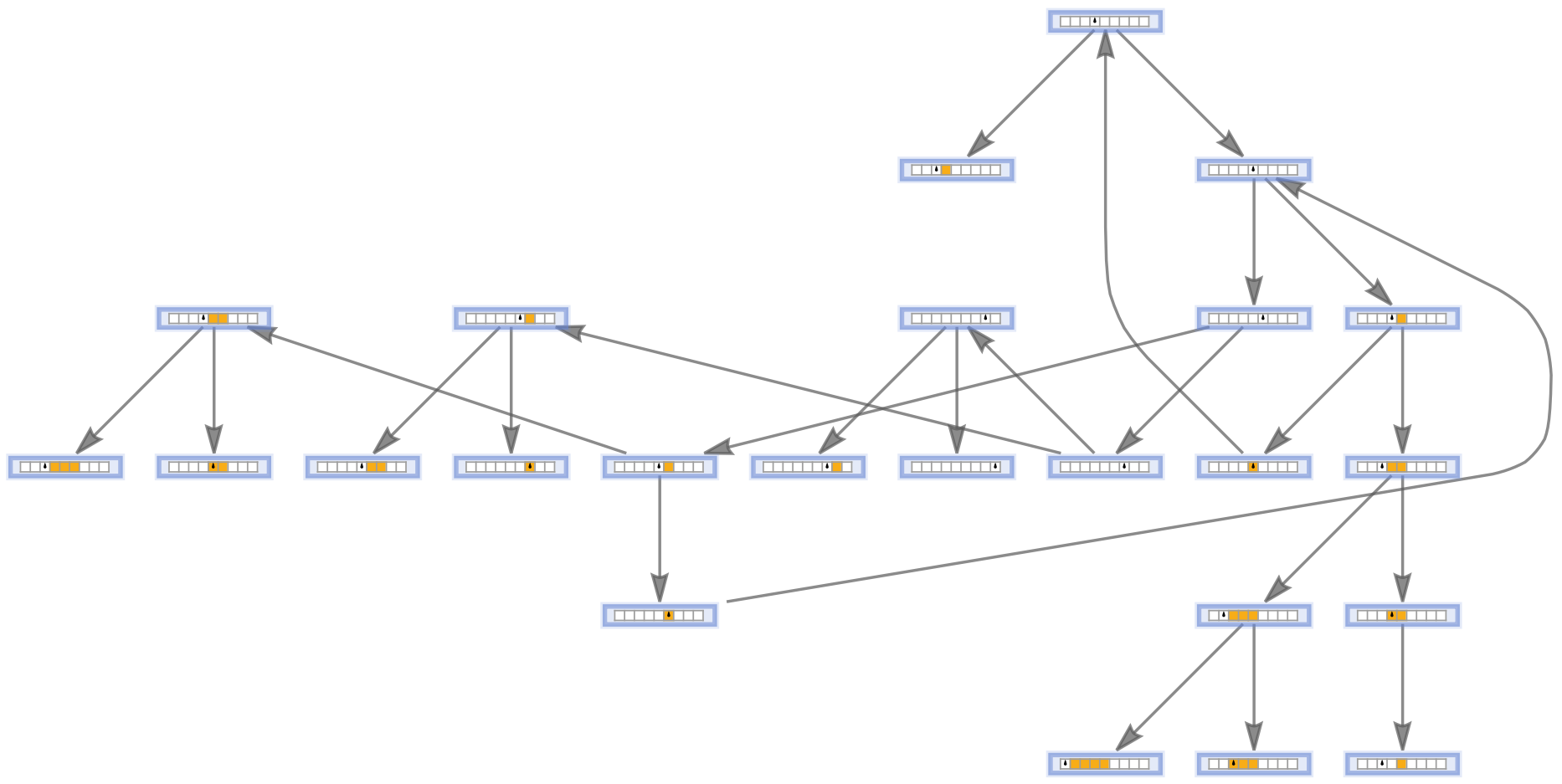

Continuing for a few more steps we get:

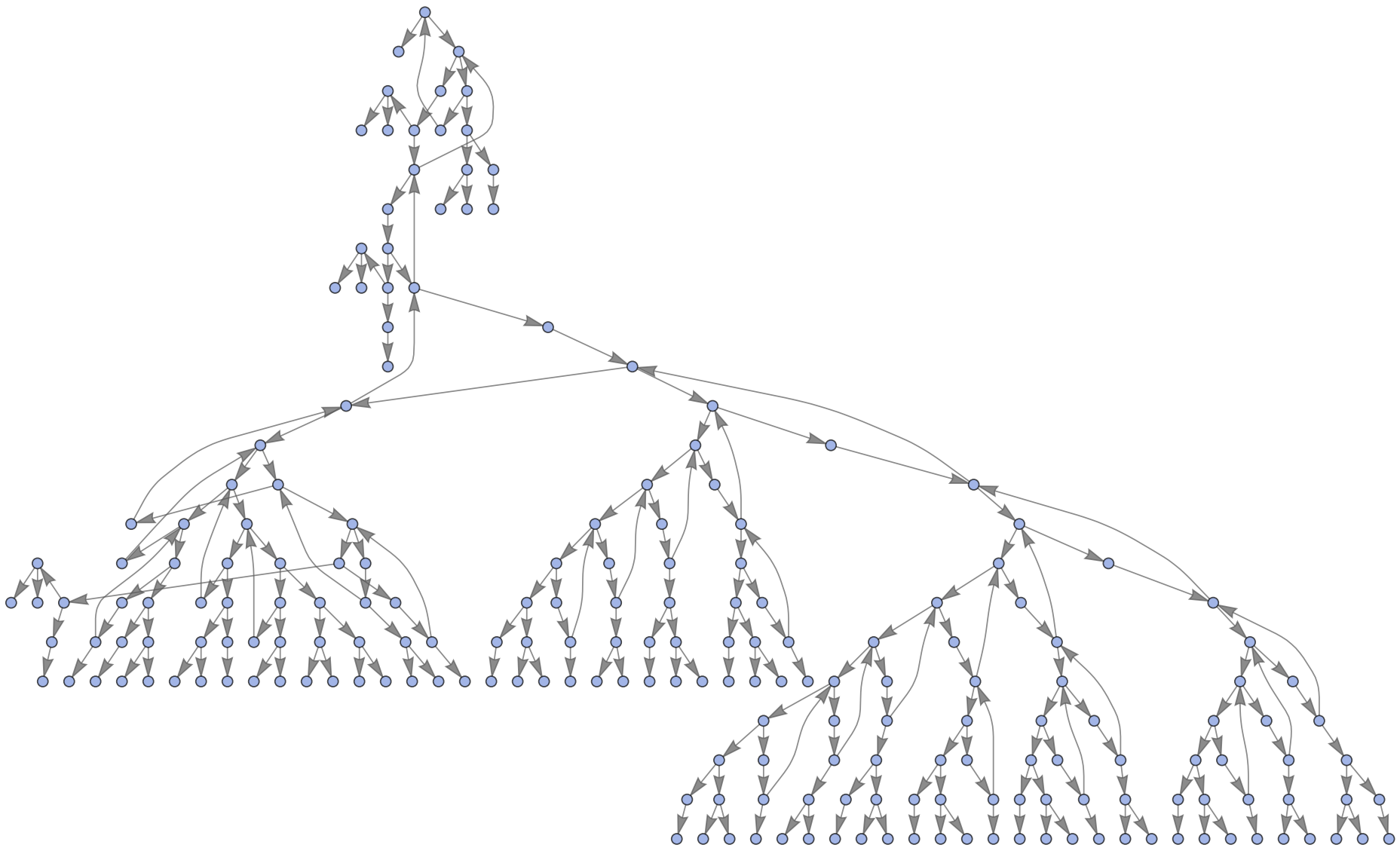

(Note that since in this rule there is no halting, the dangling ends visible here must show additional growth if the evolution is continued.)

It’s perfectly possible to get both rapid growth, and halting. Here’s an example

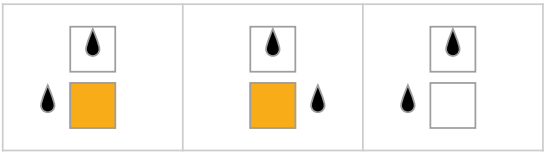

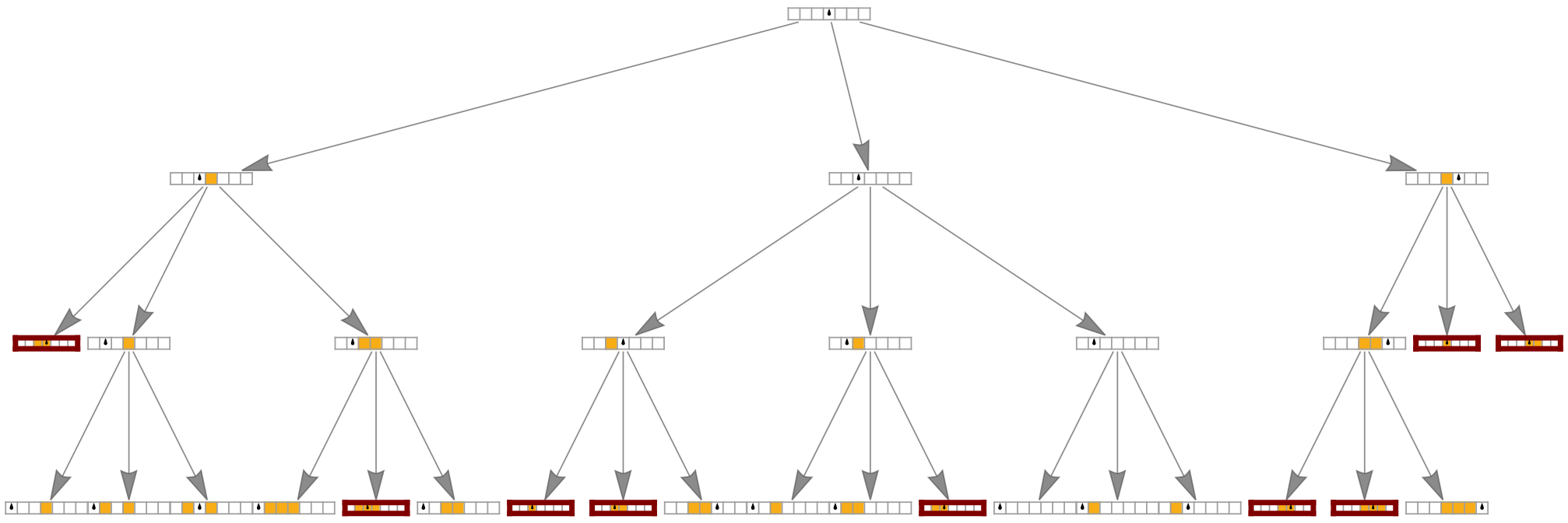

or, continuing for longer:

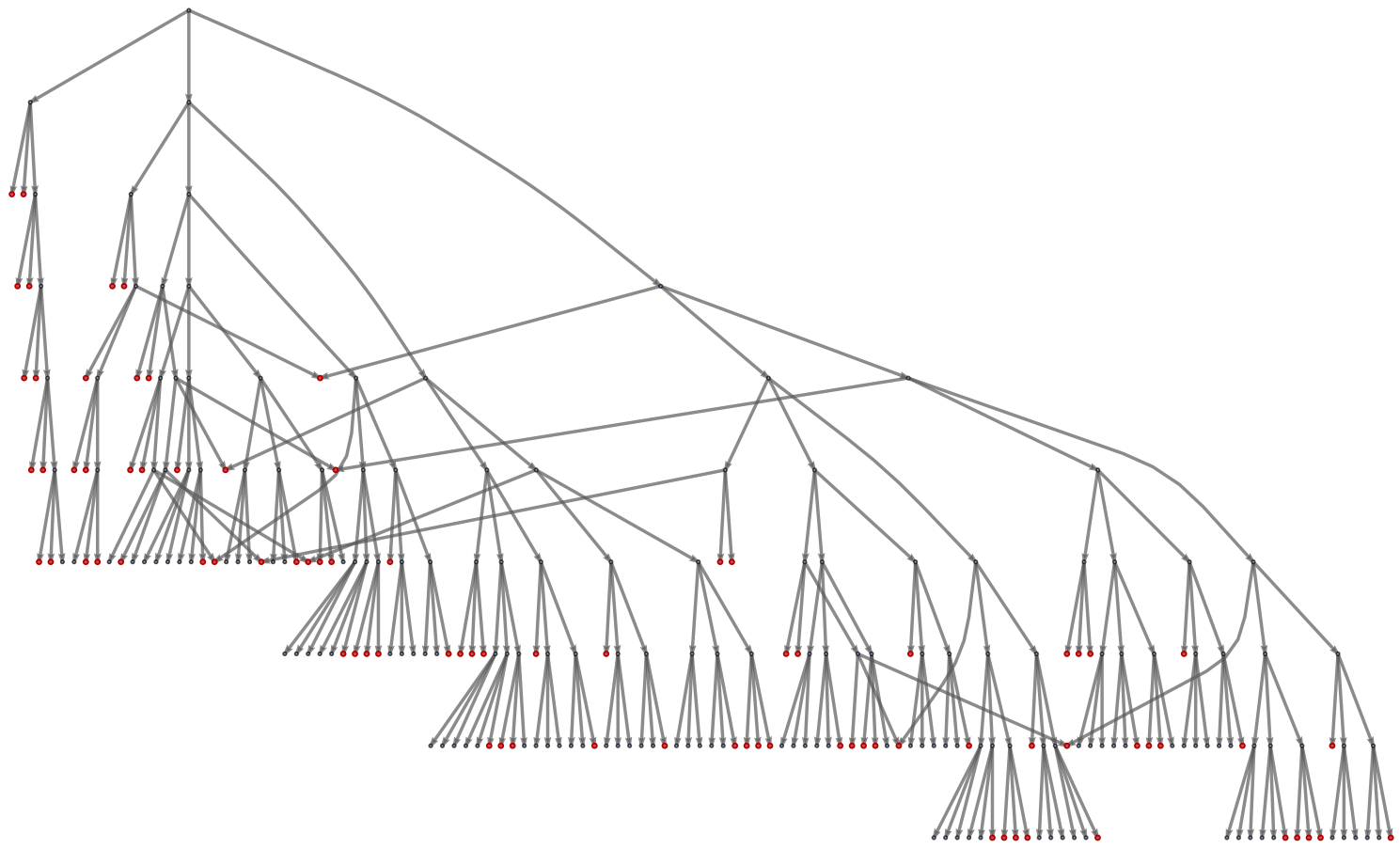

## Visualization and Multispace

It’s fairly easy to get a sense of the behavior of an ordinary Turing machine just by displaying its successive configurations down the page. But what about for a multiway Turing machine? The multiway graph shows one the relationships defined by evolution between complete configurations (states) of the system:

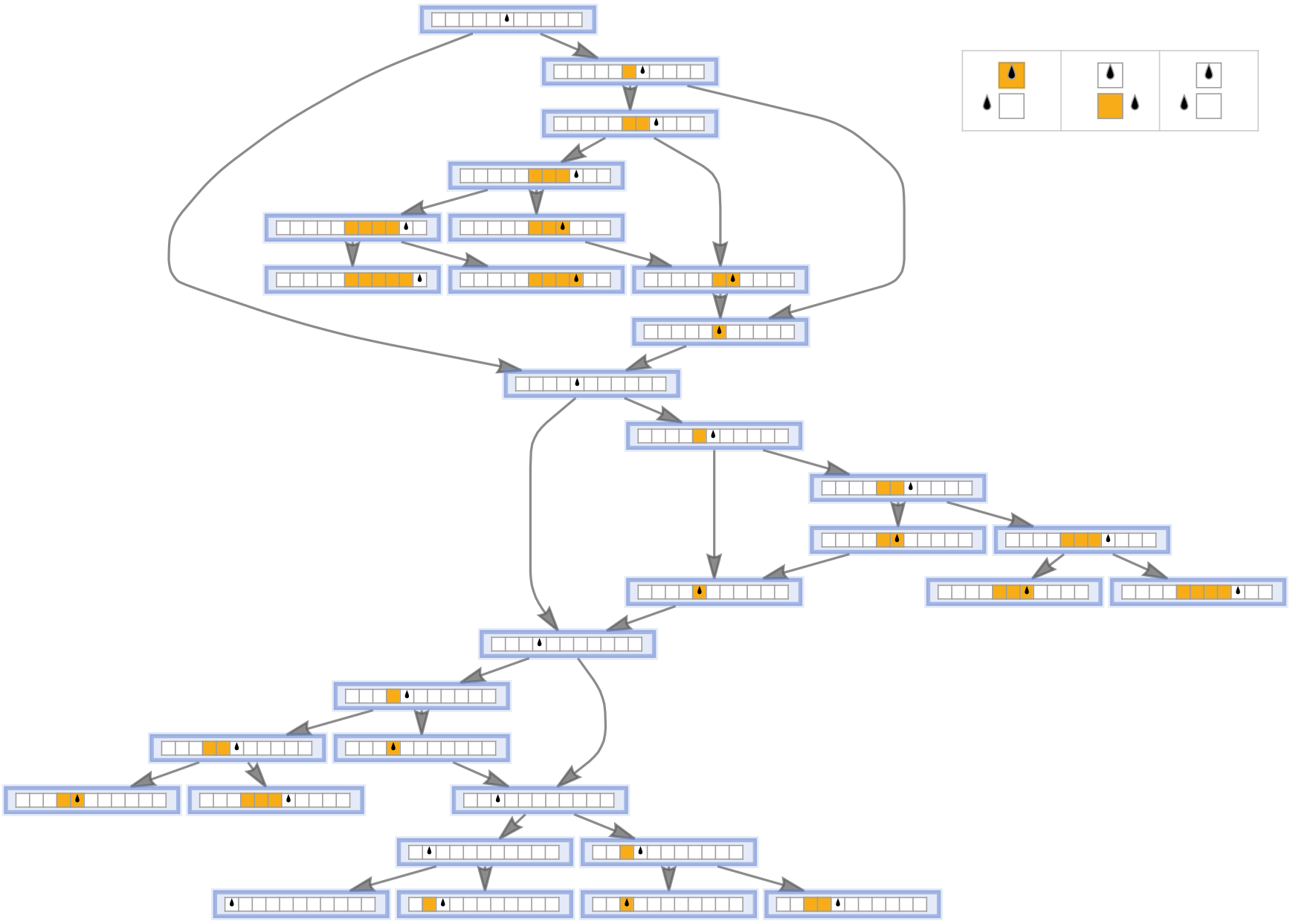

But such a picture does not make clear the relationships between the detailed forms of different configurations, and for example the "spatial motion" of the head. So how can we do this?

One possibility is in effect to display all the different possible configurations at a given step "overlaid on top of each other". For the machine above the new configurations obtained at successive steps are (though note that these do not appear on specific layers in the rendering of the multiway graph above):

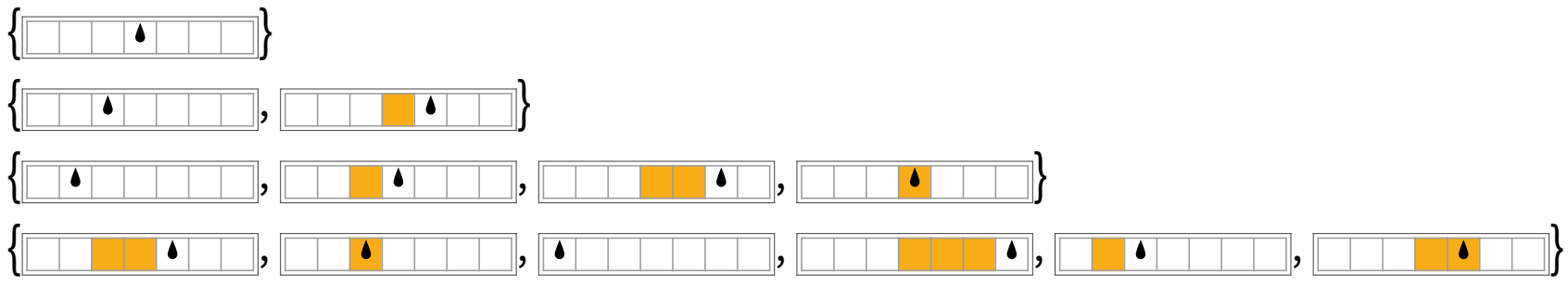

Putting these "on top of each other", and "averaging colors" we get:

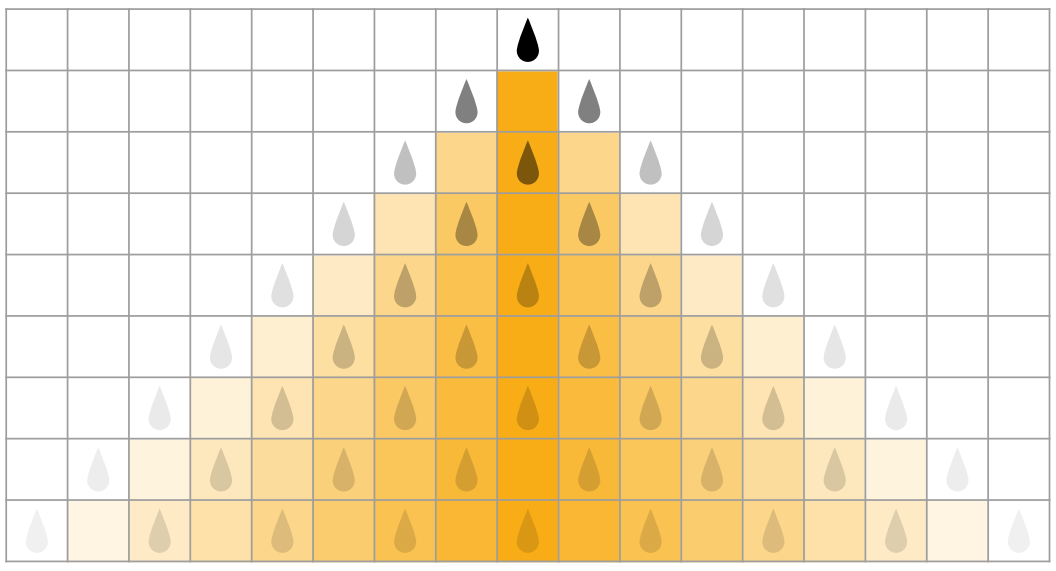

In this case, such a visualization reveals the "diffusive" character of the behavior: the motion of the head in different possible evolutions effectively corresponds to all possible paths in an ensemble of random walks, so that the probability to find the head at offset x is Binomial[t,x]/$2^t$.

We can see the lattice of paths more explicitly if we join head positions obtained at successive steps of evolution:

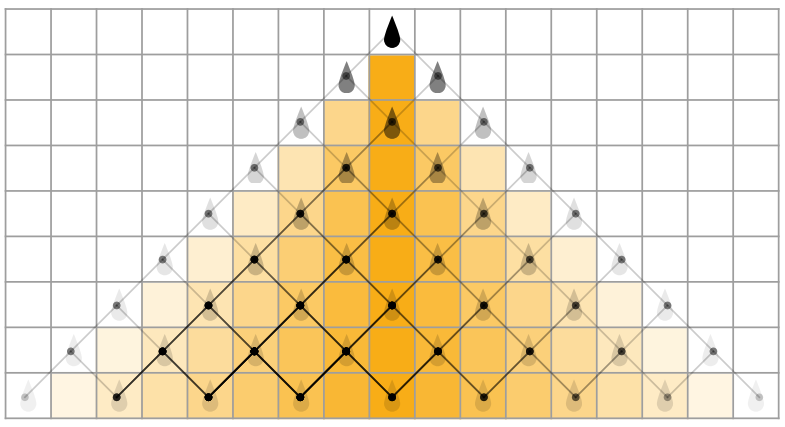

What does this approach reveal about other multiway Turing machines we've discussed? Here's one of our first examples:

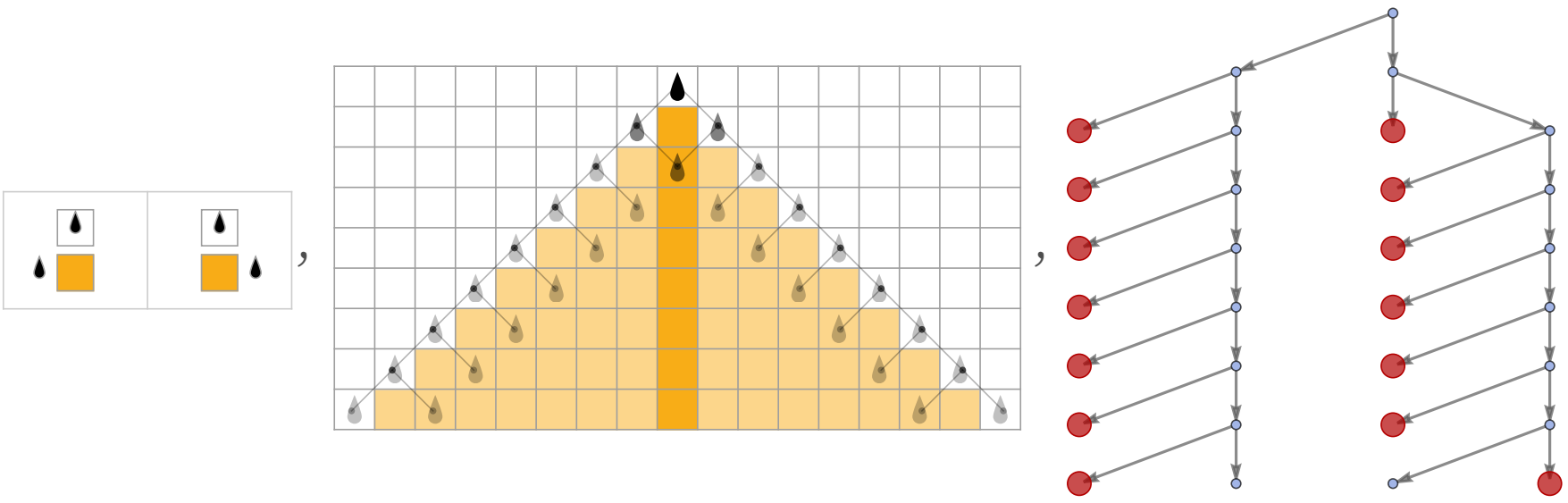

And here's another of our examples:

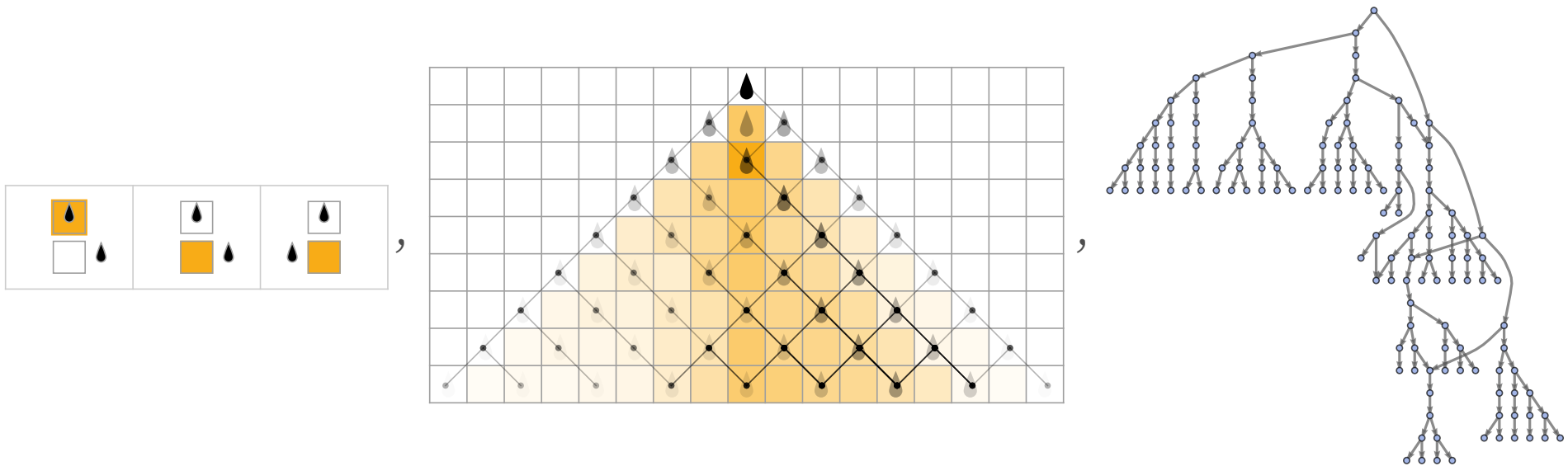

But generally this kind of averaged picture is not particularly helpful. So is there an alternative? Basically we need some way to simultaneously visualize both multiway structure, and the detailed forms of individual configurations.

In analogy to our Physics Project, this is essentially the problem of simultaneously visualizing both branchial and spatial structure. So where should we start? Since our everyday experience is with ordinary space, not branchial space, it seems potentially better to start with ordinary space, and then add in the branchial component. And thinking about this led to the concept of multispace: a generalization of space in which there is in effect at every point a stack of multiway possibilities, with these different possibilities being connected at different places in branchial space.

In our Physics Project, even visualizing ordinary space itself is difficult, because it's built up from lower-level structures corresponding to hypergraphs. But for Turing machines the basic structure of space is constrained to be very simple: there is just a one-dimensional tape with a head that can go backwards and forwards on it.

But in understanding multispace, it is probably easier to start from the branchial side. Let's for now ignore the values written on the tape of a Turing machine, and consider only the position of the head, say relative to where it started. Every state in the multiway graph can then be labeled with the position of the head in that configuration of the Turing machine. So for the first rule above we get:

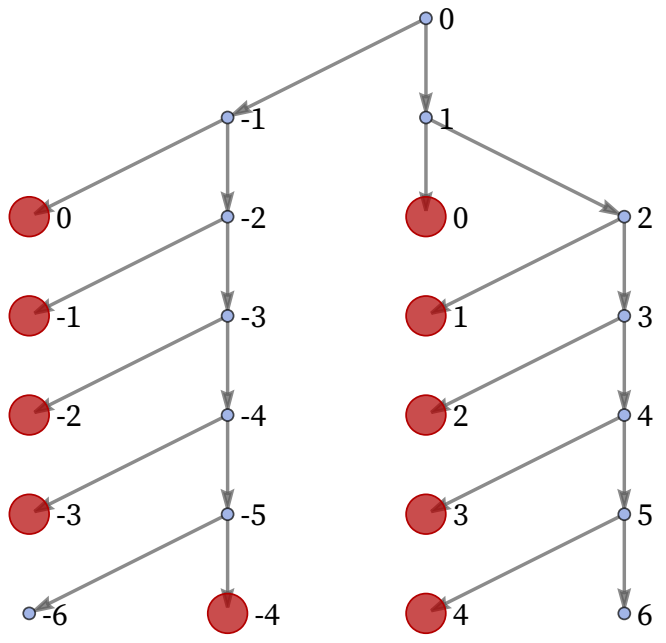

But now let's imagine setting this up in 3D, with the timelike direction down the page, the branchlike direction across the page, and the spacelike direction into the page:

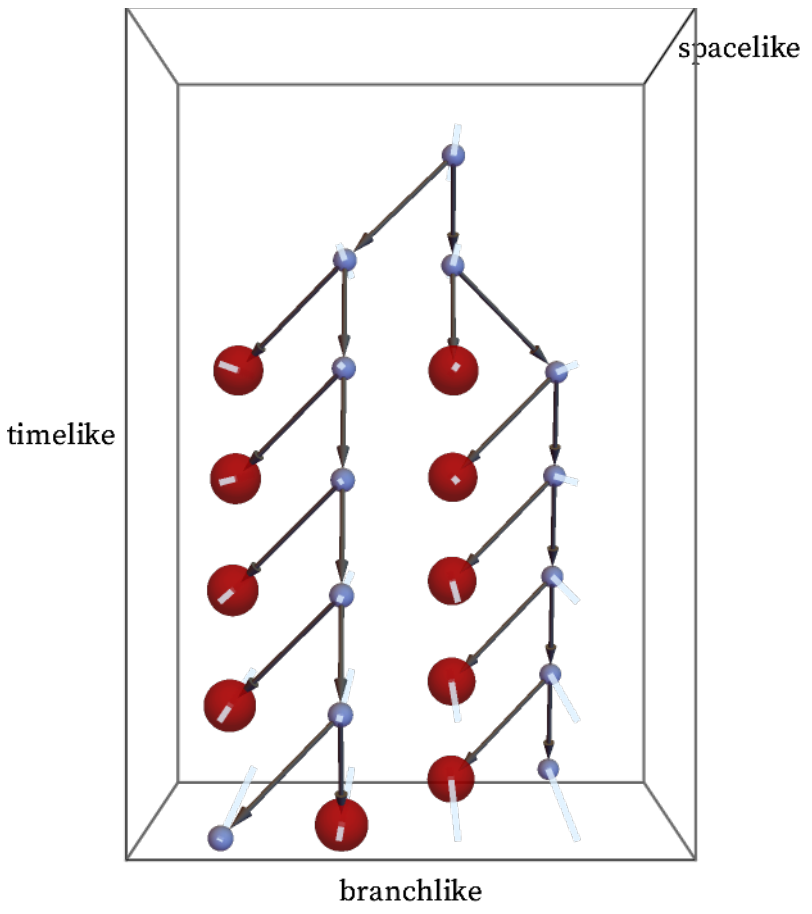

But if we now rotate this, we can see the interplay of branchial and spatial structure:

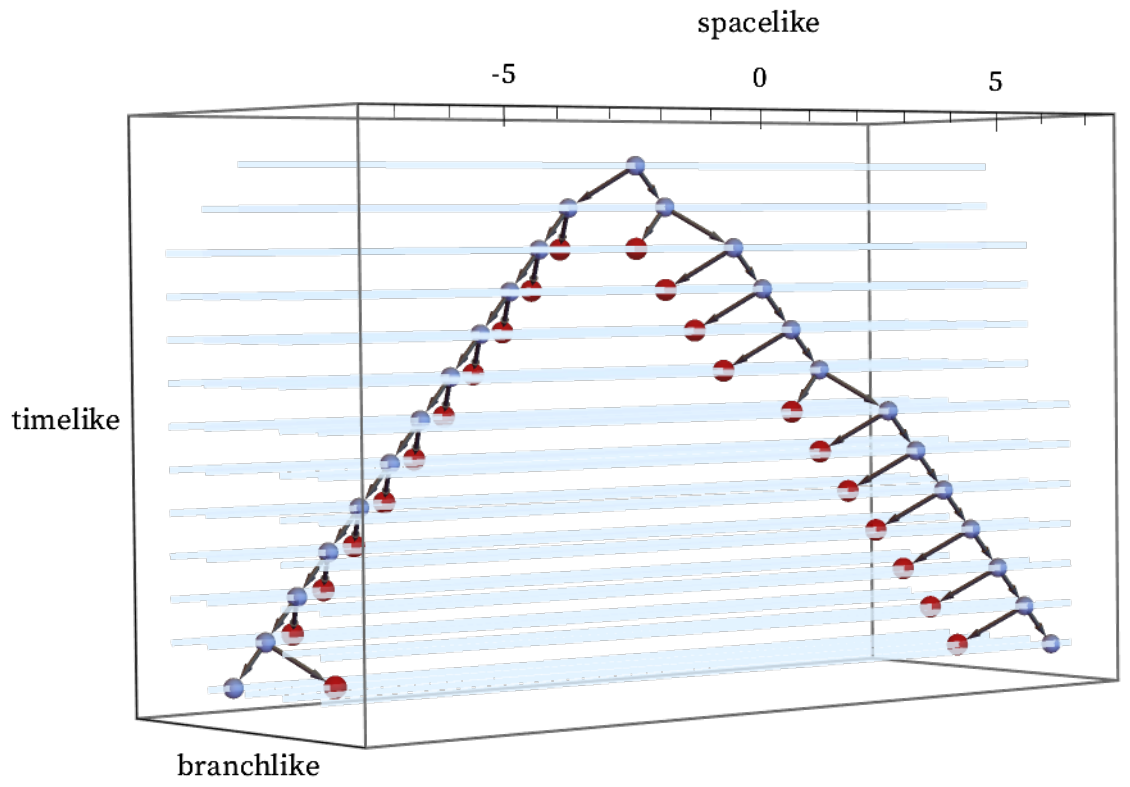

In this particular case, we see that the main branches visible in the multiway system (i.e. in the branchlike direction) correspond to progressively more separated positions of the head in the spacelike direction.

Things are not always this simple. Here's the head-position-annotated multiway graph for the Turing machine from the beginning of this section:

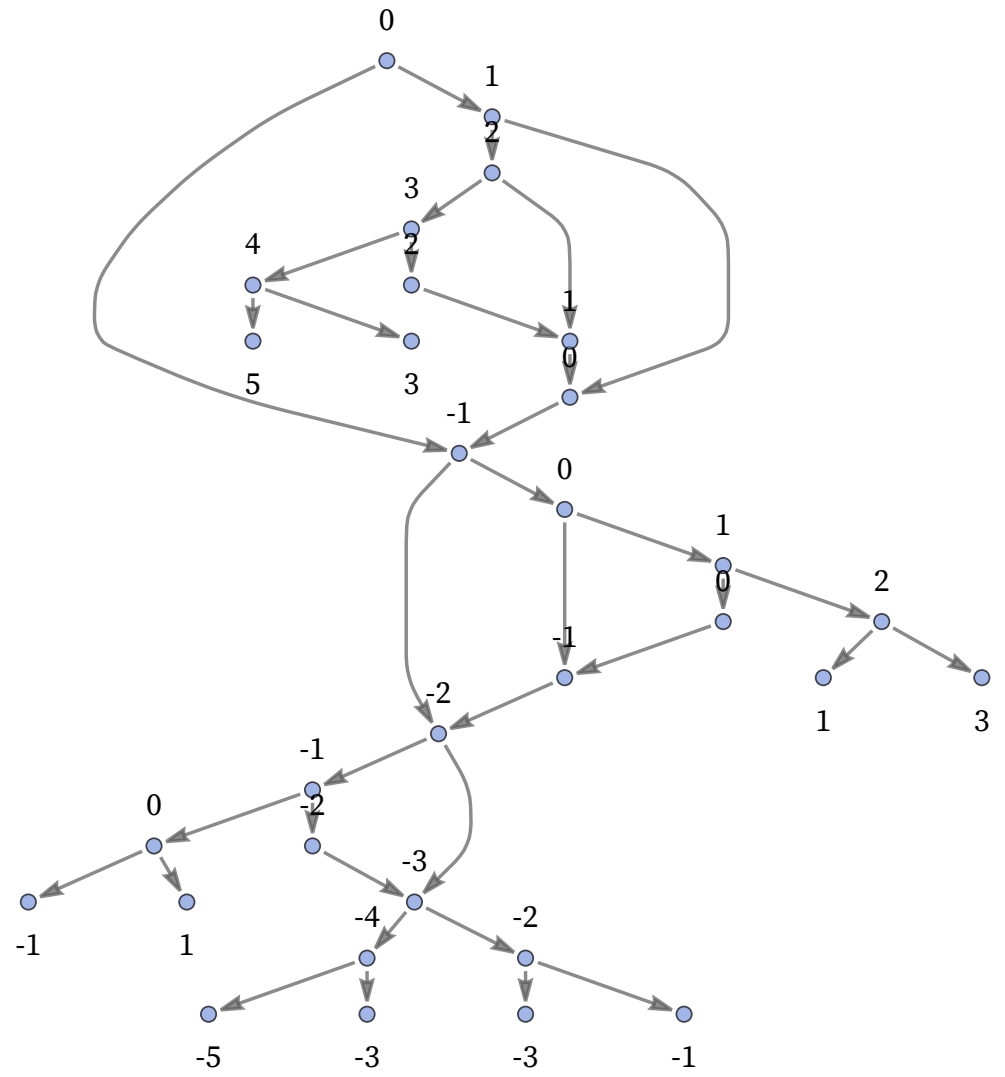

And now here's a 3D view of the evolution of this Turing machine in multispace. First we're effectively projecting into the branchtime plane. Note however that because connections can be made in 3D, the layout chosen for the rendering is different from what gets chosen when the multiway graph is rendered in 2D:

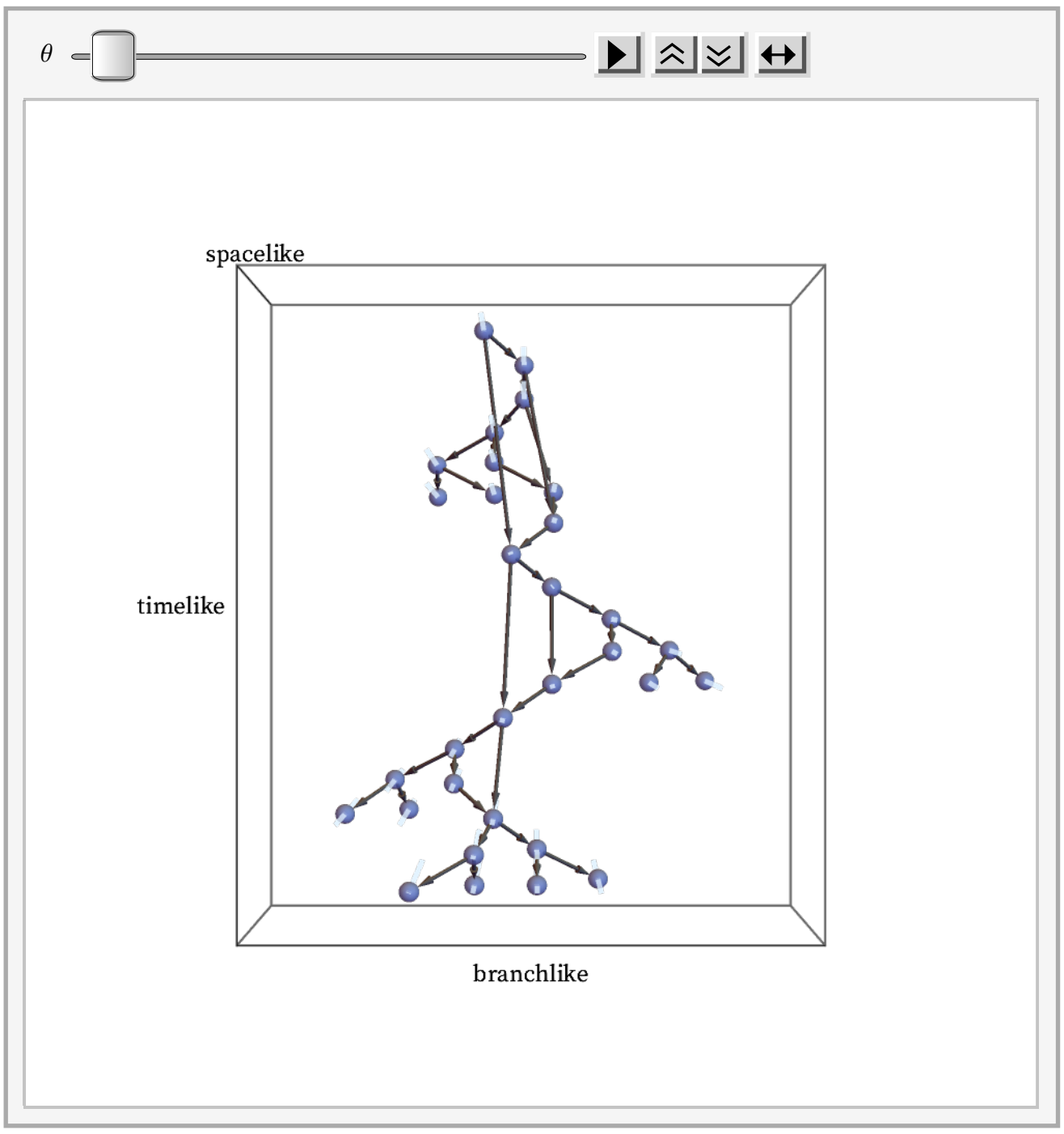

Rotating in multispace we can see the interplay between branchlike and spacelike structure:

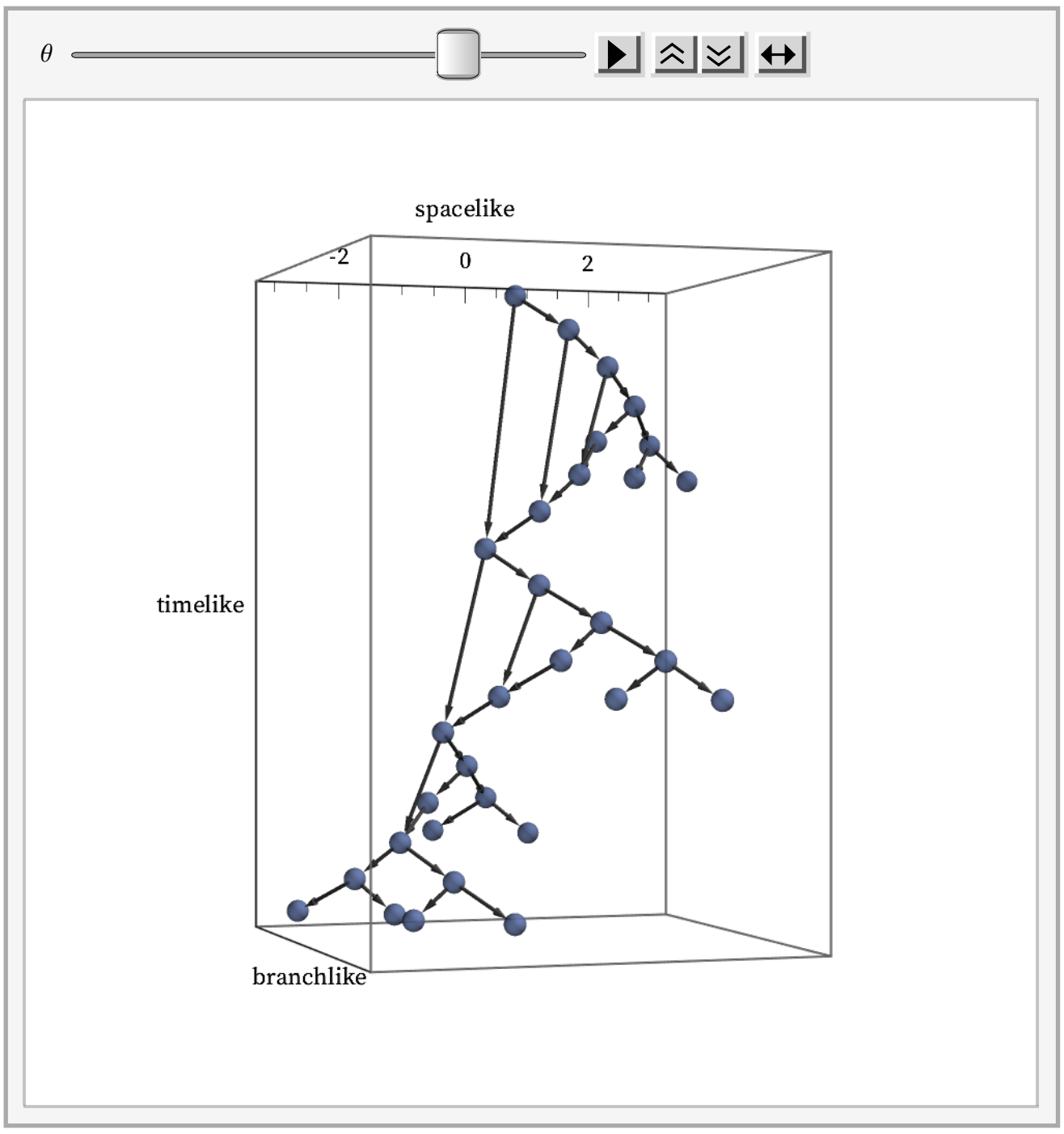

Even for this comparatively simple multiway Turing machine, it's already quite difficult to tell what's going on. But continuing for a few more steps, one can see definite patterns in multispace (the narrowing at the bottom reflects the fact that this is run only for a finite number of steps; it would fill in if one ran it for longer):

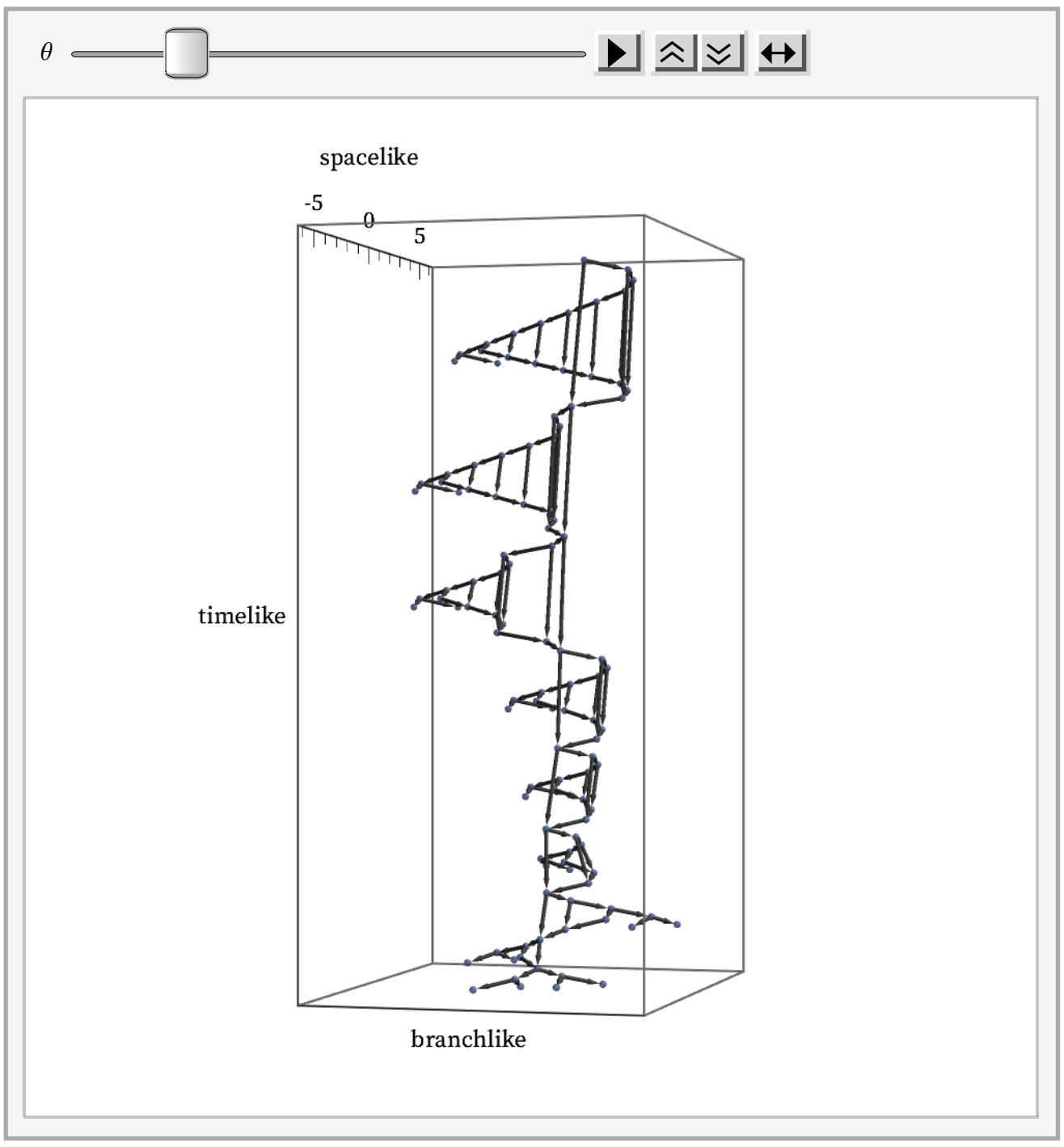

In our Physics Project, there is no intrinsic coordinatization of either physical space or branchial space—so any coordinatization must be purely emergent. But the basic structure of Turing machines implies an immediate definition of space and time coordinates—which is what we use in the pictures above to place nodes in the spacelike and timelike directions.(The timelike direction is slightly subtle: we're assigning a coordinate based on where a state appears in the layering of the multiway graph, which may or may not be the "time step" at which it first appears.) But in Turing machines there is no immediate definition of branchial coordinates—and in the pictures above we're deriving branchial coordinates from the (somewhat arbitrary) layout used in the 2D rendering of the multiway graph.

So is there a better alternative? In effect, what one has to do is to find some good way to coordinatize the complete state of the Turing machine, including the configuration of its tape. One possibility is just to treat the configuration of the tape as defining a base-k num-ber —and then for example to compute a coordinate from the log of this number. (One can also imagine shifting the number so that its "decimal" point is at the position of the head, but the head position is in a sense already "handled" through the spatial coordinate.) Here's the result from doing this for the Turing machine above:

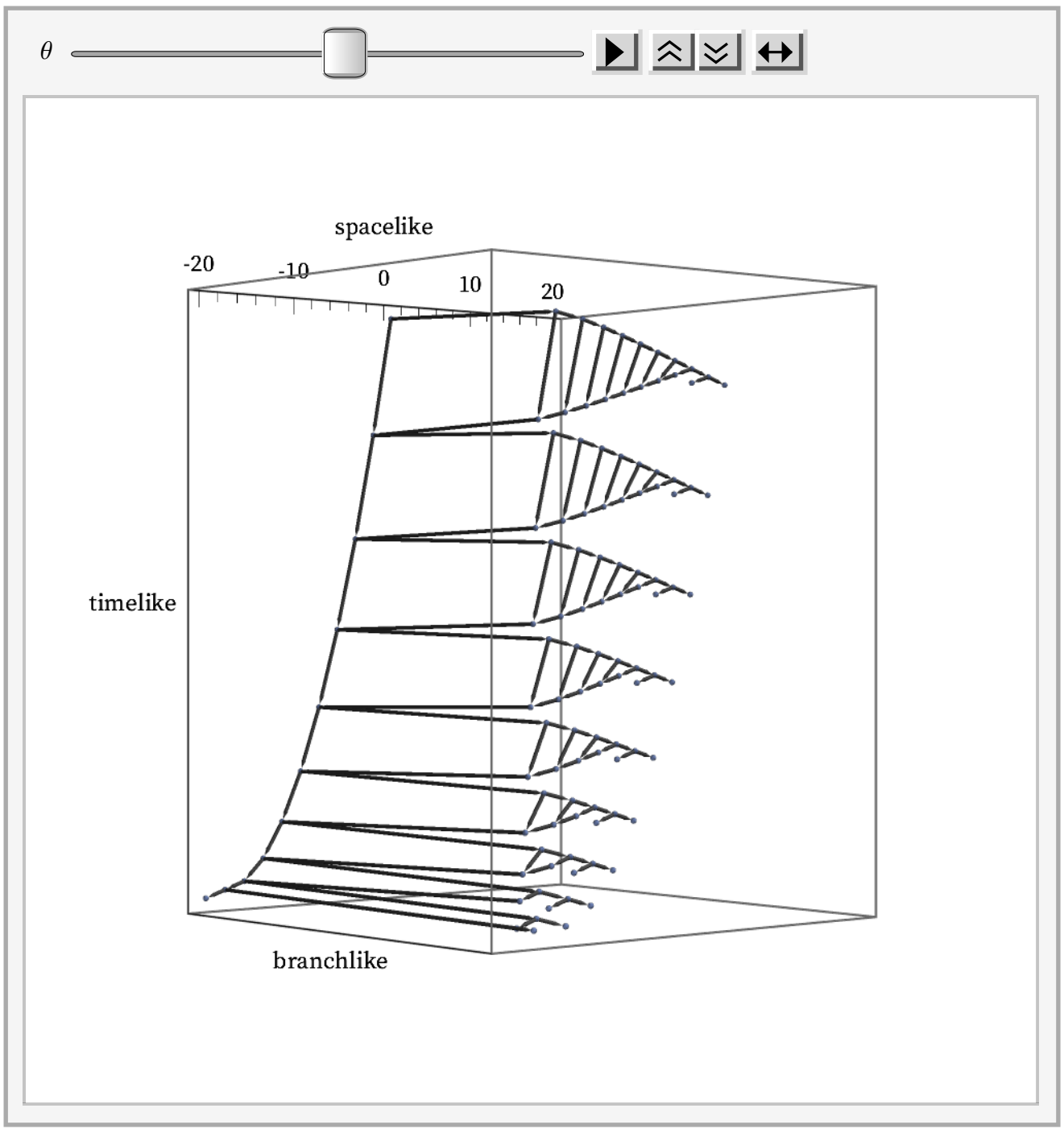

Here are a couple of other examples:

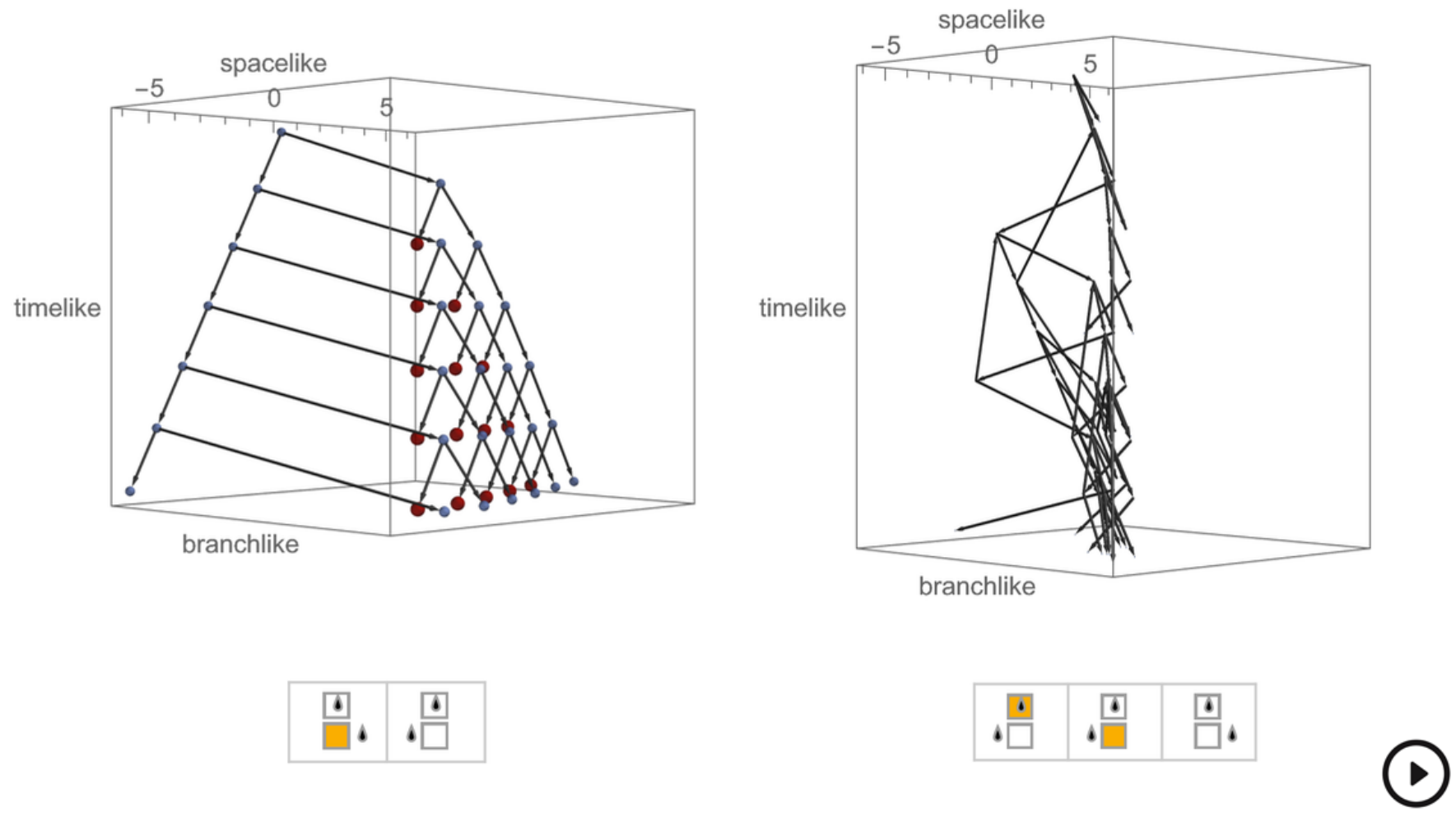

Thinking about tape configurations suggests another visualization: just stack up all tape configurations that can be generated on successive steps. In the case of the rule

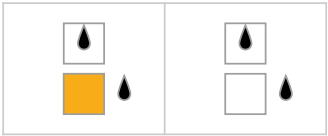

one just gets at step t the binary digits of successive integers 0 through $2^{t-1}$ on the tape:

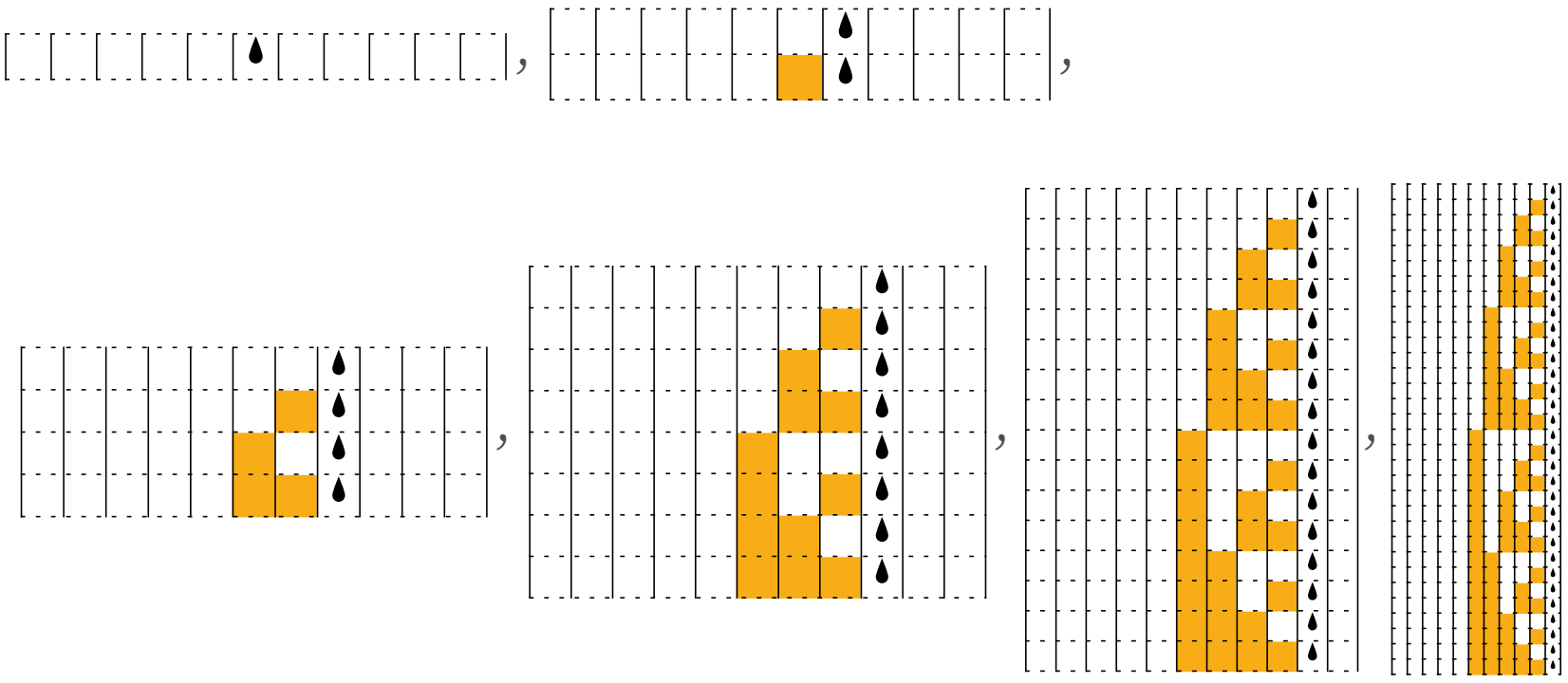

But even for a rule like

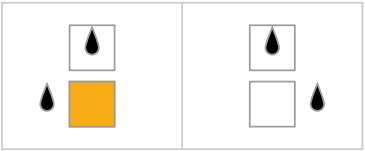

the result is more complicated

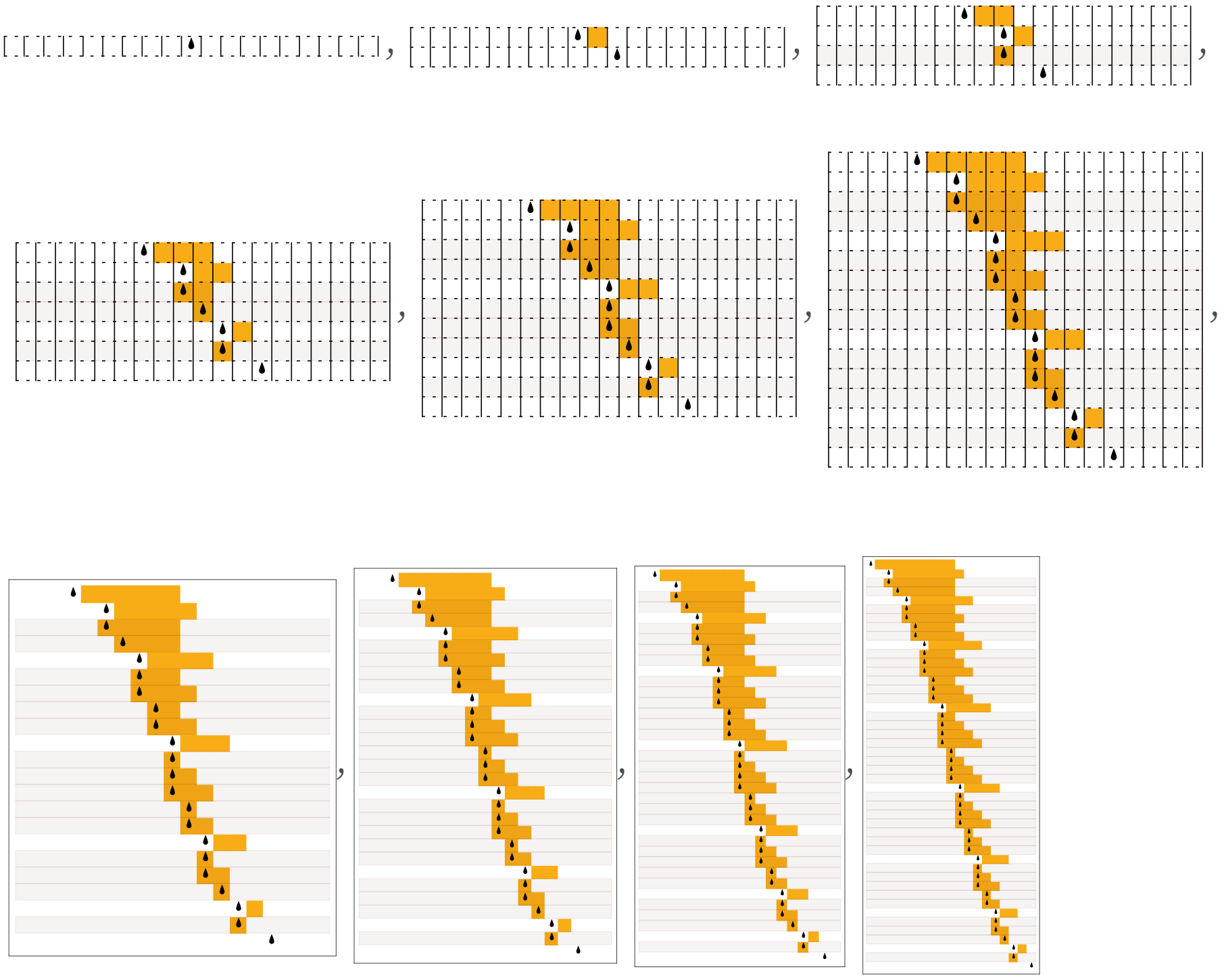

where now the grayed rows correspondto states on which the Turing machine has already halted.

When we construct a multiway graph---and render it in multispace or elsewhere---we are directly representing the evolution of states of something like a multiway Turing machine.

But particularly the study of quantum mechanics in our Physics Project emphasizes the importance of a complementary view---of looking not at the evolution of individual states, but rather at relationships (or "entanglements") between states defined by their common ancestry in the multiway graph.

Consider the rule

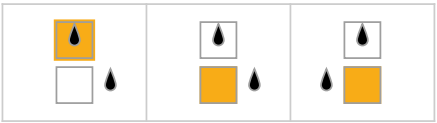

that generates a multiway graph:

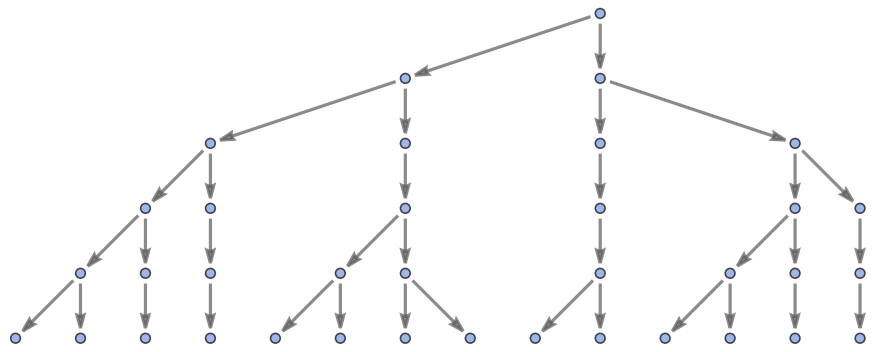

Now make a foliation of this graph:

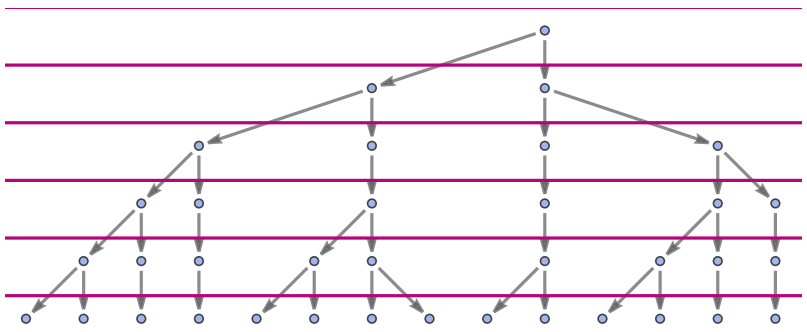

If we look at the last slice here, there are several pairs of nodes that have common ancestors—each coming from the use of different cases in the original rule. The branchial graph that joins configurations (nodes) with immediate common ancestors is, however, rather trivial:

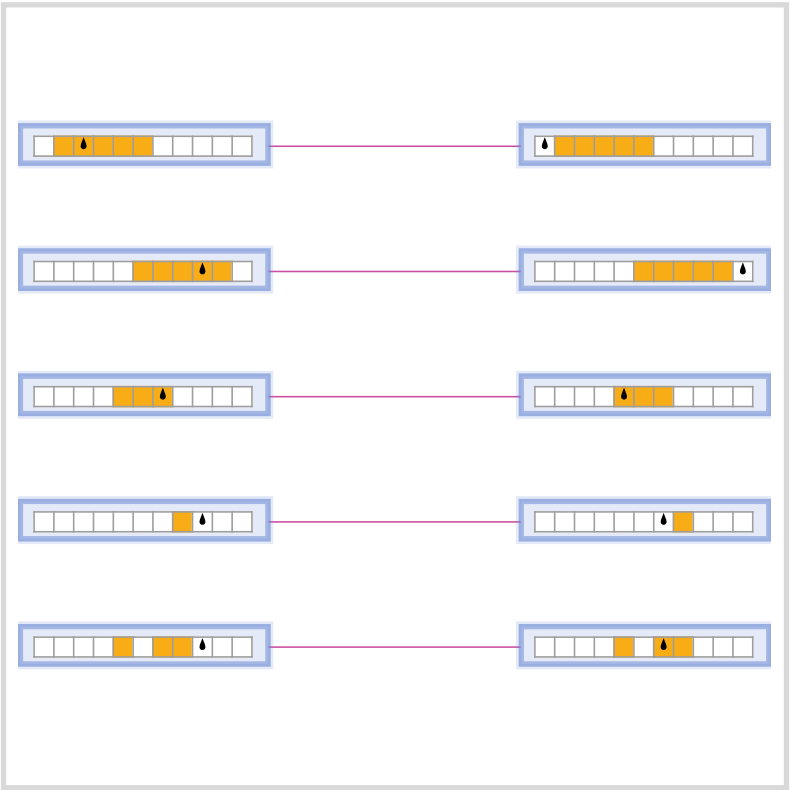

But even for the rule

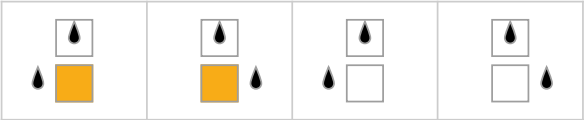

with multiway graph

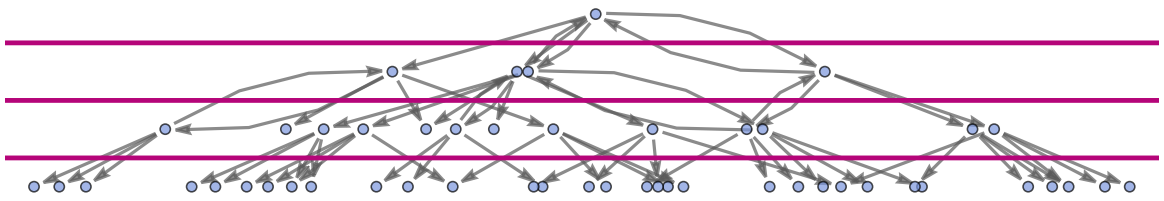

the branchial graphs on successive steps are slightly less straightforward:

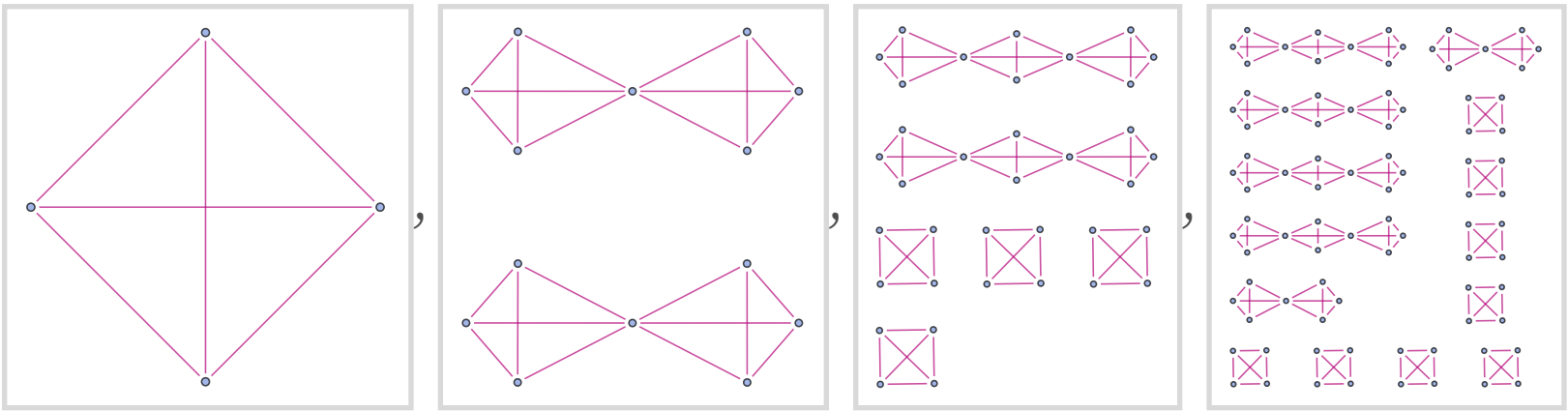

One can also construct "thickened branchial graphs" in which one looks not just at immediate common ancestors, but say at common ancestors up to $\tau$ steps back. For $\tau > 1$ the result is a hypergraph, and even in the first case here, it can be somewhat nontrivial:

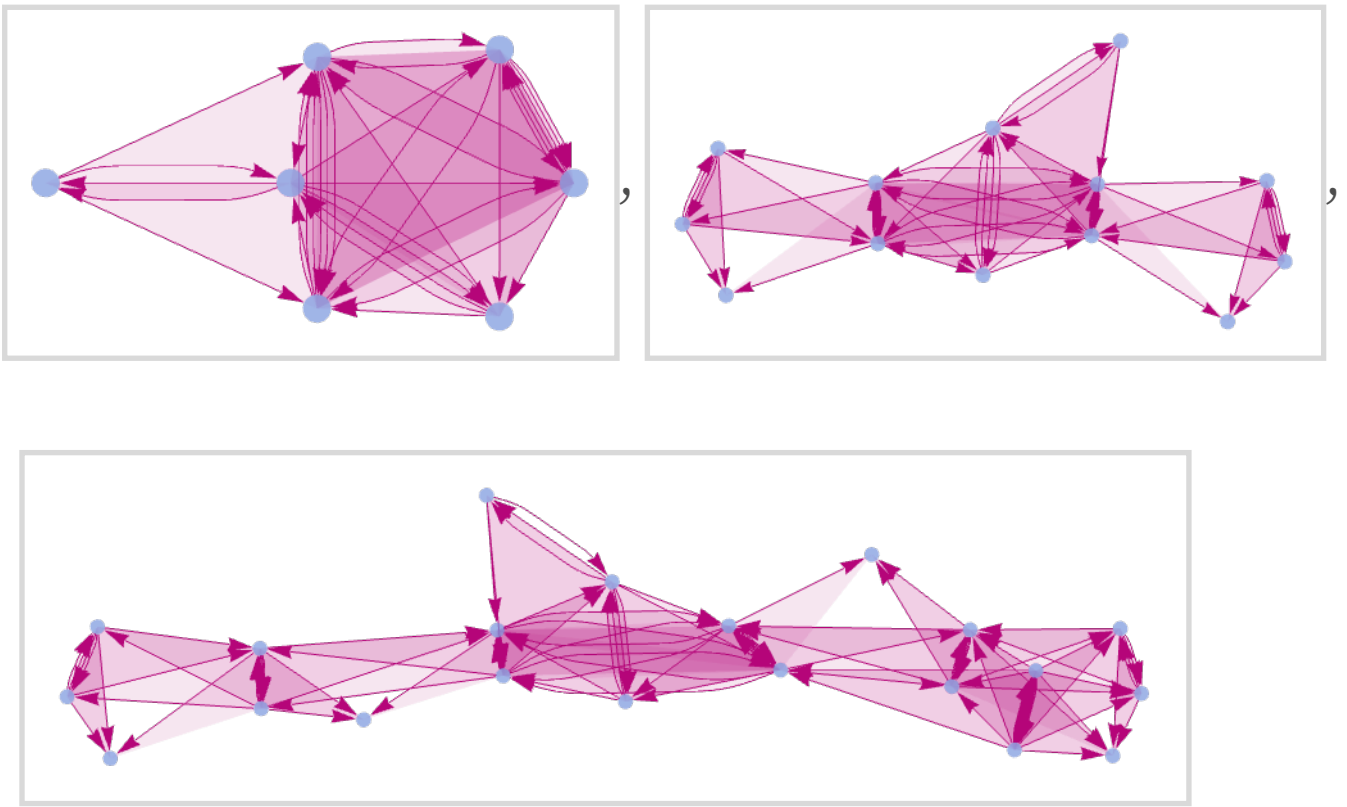

## The World of Simple Multiway Turing Machines

Unless an ordinary Turing machine has both at least s = 2 states and k = 2 colors it will always have essentially trivial behavior. But what about multiway Turing machines? With k = 1 color the system in effect cannot store anything on the tape, and it must always "do the same dance" at whatever position the head is in.

Consider for example the s = 2 rule:

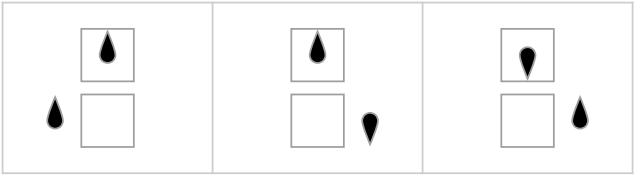

The states of this system can be represented in the form {$\sigma$, i} where $\sigma$ is the head state, and i is the position of the head. The multiway graph for the evolution of the system after 3 steps starting from a blank tape with the head in head state 1 can then be drawn as (with the coordinates of each state being determined from {$\sigma$, i}):

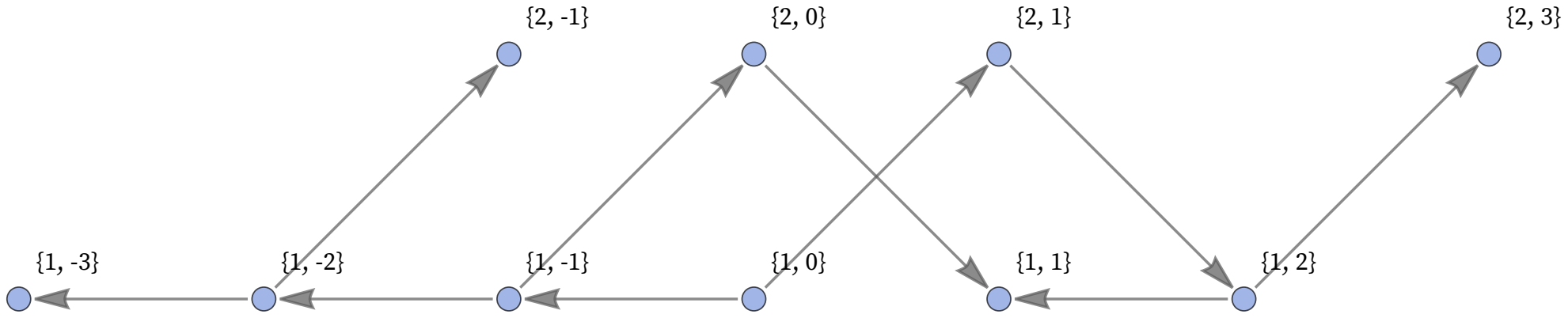

Continuing longer we get

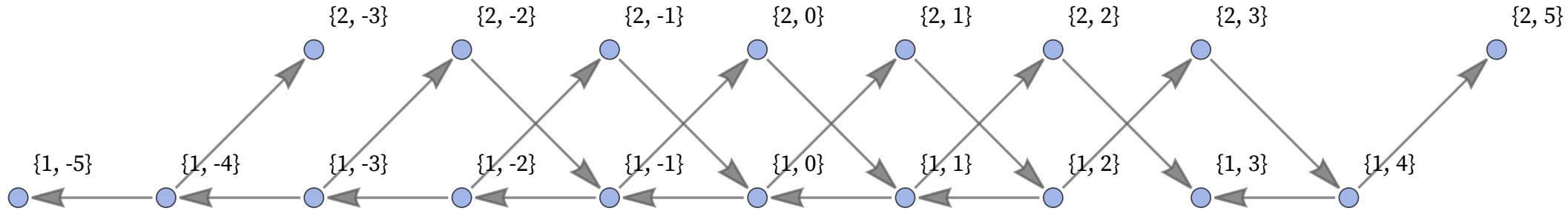

and we can see that eventually the multiway graph is effectively a repeating braid—though with "slightly frayed" ends at any given step. The braid can be constructed from a sequence of "motifs", here consisting of 3 "vectors", where each vector is directly determined by a case in the underlying rule, or in this case:

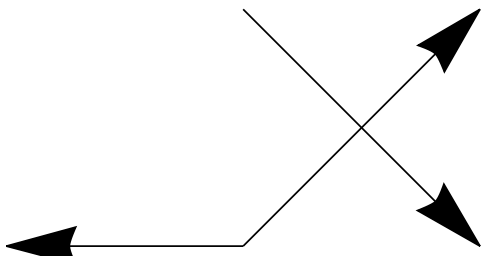

If we look at all s = 2 multiway Turing machines with 3 cases in their rules, these are the multiway graph structures we get:

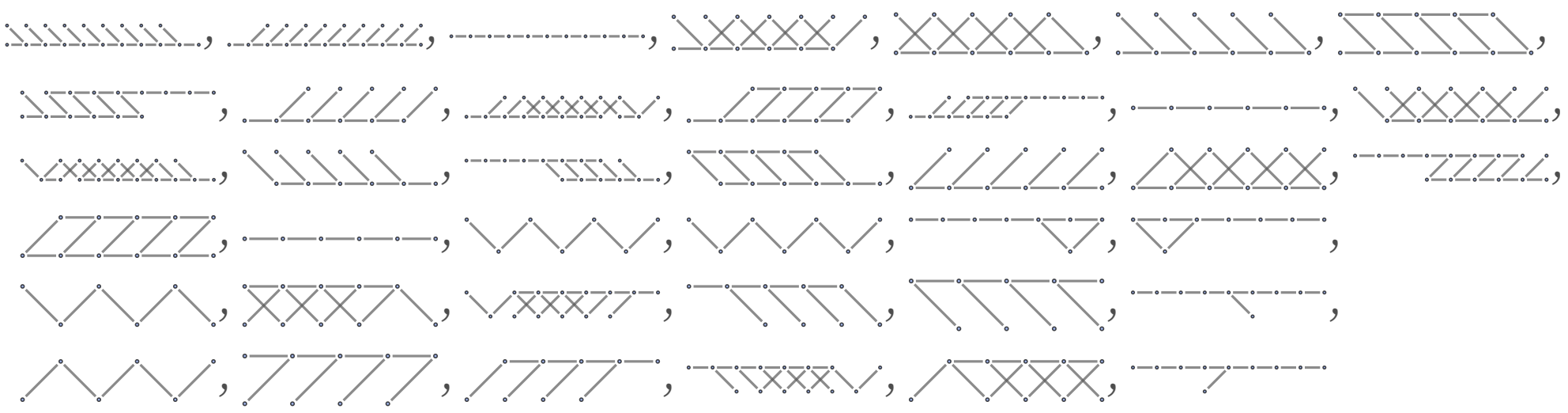

In all cases, the effective repetition length for any path is 1, 2 or 4.

With s = 3 possible head states, slightly more elaborate multiway graphs become possible. For example

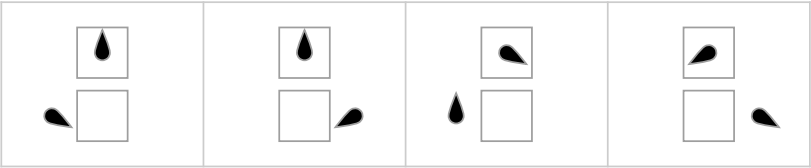

gives the “8-cycle braid”:

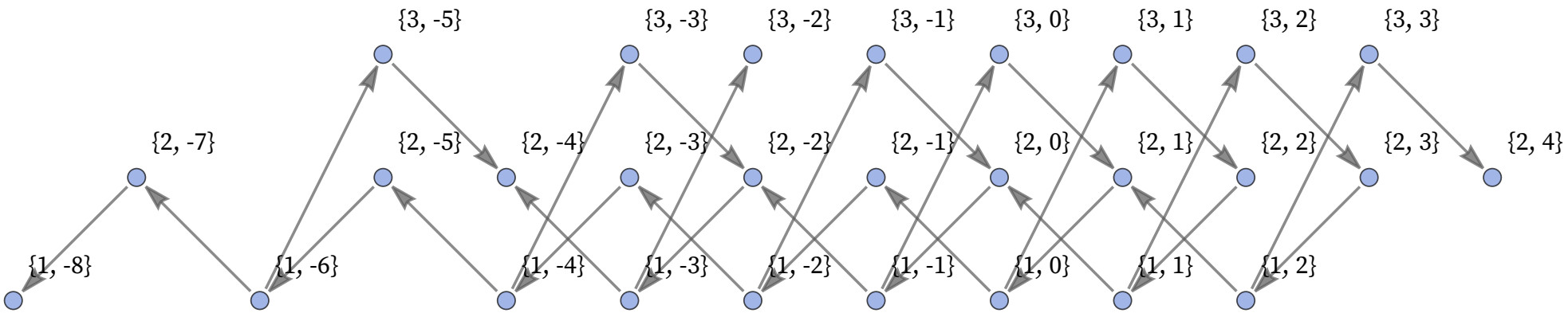

To get beyond “pure-braid” multiway graphs, we must consider k ≥ 2 colors. But what happens if we use just one head state s = 1 (so we basically have a “mobile automaton” with rules of range 0)?

For s = 1, k = 2, there are 8 possible cases that can occur in the rule

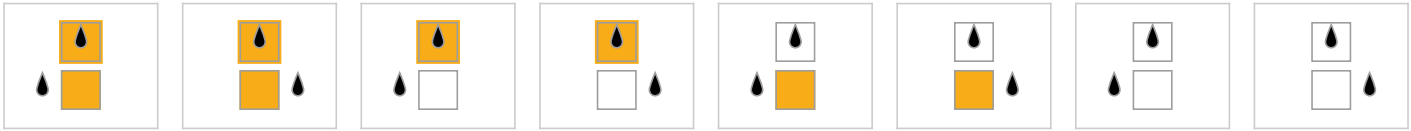

and thus 256 possible multiway Turing machines, of which 231 are not “purely deterministic”.

If we allow only p = 2 possible cases in the rule, there are 28 possible machines, of which 12 are not purely deterministic, and the distinct possible nontrivial forms of multiway graphs that we can get are just (where in the first rule the second color is not used):

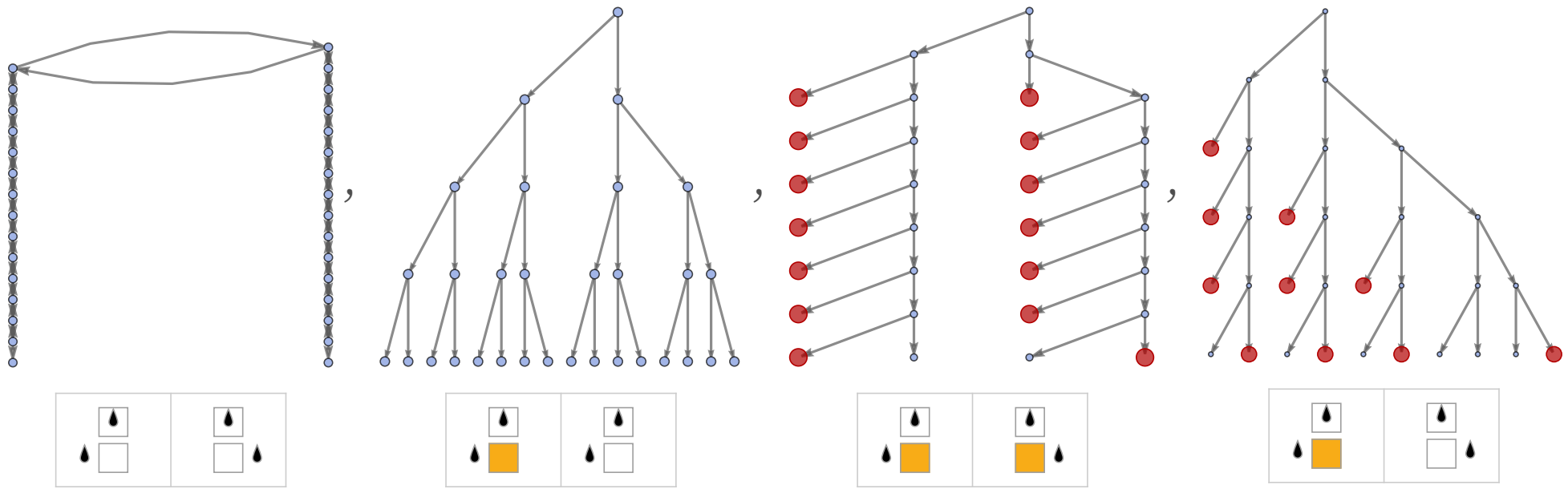

With p = 3 possible cases in the rule, there are 56 possible machines, none purely deterministic. The distinct multiway graphs they generate are:

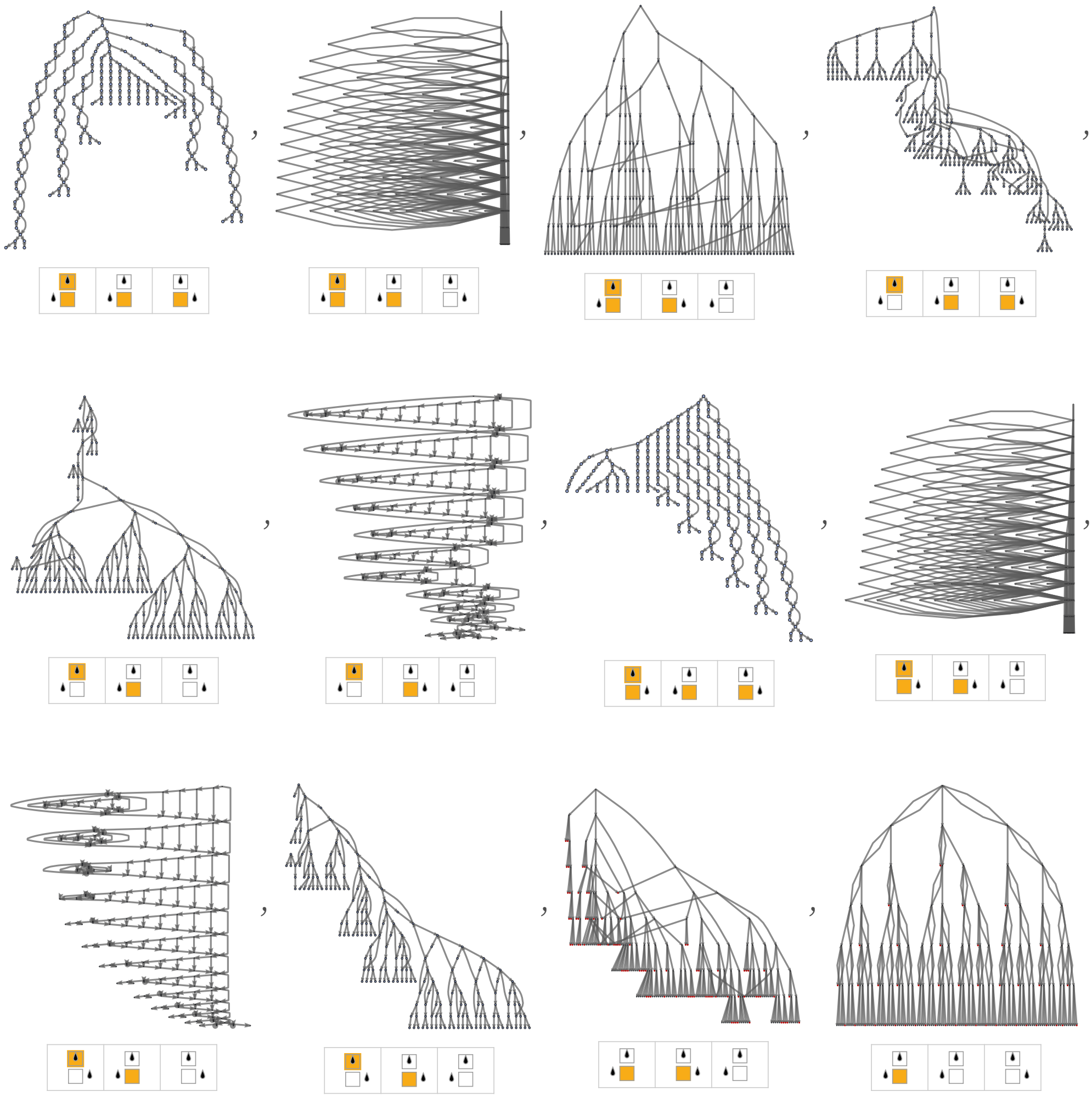

Some of these we have already seen above. But it is remarkable how complex these graphs appear to be.

Laying the graphs out in multispace, however, some show very clear regularities:

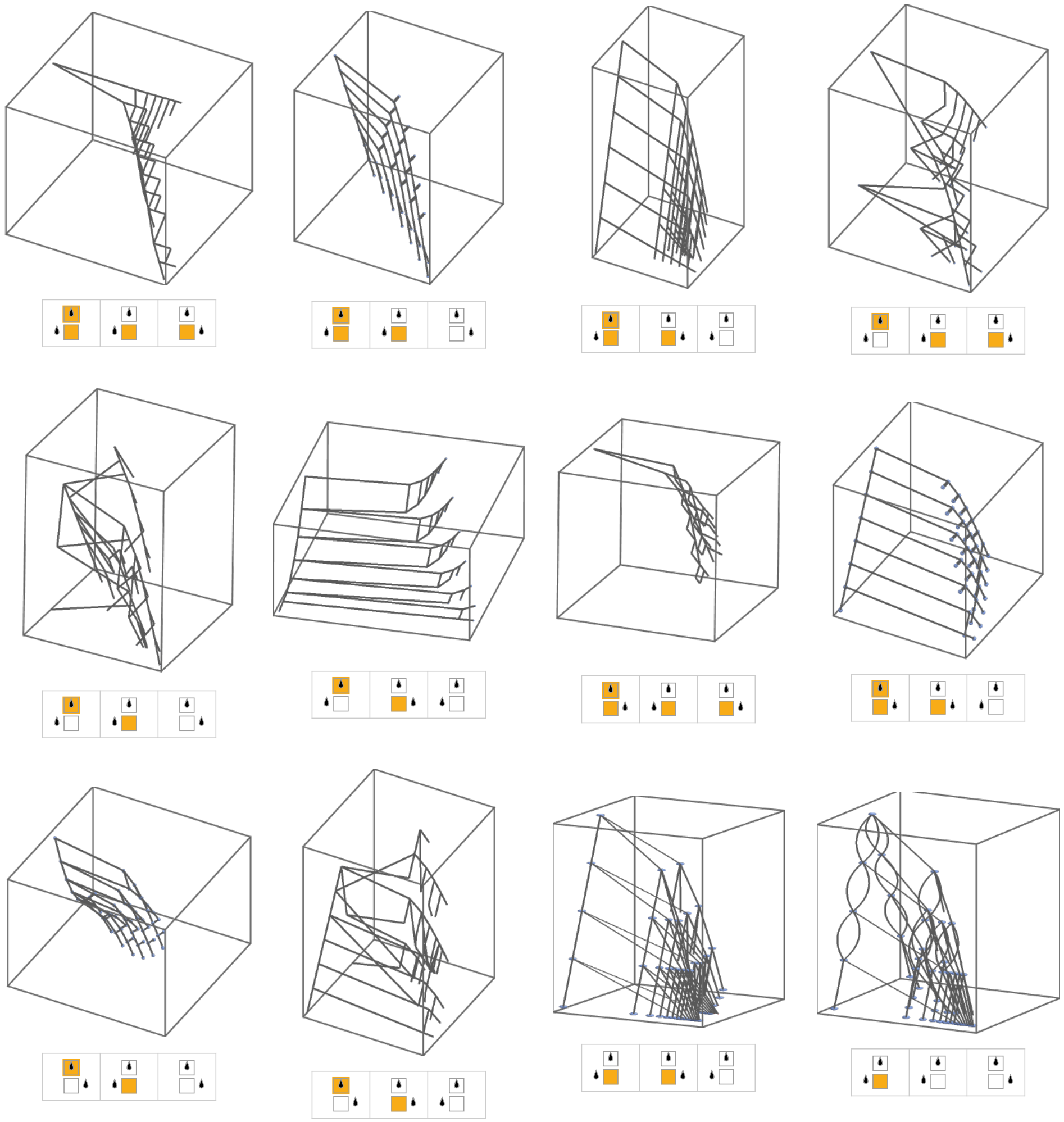

If we look at the complicated cases for more steps, we still see considerable complexity:

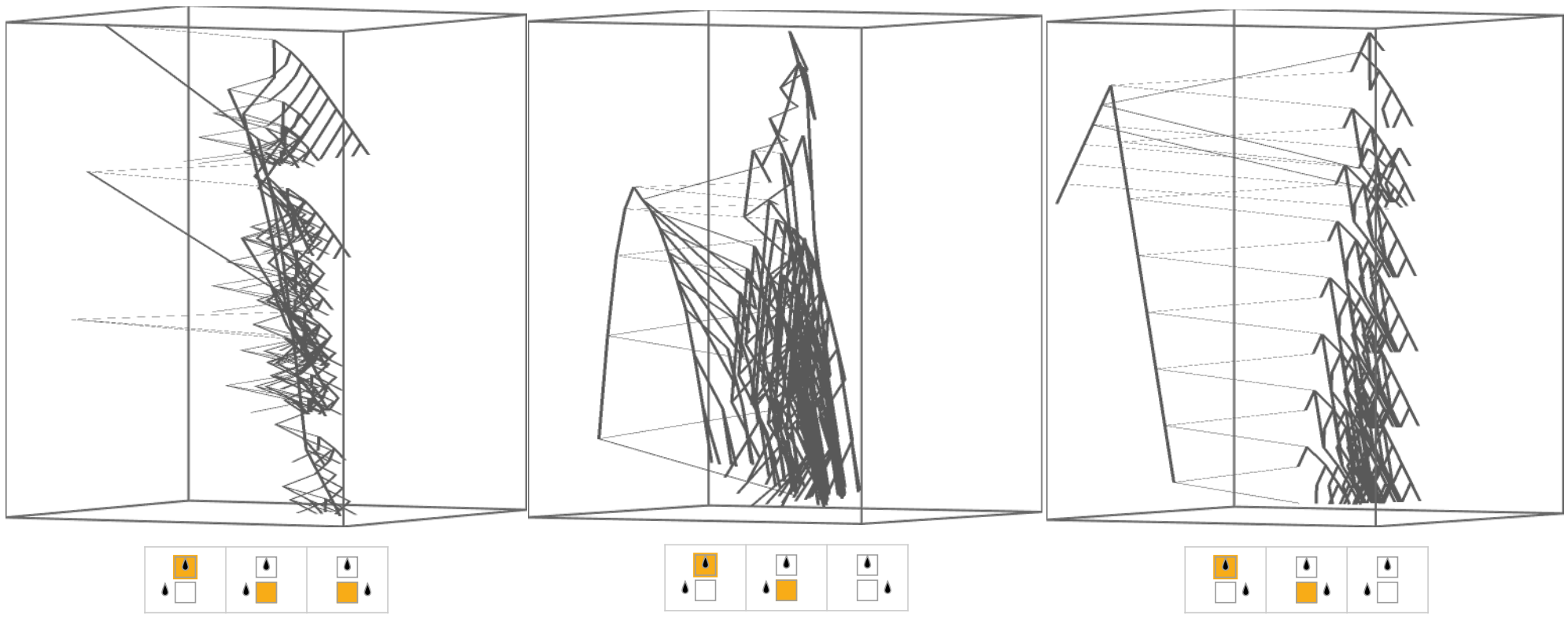

The number of states reached at successive steps for the first rule is

{1, 3, 7, 13, 22, 36, 56, 86, 131, 195, 288, 425, 623, 912, 1335, 1949}

while for the other two it is:

{1, 3, 7, 14, 24, 39, 61, 93, 140, 209, 310, 458, 675, 993, 1459, 2142}

Both of these appear to grow exponentially, but with slightly different bases, both near 1.46.

Looking at the collections of states generated also does not give immediate insight into the behavior:

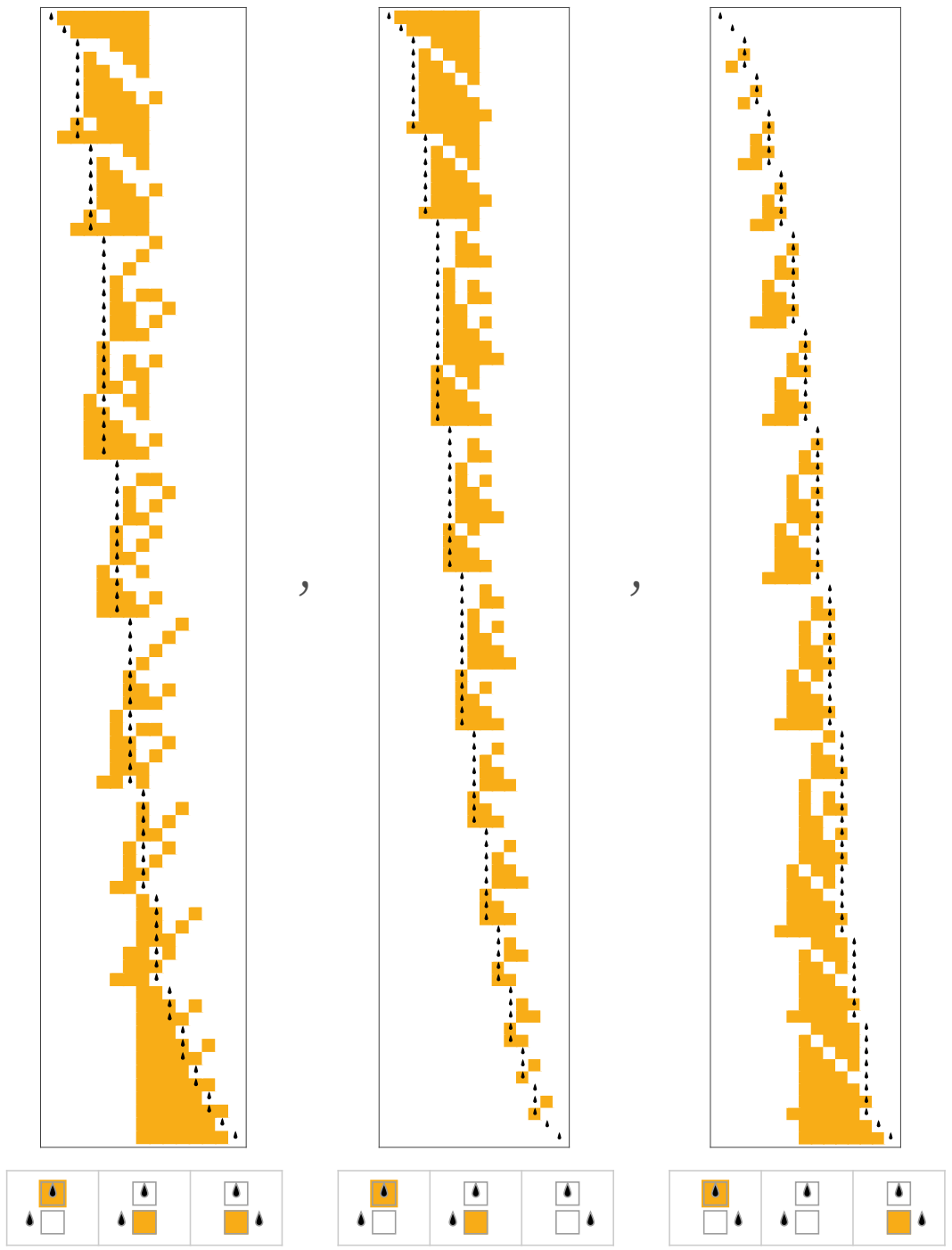

So what happens if we allow p > 3 cases in the rule? With p = 4 cases, there are 70 possible machines, none purely deterministic. Here are some examples of the more complicated behavior that is seen:

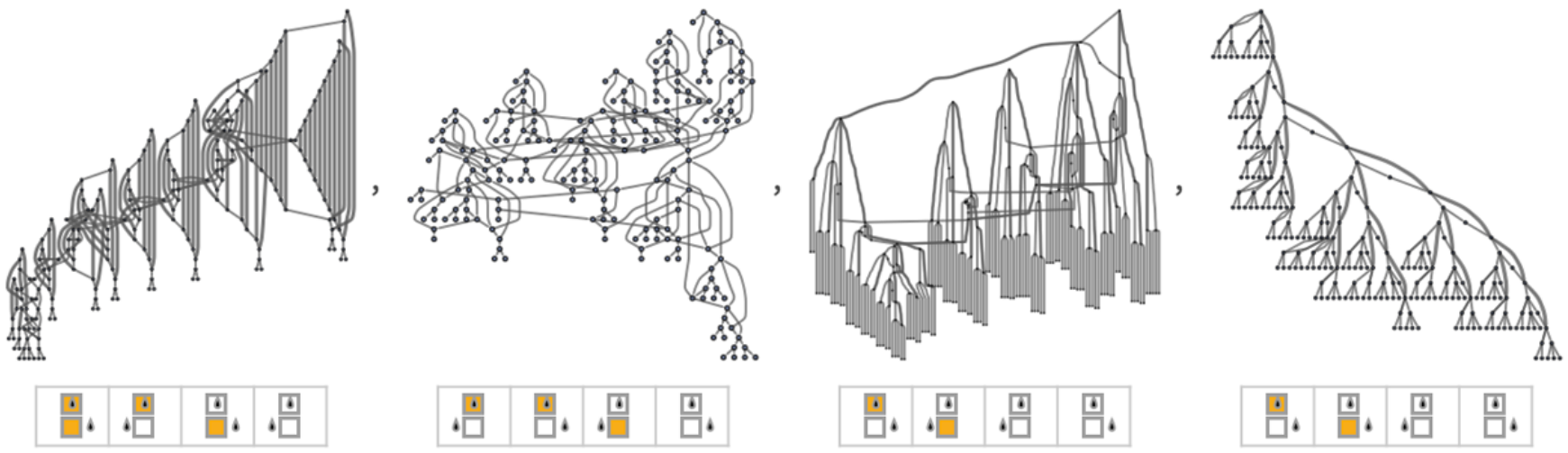

We can go on and look at machines with larger p. Here are examples with p = 5:

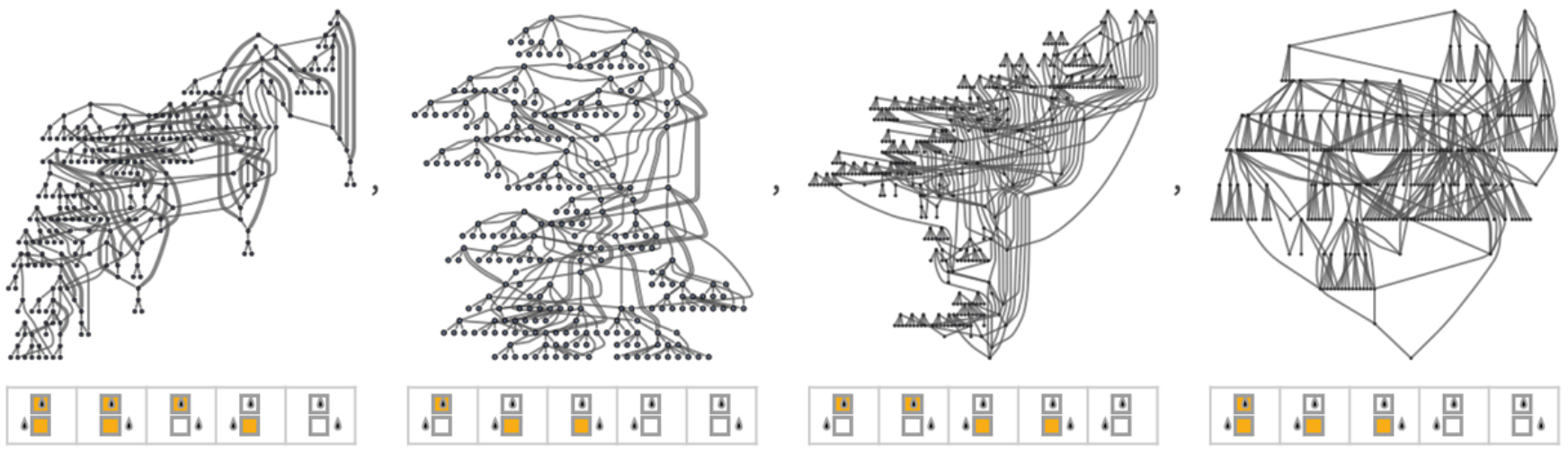

We can also ask what happens if we allow s = 2 instead of s = 1, with k = 2. With p = 3 cases in the rule, the most complicated multiway graphs are just:

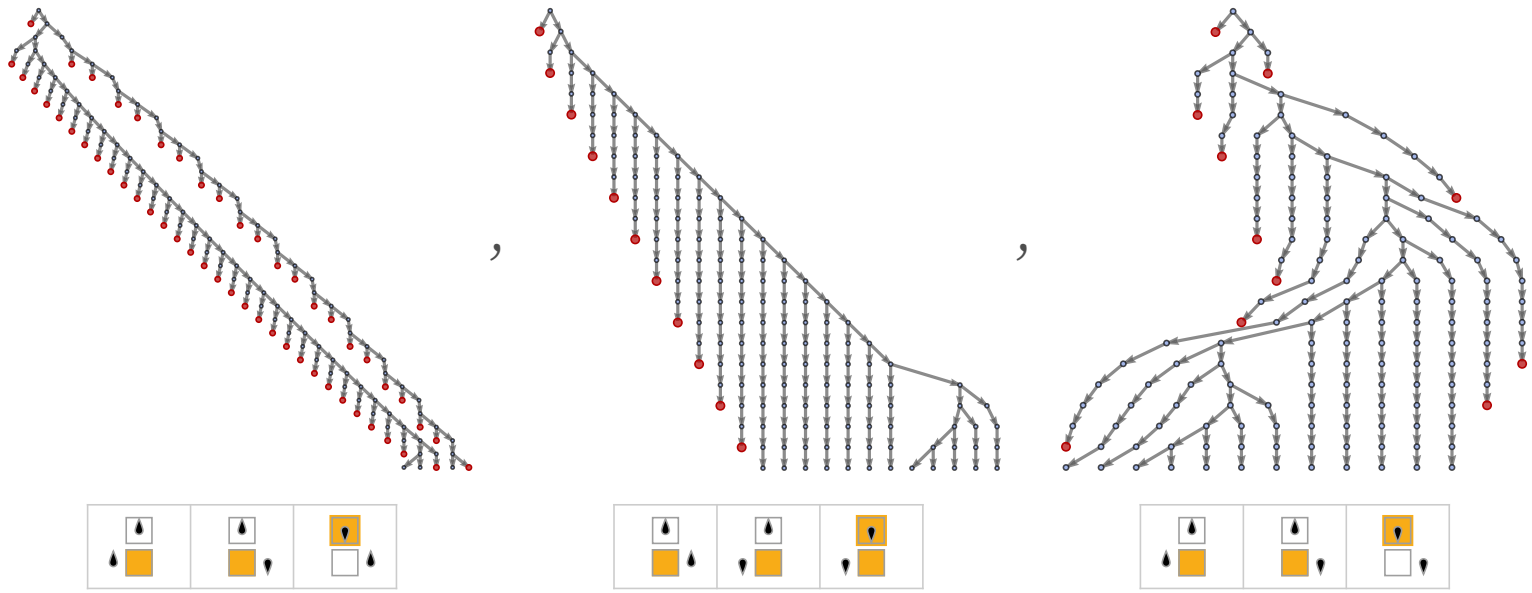

With p = 4 cases in the rule, the behavior can be slightly more complicated, but does not appear to go beyond what we already saw with a single head state s = 1:

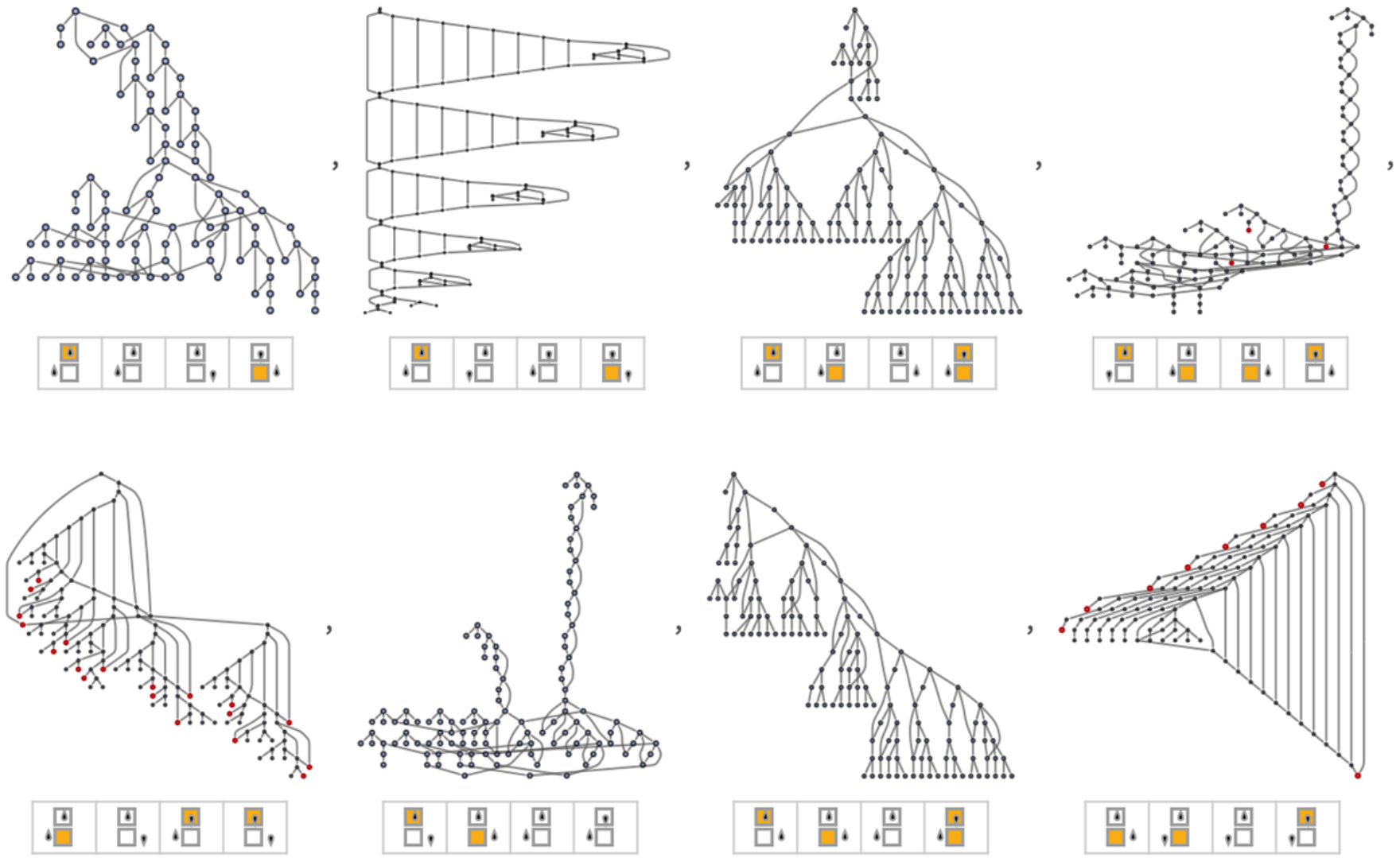

## What Is the Simplest Multiway Turing Machine?

We might have assumed that to get complicated behavior in a multiway Turing machine it would need complicated rules, and that as the rules got more complicated the behavior would somehow get correspondingly more complicated. But that's not what we saw in the previous section. Instead, even multiway Turing machines with very simple rules were already capable of apparently complex behavior, and the behavior didn't seem to get fundamentally more complicated as the rules got more complicated.

This might seem surprising. But it's actually very much the same as we've seen all over the computational universe—and indeed it's just what my Principle of Computational Equivalence implies should happen. Because what the principle says is that as soon as one goes beyond systems with obviously simple behavior, one gets to systems that are equivalent in the sophistication of the computations they perform. Or, in other words, that above a very low threshold, one sees systems that all achieve a maximal level of computational sophistication that is always the same.

The Principle of Computational Equivalence talks about "typical computations" that a system does, and implies, for example, that these computations will tend to be computationally irreducible, in the sense that one cannot determine their outcomes except by computations that are essentially just as long as the ones the system itself is performing. But another consequence of the Principle of Computational Equivalence is that as soon as the behavior of a system is not obviously simple, it is likely to be computation universal. Or, in other words, once a system can exhibit complicated behavior, it is essentially always possible to find a particular initial condition that will make it emulate any other system—at least up to a translation given by some kind of bounded computation.

And we have good evidence that this is how things work in simple cellular automata, and in ordinary, deterministic Turing machines. With $s = 2$, $k = 2$ such Turing machines always show simple, highly regular behavior, and no such machines are capable of universal computation. But with $s = 2$, $k = 3$ there are a few machines that show more complex behavior. And we know that these machines are indeed computation universal, establishing that among ordinary Turing machines the threshold of universality is in a sense as low as it could be.

So what about multiway Turing machines? Where might the threshold for universality for such machines be? Based on our experience with ordinary Turing machines (and many other kinds of systems) the results in the previous section might tend to suggest that $s = 1$, $k = 2$, $p = 3$ would already be sufficient. With ordinary Turing machines, universality is achieved when $s = 2$, $k = 3$, corresponding to 6 cases in the rule. Here universality seems possible even with just 3 cases in the rule—and with $s = 1$, $k = 2$.

Perhaps the most surprising part of this is the implication that universality might be possible even with just a single head state $s = 1$. And this implies that we do not even need to have a Turing machine with any head states at all; we can just use a mobile automaton, and actually one with "radius 0" (i.e. with rules that are not directly affected by any neighbors)---that we can indicate for example as:

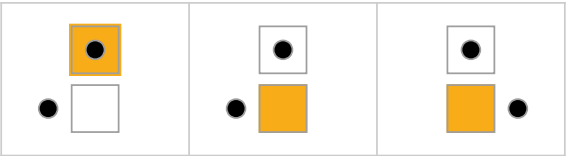

Turing machines are often viewed as extensions of finite automata, in which a tape is added to provide potentially infinite memory. Ordinary deterministic Turing machines are then extensions of deterministic finite automata (DFAs), and multiway Turing machines of nondeterministic finite automata (NDFAs). And finite automata with just a single state s = 1 are always trivial—whether they are deterministic or nondeterministic. We also know that ordinary deterministic Turing machines with s = 1 are always trivial, regardless of how many tape colors k they have; s = 2 is the minimum to achieve nontrivial behavior (and indeed we know that s = 2, k = 3 is sufficient to achieve in a sense maximally complex behavior).

But what we saw above is that multiway Turing machines can show nontrivial behavior even when s = 1. Quite likely there is a direct construction that can show how an s = 1 multiway Turing machine can emulate an s > 1 one. But here is some intuition about why this might be possible. In an ordinary Turing machine, the head moves "in space" along the tape, and the head state carries state information with it. But in a multiway Turing machine, there is in effect motion not only in space, but also in branchial space. And even though without multiple head states the head may not be able to carry information in ordinary space, the collection of nearby configurations in branchial space can potentially carry information in branchial space. In other words, in branchial space one can potentially think of there being a "cloud" of nearby configurations that represent some kind of "super head" that can in effect assume multiple possible head states.

Crucial to this picture is the fact that multiway systems are set up to merge identical states. If it were not for this merging, the multiway graph for a multiway Turing machine would just be a tree. But the merging "knits together" branchial space, and allows one to make sense of concepts like distance and motion in it.

In an ordinary Turing machine, one can think of the evolution as progressively transforming the state of the system. In a multiway Turing machine it may be better to think of the evolution as knitting together elementary pieces of the multiway graph---with the result that it matters less what the individual states of the Turing machine are like, and more how different states are related in branchial space, or in the multiway system.

But what exactly do we mean by universality anyway? In an ordinary deterministic system the basic definition is that a universal system—if given appropriate initial conditions—must be able to emulate any other computational system (say any Turing machine). Inevitably, there will be encoding and decoding required. Given a specification of the system one is trying to emulate (say, the rules for a Turing machine), one must have a procedure for setting up appropriate initial conditions—or, in effect, for “compiling” the rules to the appropriate “code” for the universal machine. Then when one runs the universal machine, one must have a way of “decoding” its evolution to identify the steps in the evolution of the system one is emulating.

For the idea of universality to be meaningful, it’s important of course that the processes of encoding and decoding don’t do too much computation. In particular, the amount of computation they need to do must be bounded: so even if the “main computation” is computationally irreducible and needs to run progressively longer, these do not. Or, put another way, any feature of the encoding and decoding should be decidable, whereas the main computation can show computational irreducibility, and undecidability.

We have talked here about universality being a story of one machine emulating another. And ultimately this is what it is about. But in traditional computation theory (which was developed before actual digital computers were commonplace) there are typically two additional features of the setup. First, one often thinks about the Turing machine as just “computing a function”, so that when fed certain input it gives certain output—and we don’t ask “what it’s doing inside”. And second, one thinks of the Turing machine as “halting” when it has produced its answer. But particularly when one wants to get intuition about computational processes, it’s crucial to be able to “see the process”. And in keeping with the opera-tion of modern digital computers, one is less concerning with "halting", and more just with knowing how decode the behavior (say by looking for “signals” that indicate that something should be “considered to be a result”).

(Even beyond this, Turing machines are sometimes thought of purely as “decision machines”: given a particular input they eventually give the result “true” or “false”, rather than generating different forms of result.)

In studying multiway Turing machines I think the best way to define universality is the one that is most directly analogous to what I’ve used for ordinary Turing machines: the universal machine must be able to emulate the multiway graph for any other machine. In other words, given a universal multiway Turing machine there must be a way to set up the initial conditions so that the multiway graph it generates can be “decoded” to be the multiway graph for any other system (or, in particular, for any other multiway Turing machine).

It is worth realizing that the “initial conditions” here may not just be a single Turing machine configuration; they can be an ensemble of configurations. In an ordinary determin-istic Turing machine one gives initial conditions which are “spread out in space”, in the sense that they specify colors of cells at different positions on the tape. In a multiway Turing machine one can also give initial conditions which are “spread out in branchial space”, i.e. involve different configurations that will initiate different branches (which might merge) in the multiway system. In an ordinary Turing machine—in physics terms—it is as if one specifies one’s initial data on a “spacelike hypersurface”. In a multiway Turing machine, one also specifies one’s initial data on a “branchlike hypersurface”.

By the way, needless to say, a universal deterministic Turing machine can always emulate a multiway Turing machine (which is why we can run multiway Turing machines on practical computers). But at least in the most obvious approach it can potentially require exponentially many steps of the ordinary Turing machine to follow all the branches of the multiway graph for the multiway Turing machine.

But what is the multiway analog of an ordinary Turing machine "computing a function"? The most direct possibility is that a multiway Turing machine defines a mapping from some set of configurations to some other set. (There is some subtlety to this, associated with the foliation of the multiway graph—and specifying just what set of configurations should be considered to be "the result".) But then we just require that with appropriate encoding and decoding a universal multiway Turing machine should be able to compute any function that any multiway Turing machine can compute.

There is another possibility, however, suggested by the term "nondeterministic Turing machine": require just that there exists some branch of the multiway system that has the output one wants. For example, if one is trying to determine the answer to a decision problem, just see if there is any "yes" branch, and, if so, conclude that the answer is yes.

As a simple analog, consider the string substitution with transformations {"()"→"","()"→"|"}. One can compute whether a given sequence of parentheses is balanced by seeing whether its evolution will eventually generate an empty string ("") state—but even if this state is generated, all sorts of "irrelevant" states will also be generated:

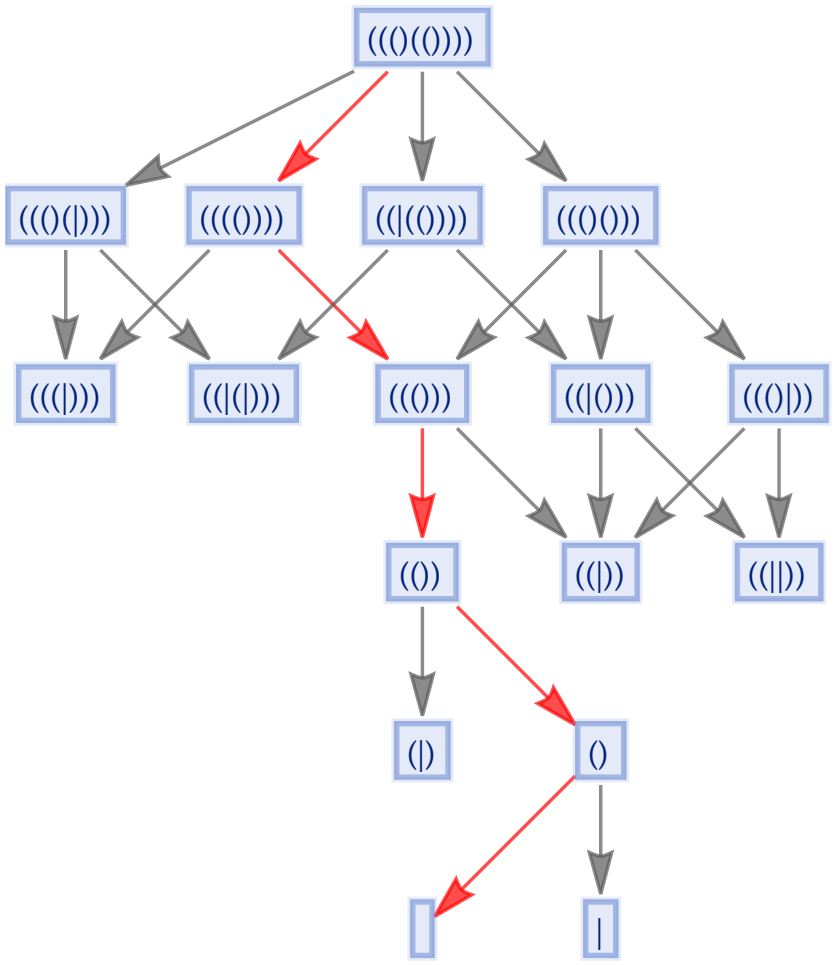

There is presumably always a way to translate the concept of universality from our "emulate the whole multiway evolution" to the "see if there's any path that leads to some specific configuration". But for the purposes of intuition and empirical investigation, it seems much better to look at the whole evolution, which can for example be visualized as a graph.

So what is the simplest universal multiway Turing machine? I am not certain, but I think it quite likely that it has s = 1, k = 2, p = 3. Two candidates are:

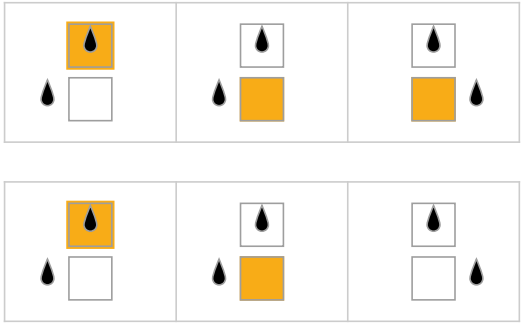

To prove universality one would have to show that there exists a procedure for setting up initial conditions that will make a particular rule here yield a multiway graph that corresponds (after decoding) to the multiway graph for any other specified system, say any other multiway Turing machine. To get some sense of how this might be possible, here are the somewhat diverse multispace graphs generated by the first rule above, starting from a single initial configuration whose tape contains the digits of successive binary numbers:

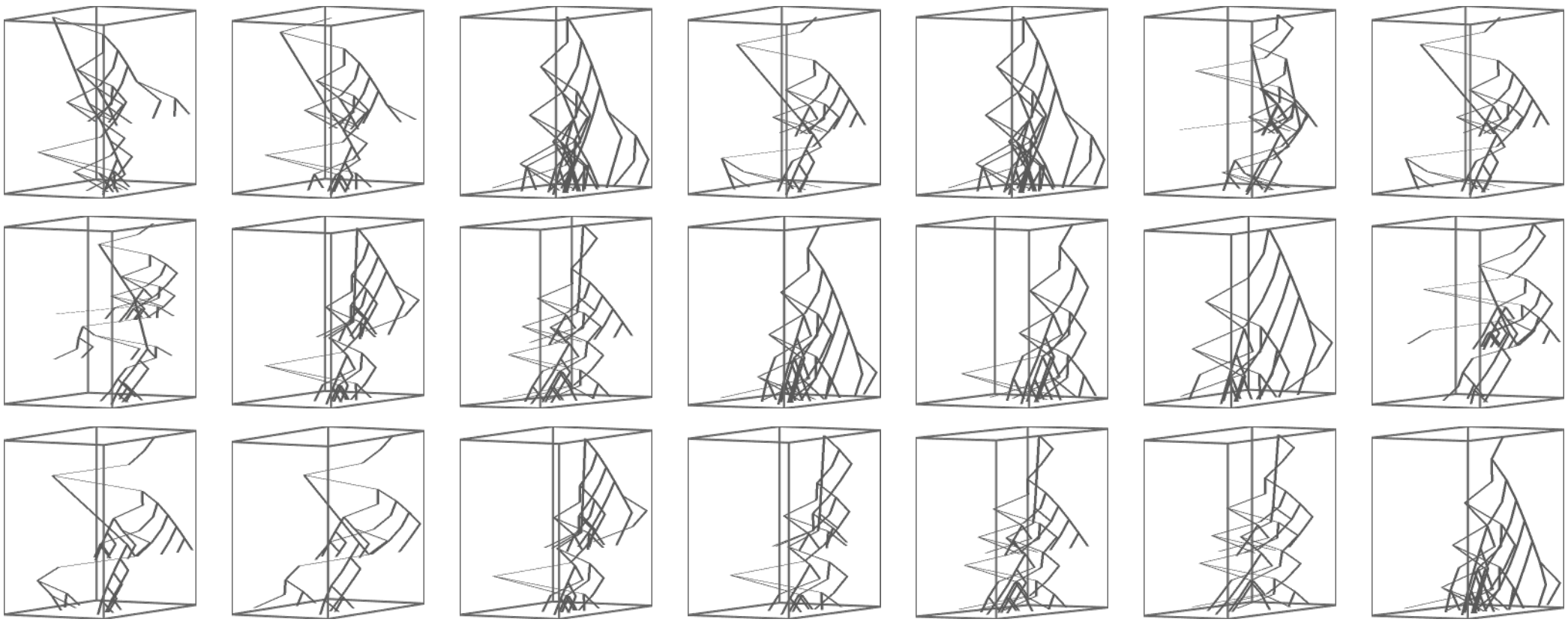

## The Halting Problem and Busy Beavers

If a system is computation universal, it is inevitable that there will be computational irreducibility associated with determining its behavior, and that to answer questions about what it will do after an arbitrarily long time may require unbounded amounts of computation, and must therefore in general be considered formally undecidable. A classic example is the halting problem—of asking whether a Turing machine starting from a particular initial condition will ever reach a particular "halt" state.

Most often, this question is formulated for ordinary deterministic Turing machines, and one asks whether these machines can reach a special "halting" head state. But in the context of multiway Turing machines, a more natural version of this question is just to ask, as we have done above, whether the multiway system will reach a configuration where none of its possible rules apply.

And in the simplest case, we can consider “deterministic but incomplete rules”, in which there is never more than one possible successor for a given state, but there may be no successor at all. An example is the rule

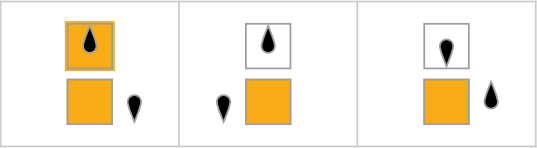

which contains no case for 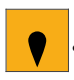. Starting this rule from a blank tape it can evolve for 5 steps, but then reaches a state where none of its rules apply:

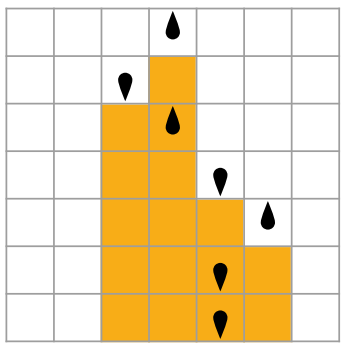

As we saw above, for s = 2, k = 2 rules, 6 steps is the longest any ordinary Turing machine rule will “survive” starting from a blank tape. For s = 3, k = 2 it is 21 steps, for s = 2, k = 3 it is 38 and for s = 4, k = 2 it is 107:

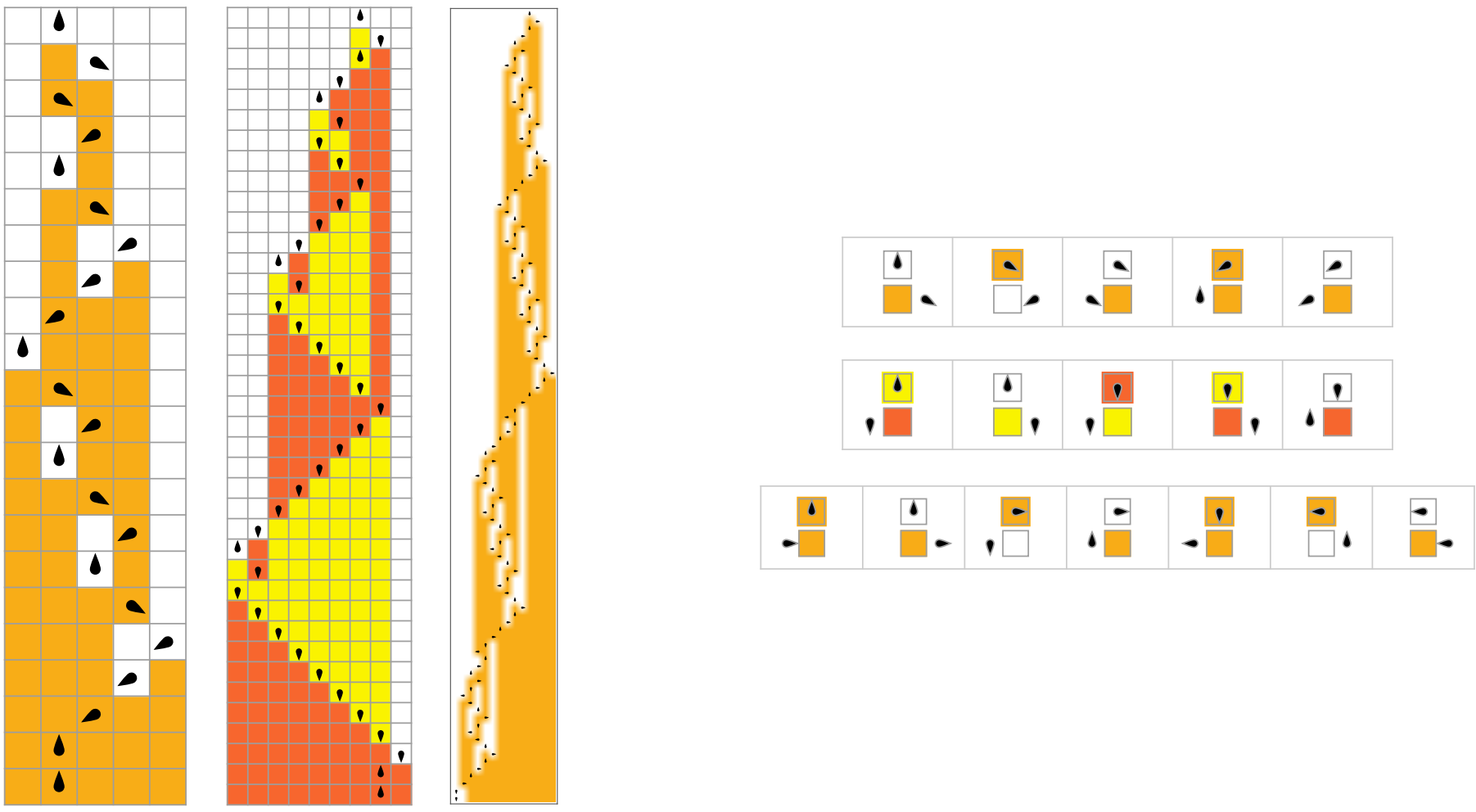

For larger s and k the survival times for the best-known “busy beaver” machines rapidly become very long; for s = 3, k = 3 it is known to be at least $10^{17}$.

So what about multiway Turing machines in general---where we allow more than one successor for a given state (leading to branching in the multiway graph)? For s = 1 or k = 1 the results are always trivial. But for s = 2, k = 2 things start to be more interesting.

With only p = 2 possible cases in the rule, the longest halting time is achieved by the deterministic rule

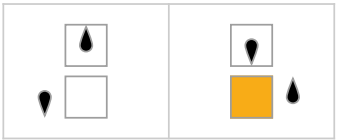

which halts in 4 steps

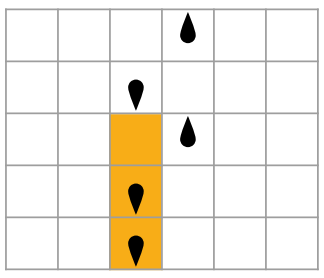

as represented by the (single-path) multiway graph:

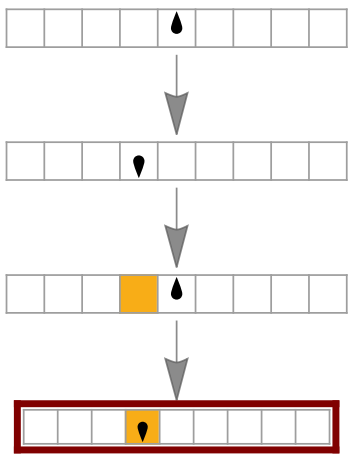

The p = 3 case includes standard deterministic s = 2, k = 2 Turing machines with a single halt state, and the longest halting time is achieved by the deterministic machine we saw above:

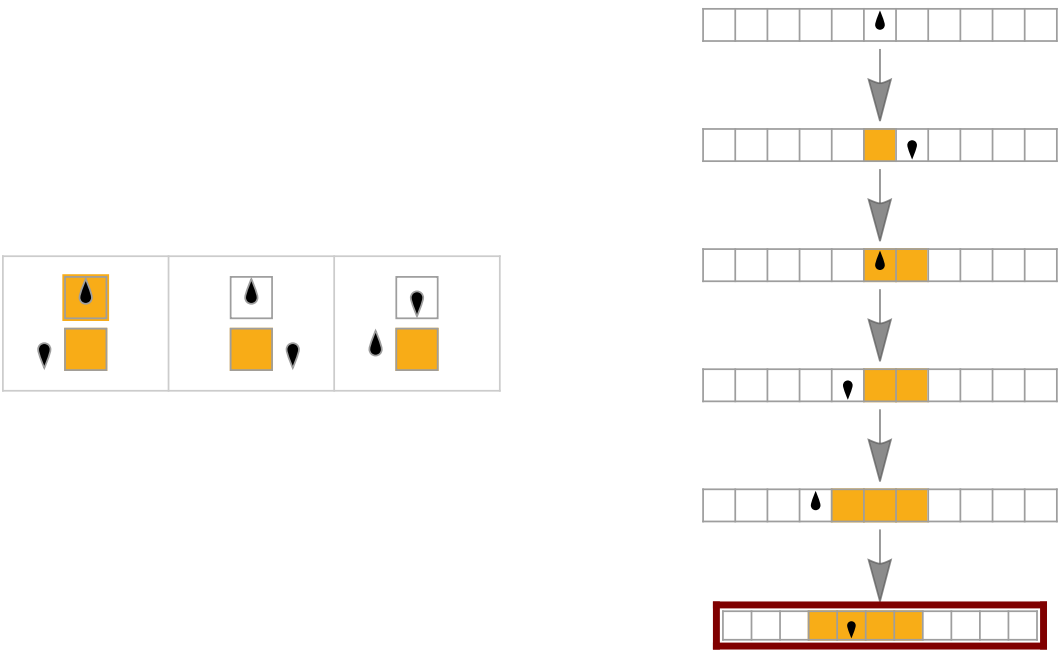

But if we look at slightly shorter halting times, we start seeing branching in the multiway graph:

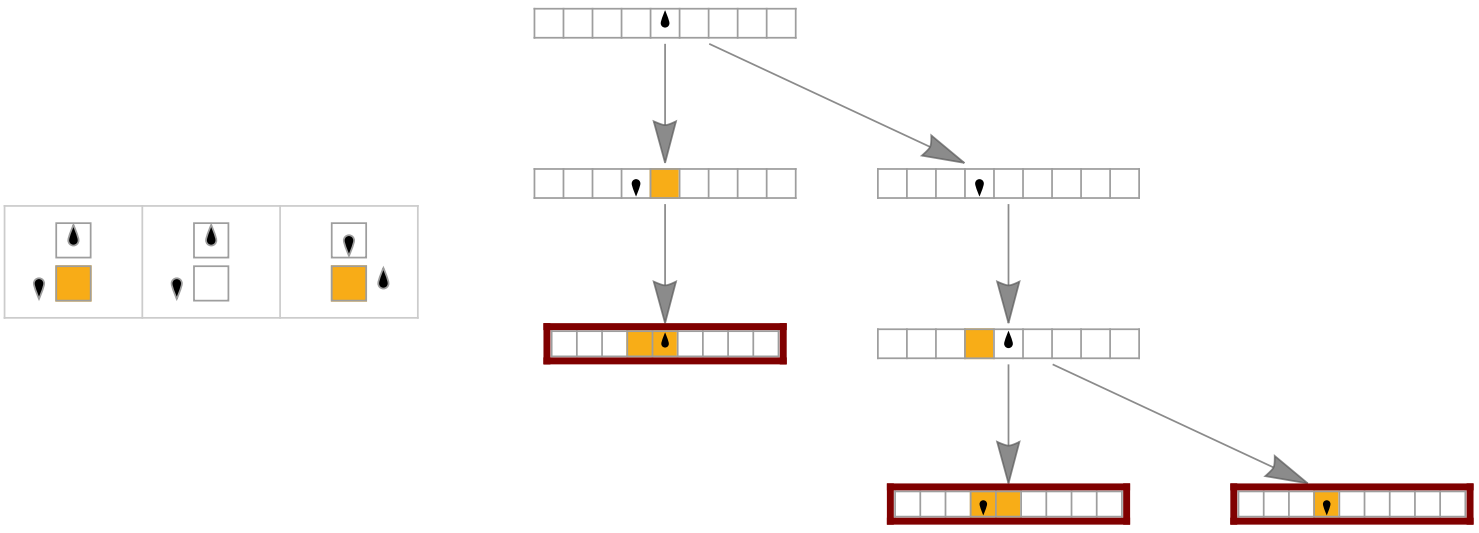

When we think in terms of multiway graphs, it is not so natural to distinguish "no-rule-applies" halting from other situations that lead to finite multiway graphs. An example is

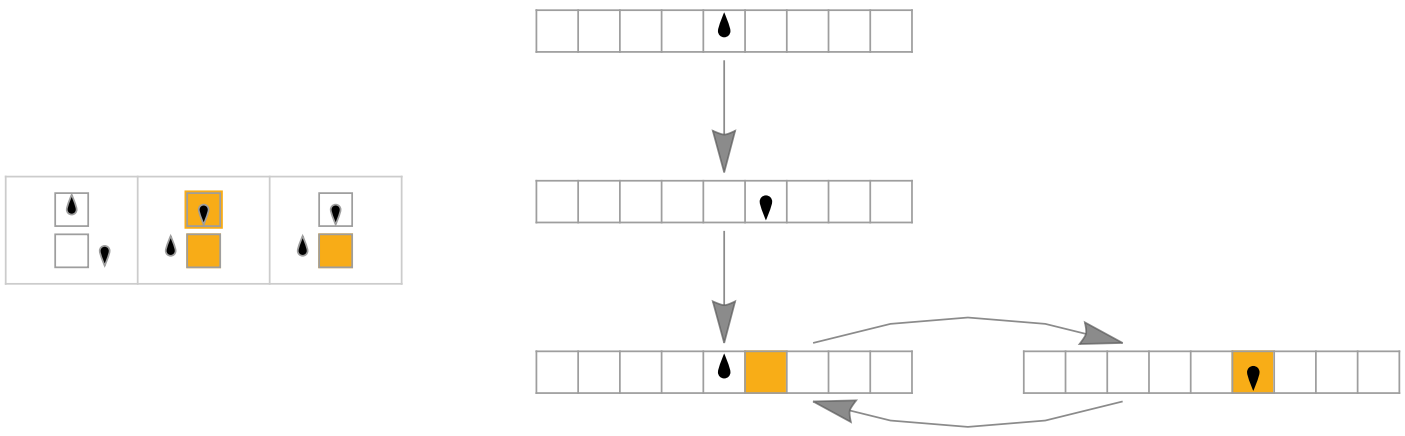

which is a deterministic machine that after 4 steps ends up in a loop:

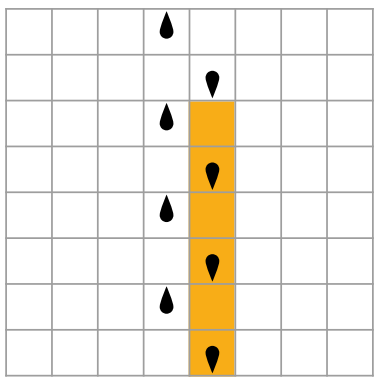

With p = 4 total cases in the rule we can have a "full deterministic s = 2, k = 2 Turing machine", with no possible cases omitted. The longest "halting time" (of 8 steps) is achieved by a machine which enters a loop:

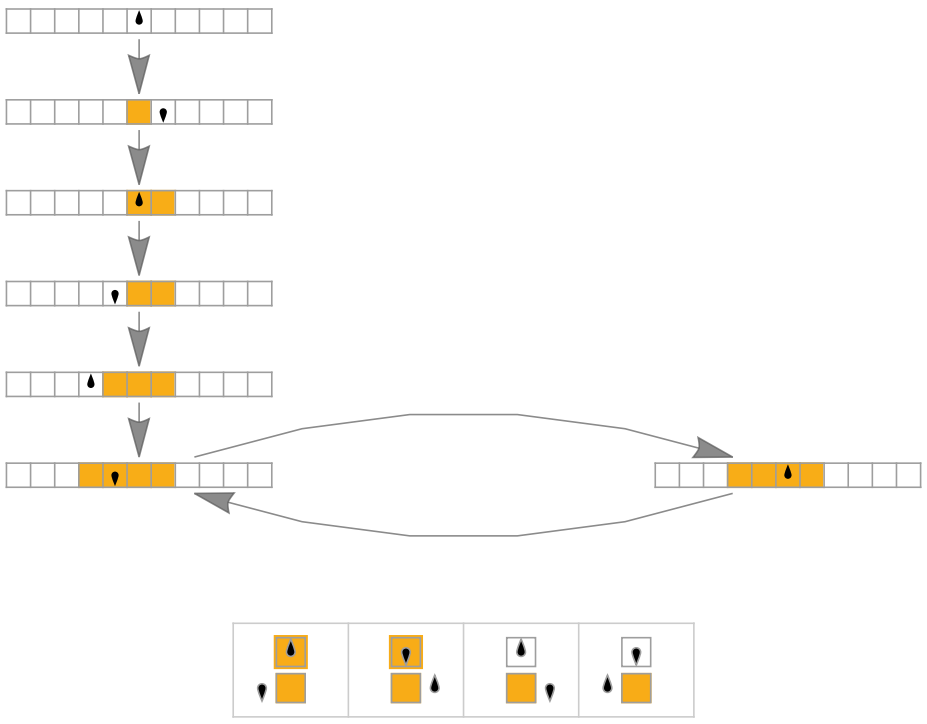

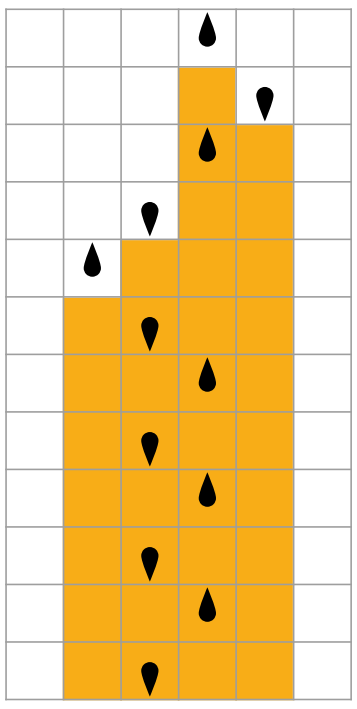

But if we consider shorter maximal halting times, then we get both “genuine halting” and branching in the multiway system:

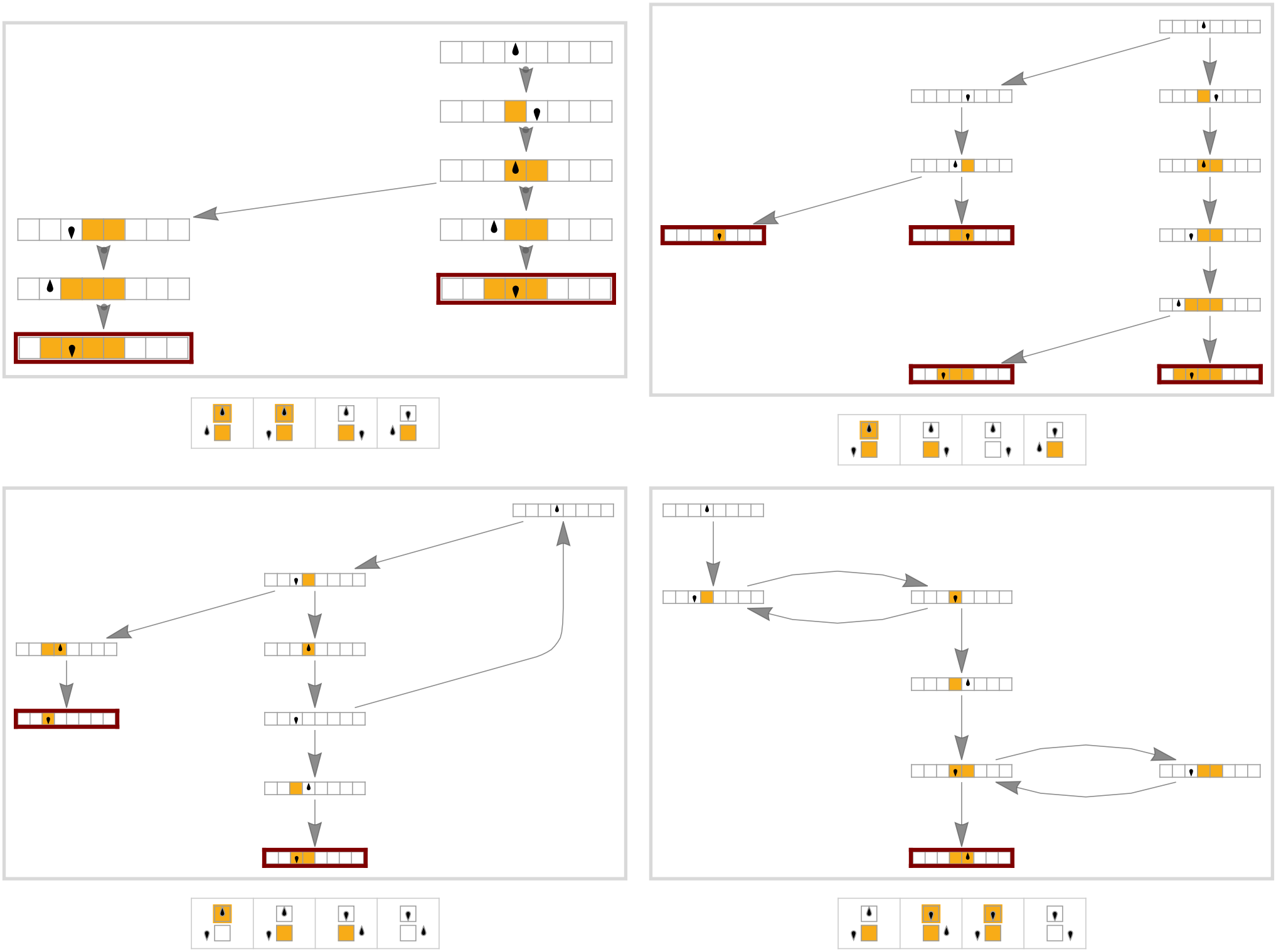

Note that unlike in the deterministic case where there is a single, definite halting time from a given initial state, a multiway Turing machine can have different halting times on different branches in its multiway evolution. So this means that there are multiple ways to define to define the multiway analog of the busy beaver problem for deterministic Turing machines.

One approach perhaps the closest to the spirit of the deterministic case is to ask for machines that maximize the maximal finite halting time obtained by following any branch. But one can also ask for machines which maximize the minimal halting time. Or one can ask for machines that give multiway graphs that involve as many states as possible while still being finite (because all their branches either halt or cycle). (Yet another criterion could be to ask for the maximum finite number of distinct halting states.)

For the p = 2 case, the maximum-halting-time and maximum-states criteria turn out to be satisfied by the same deterministic machine shown above (with 4 total states). For p = 3, though, the maximum-halting-time criterion is satisfied by the first machine we showed, which evolves through 6 states, while the maximum-states-criterion is instead satisfied by the second machine, which has 7 states in its multiway graph.

For the k = 2, s = 2, p = 4 case above the maximum-halting-time machine is again deterministic (and goes through 7 states), but maximum-states machines have 11 states:

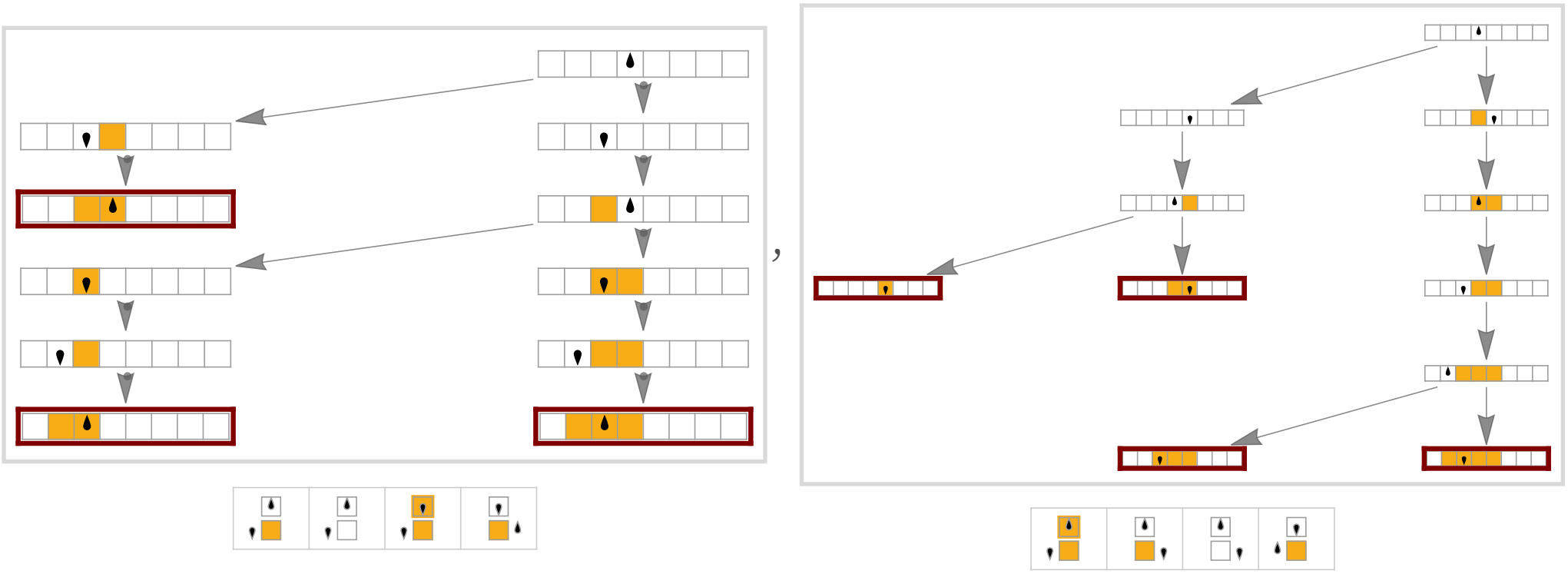

Even for ordinary deterministic Turing machines, finding busy beavers involves direct confrontation with the halting problem, and computational irreducibility. Let's say a machine has been running for a while. If it halts, well, then one knows it halts. But what if it doesn't halt? If its behavior is sufficiently simple, then one may be able to recognize—or somehow prove—that it can never halt. But if the behavior is more complex one may simply not be able to tell what will happen. And the problem is even worse for a multiway Turing machine. Because now it is not just a question of evolving one configuration; instead there may be a whole, potentially growing, multiway graph of configurations one has to consider.

Usually when the multiway graph gets big, it has a simple enough structure that one can readily determine that it will never "terminate", and just grow forever. But when the multiway graph gets complex it can be extremely difficult to be sure what will eventually happen. Still, at least in most cases, one can be fairly certain of the results.

With p = 5, the rule whose behavior allows the longest halting (or, in this case, cycling) time (8 steps) is:

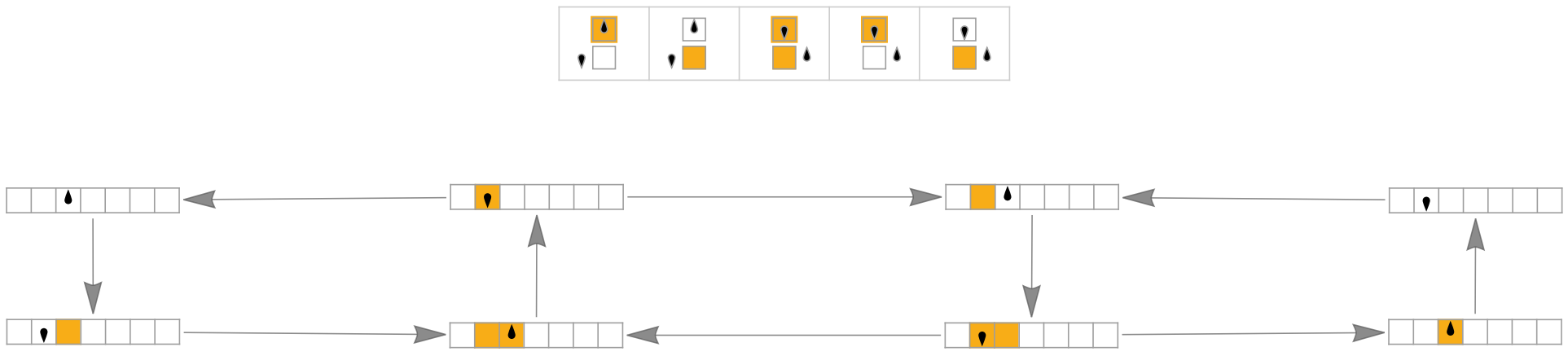

With p = 5, many other types of "finite behavior" can also occur, such as:

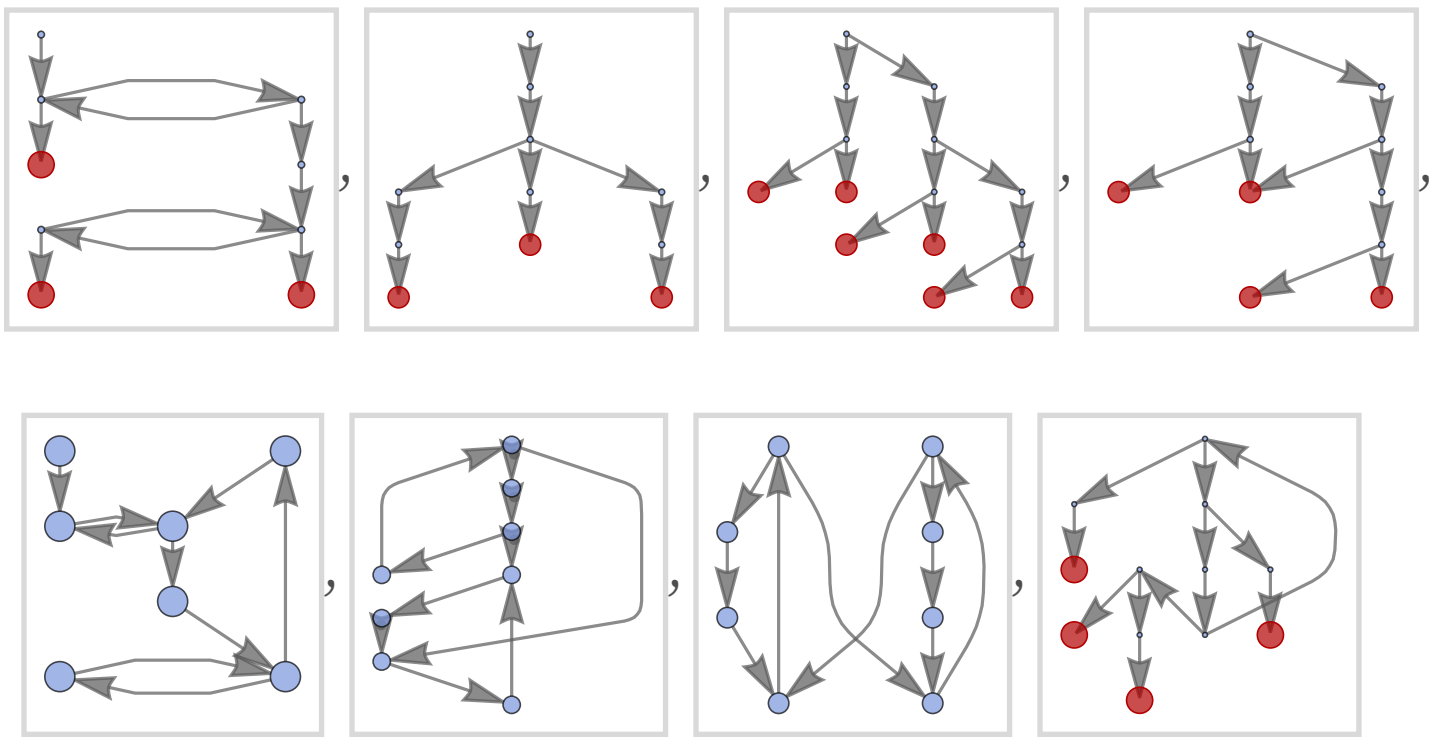

With p = 6 there is basically just more of the same—with no extension in the maximum halting time, though with larger total numbers of states:

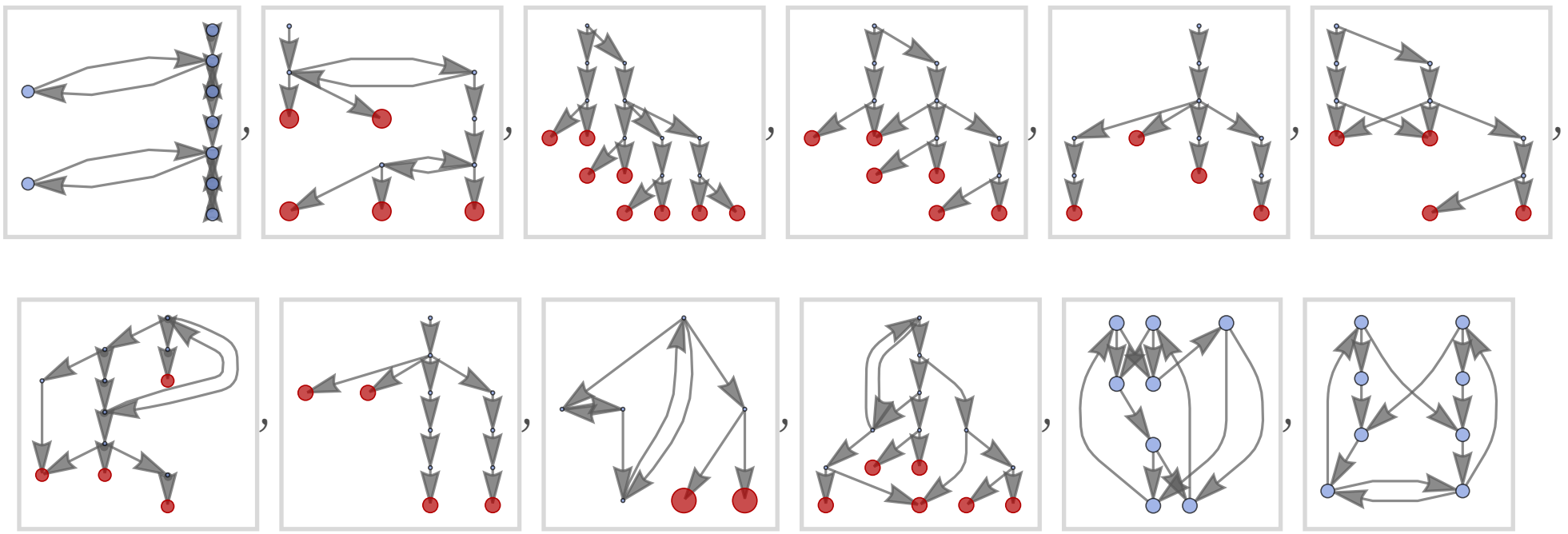

The maximum halting times and finite numbers of states reached by s=2, k=2 machines with p cases is as follows

| *p* | *steps* | *states* |
|---|---|---|
| 1 | 2 | 2 |
| 2 | 4 | 4 |
| 3 | 6 | 7 |
| 4 | 7 | 11 |
| 5 | 8 | 15 |
| 6 | 7 | 18 |
| 7 | 6 | 18 |

and the distributions of halting times and finite numbers of steps (for p for up 7) are:

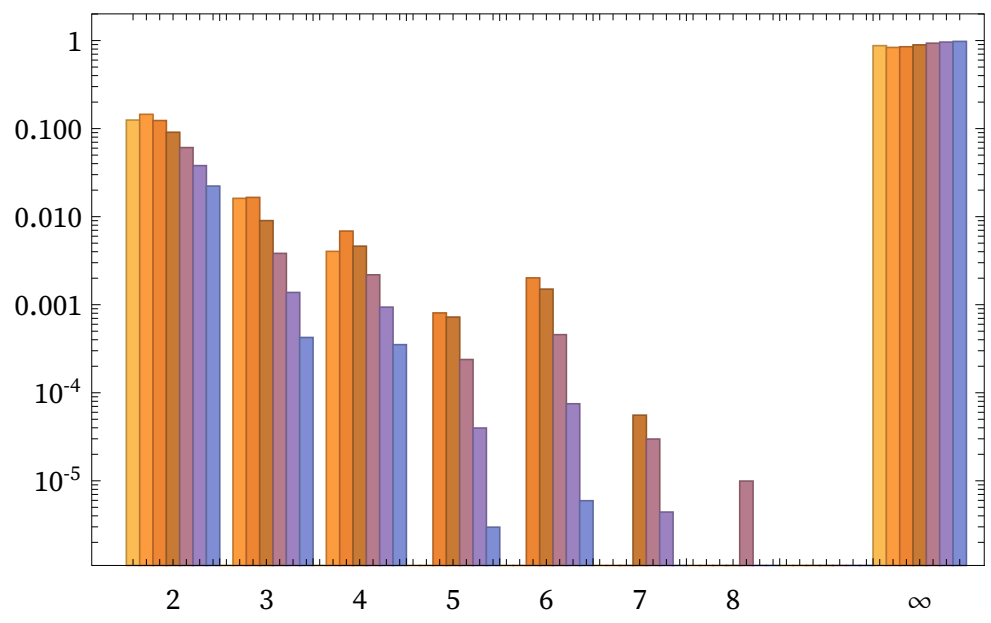

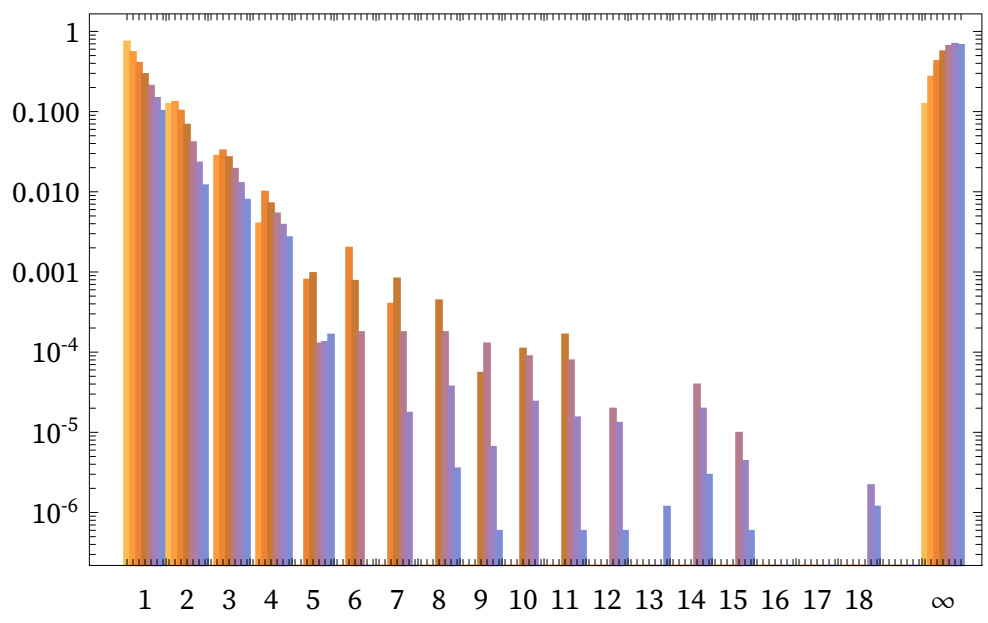

What if one considers, say, s = 3, k = 2? For p = 2 and p = 3 there are not enough cases in the rule to sample the rule range of head states, so the longest-surviving rules are basically the same as for s = 2, k = 2. Meanwhile, for p = 5 one has the 21-step “deterministic” busy beaver shown above, while for p = 4 there is a 17-step-halting-time rule:

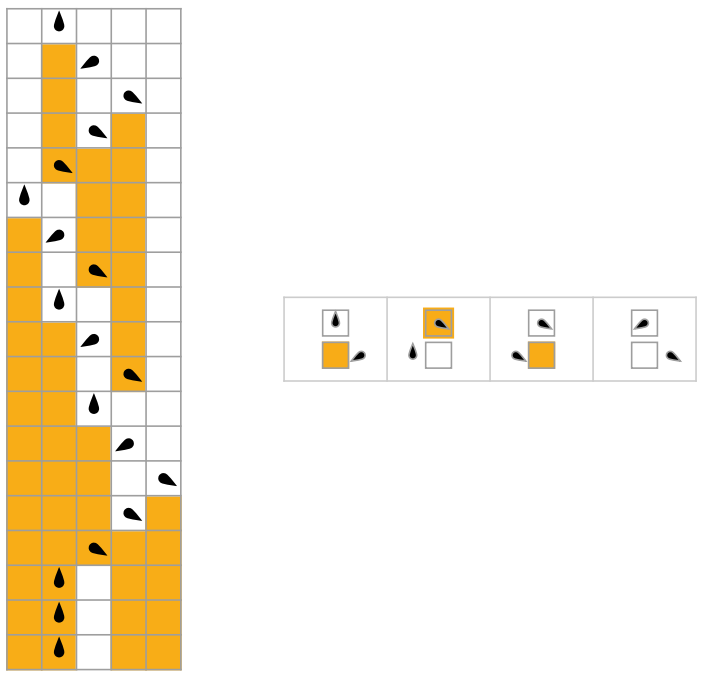

In the p = 5 case, the maximum-state machine is the following, with 30 states:

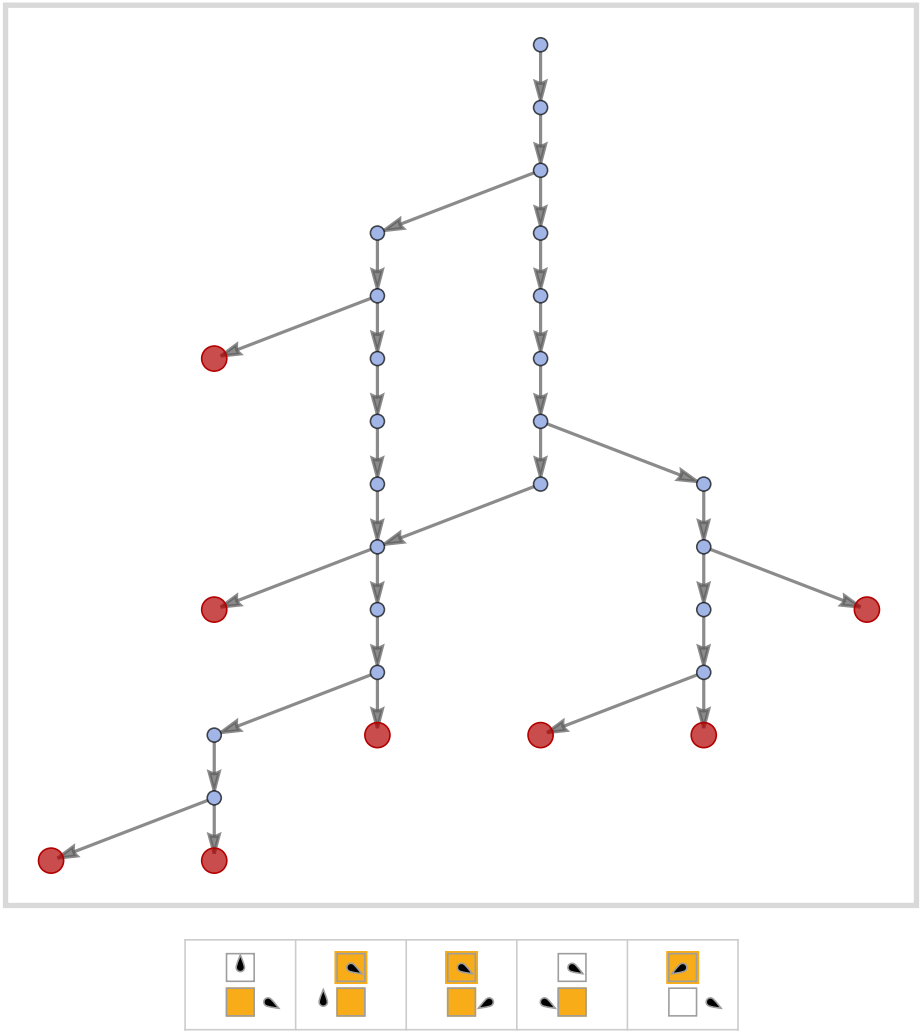

There are several interesting potential variants of the things we have discussed here. For example, instead of considering multiway Turing machines where all branches halt (or cycle), we can consider ones where some branches continue, perhaps even branching forever—but where either, say, one or all of the branches that do halt survive longest.

We can also consider Turing machines that start from non-blank tapes. And—in analogy to what one might study in computational complexity theory—we can ask how the number (or size of region) of non-blank cells affects halting times. (We can also study the "functions" computed by Turing machines by looking at the transformations they imply between initial tapes and final outputs.)

## Causal Graphs

Our main concern here so far has been in mapping out the successions of states that can be obtained in the evolution of multiway Turing machines. But one of the discoveries of our Physics Project is that there is a complementary way to understand the behaviors of systems, by looking not at successions of states but at causal relationships between events that update these states. And it is the causal graph of these relationships that is of most relevance if one wants to understand what observers embedded within a system can perceive.

In our multiway Turing machines, we can think of each application of a Turing machine rule as an "event". Each such event takes certain "input" (the state of the head and the color of the cell under it), and generates certain "output" (the new state of the head, the color under it, and its position). The causal graph maps out what outputs from other events the input to a given event requires, or, in other words, what the causal dependence is of one event on others.

For example, for the ordinary Turing machine evolution

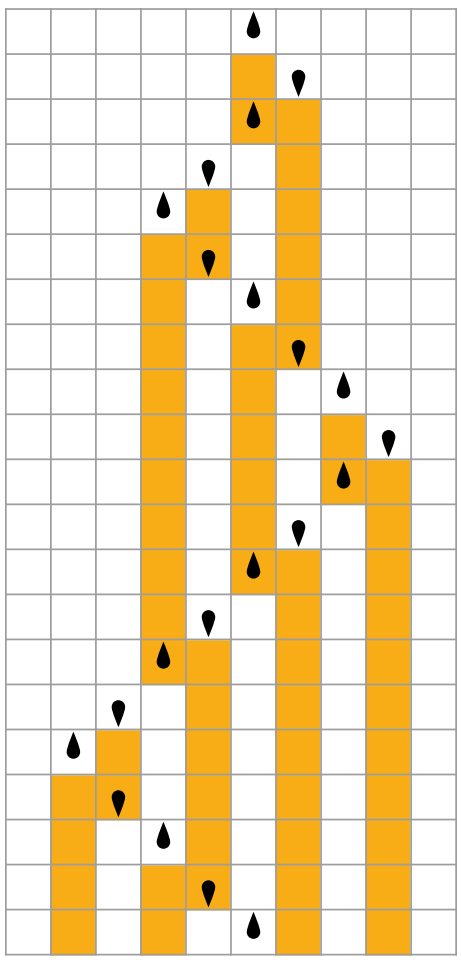

the network of causal relationships between updating events is just

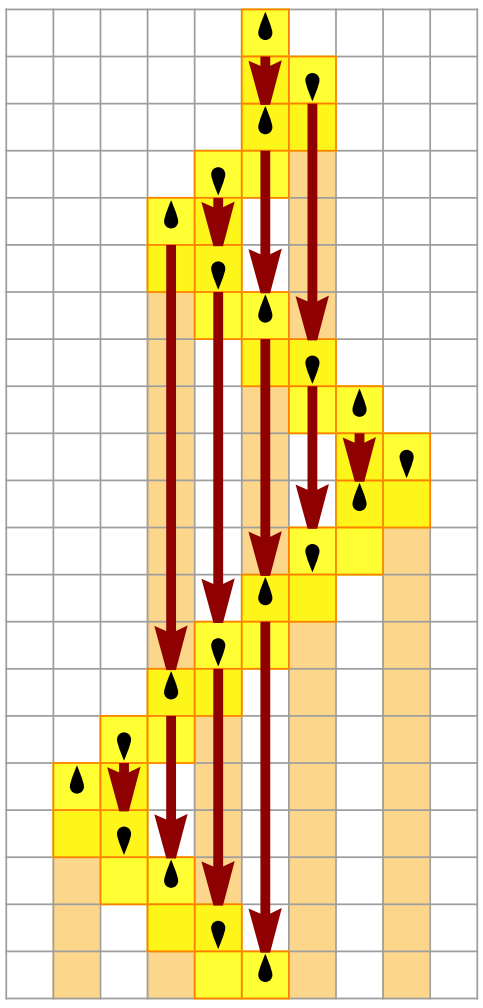

yielding the causal graph:

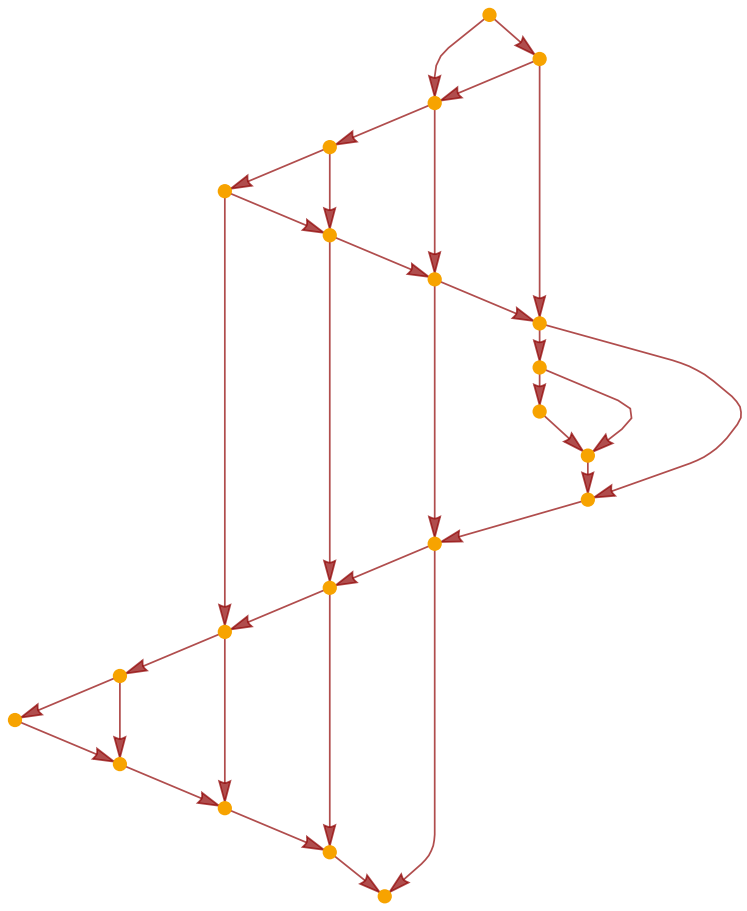

Continuing longer gives

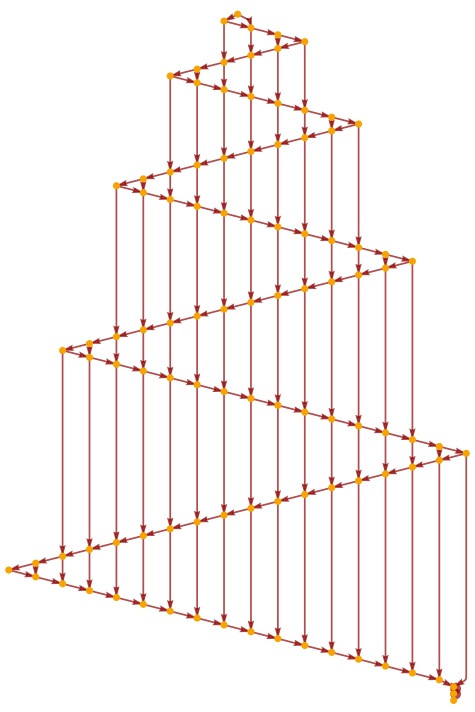

or with a different graph rendering just:

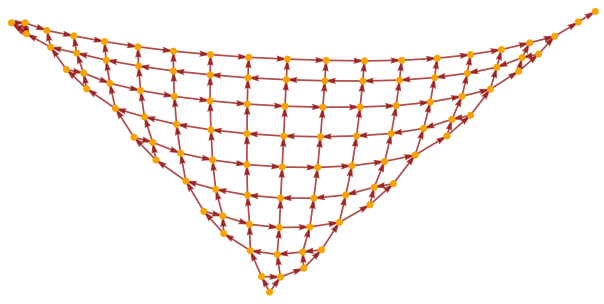

One can also define causal graphs for multiway Turing machines. As an example, look at a rule we considered above:

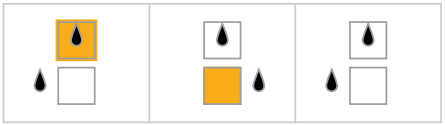

The multiway graph that describes possible successions of states in this case is:

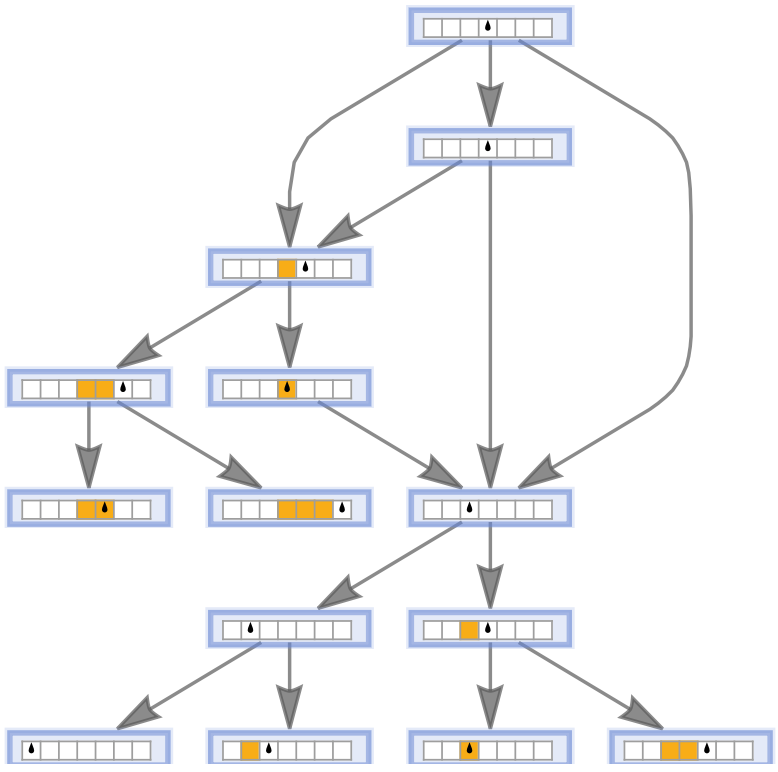

Now let's explicitly show the update event that transforms each state to its successor:

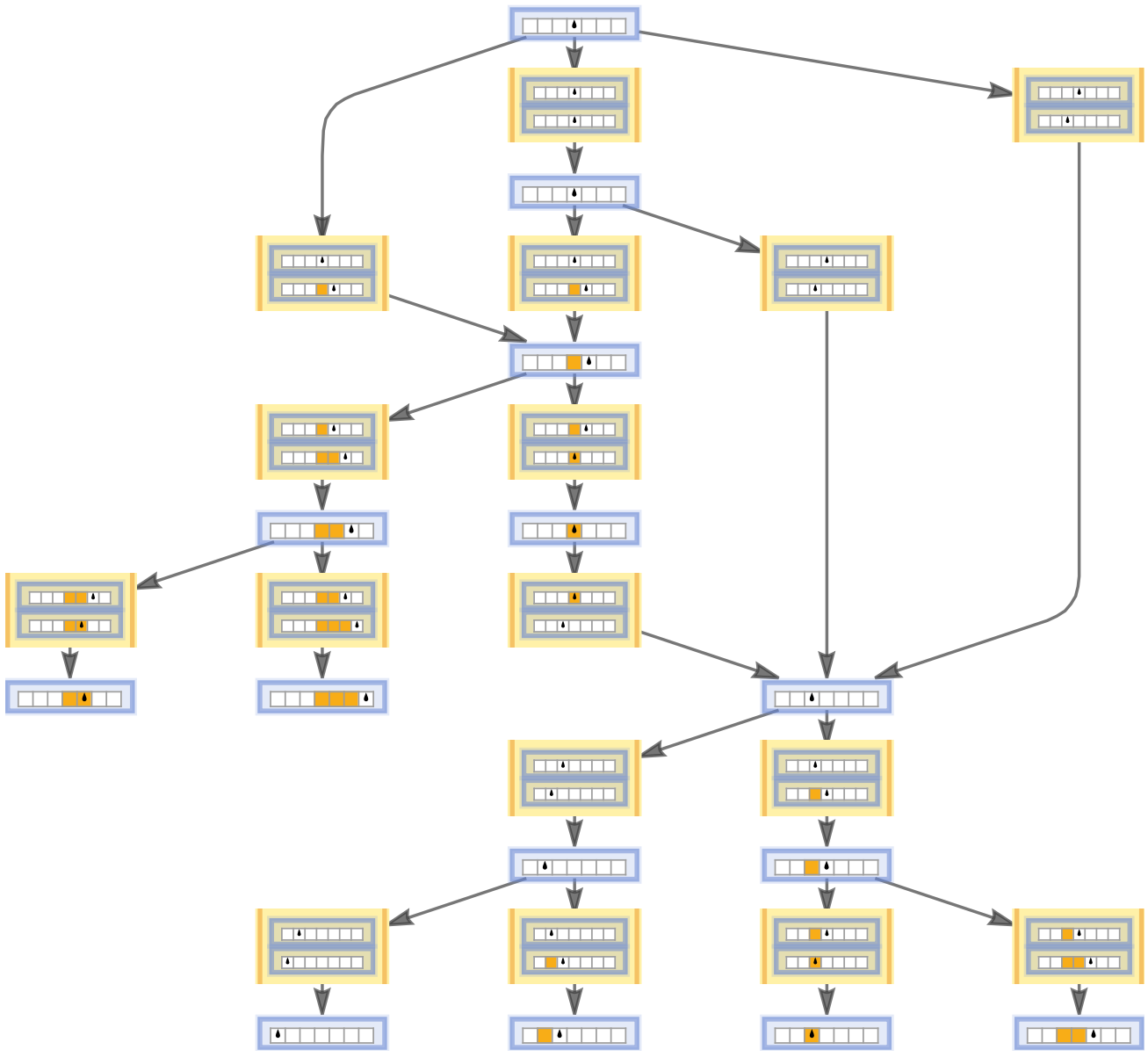

Just as for the deterministic case, we can identify the causal relationships between these update events, here indicated by orange edges:

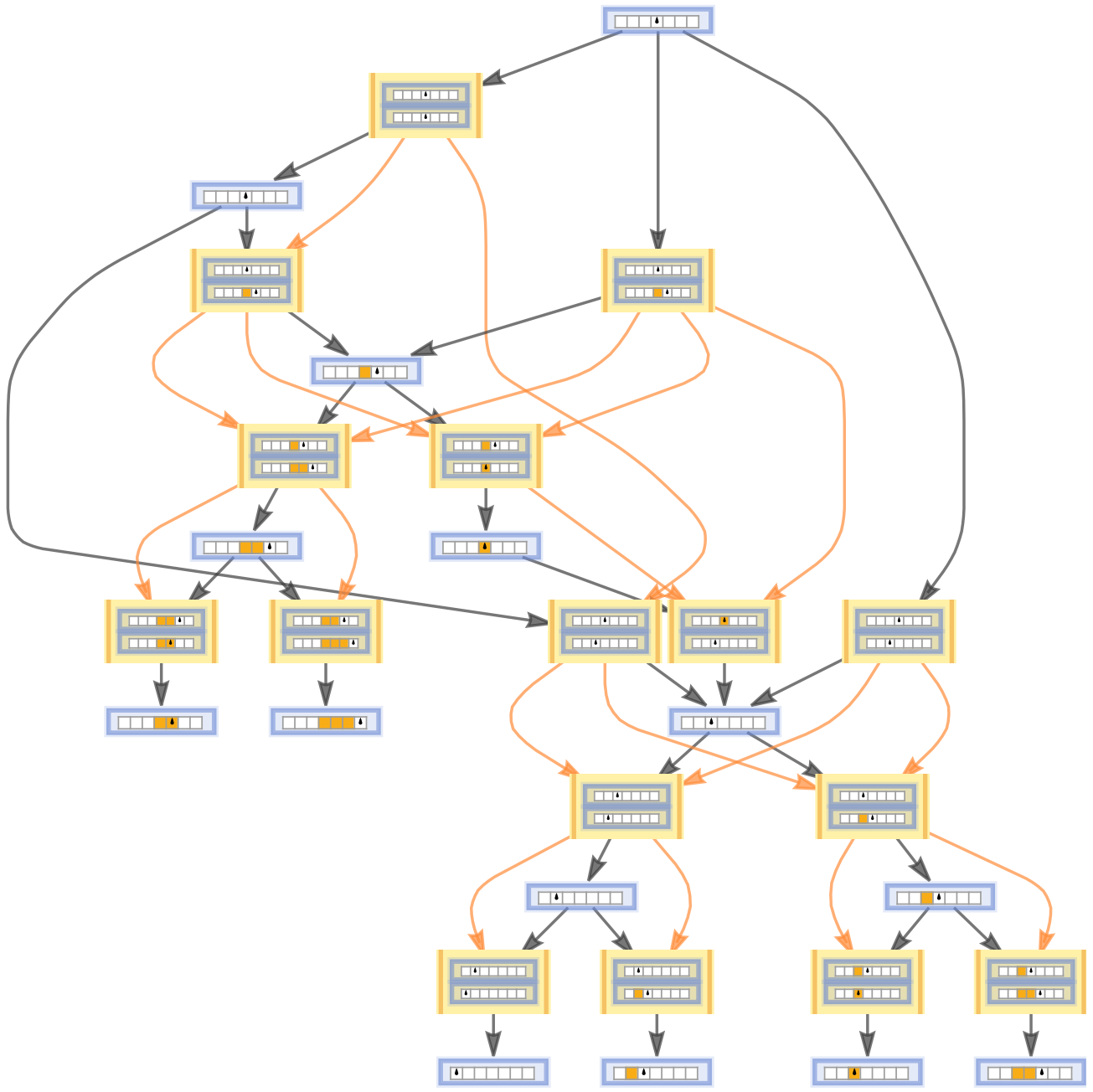

Keeping only the events and their causal relationships we get the final multiway causal graph:

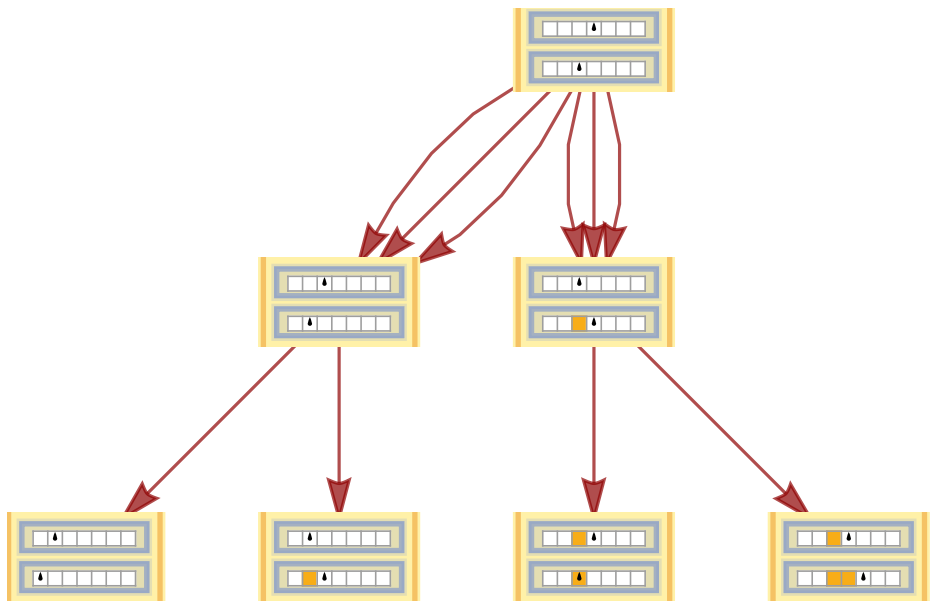

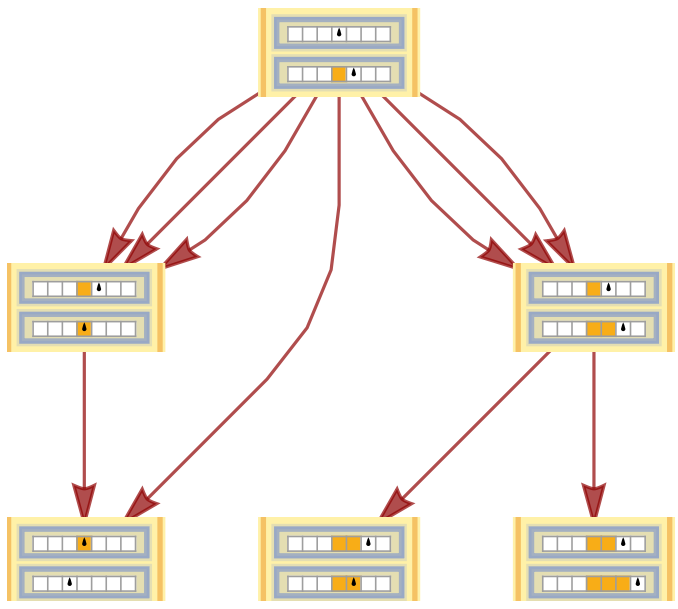

Continuing for more steps we get:

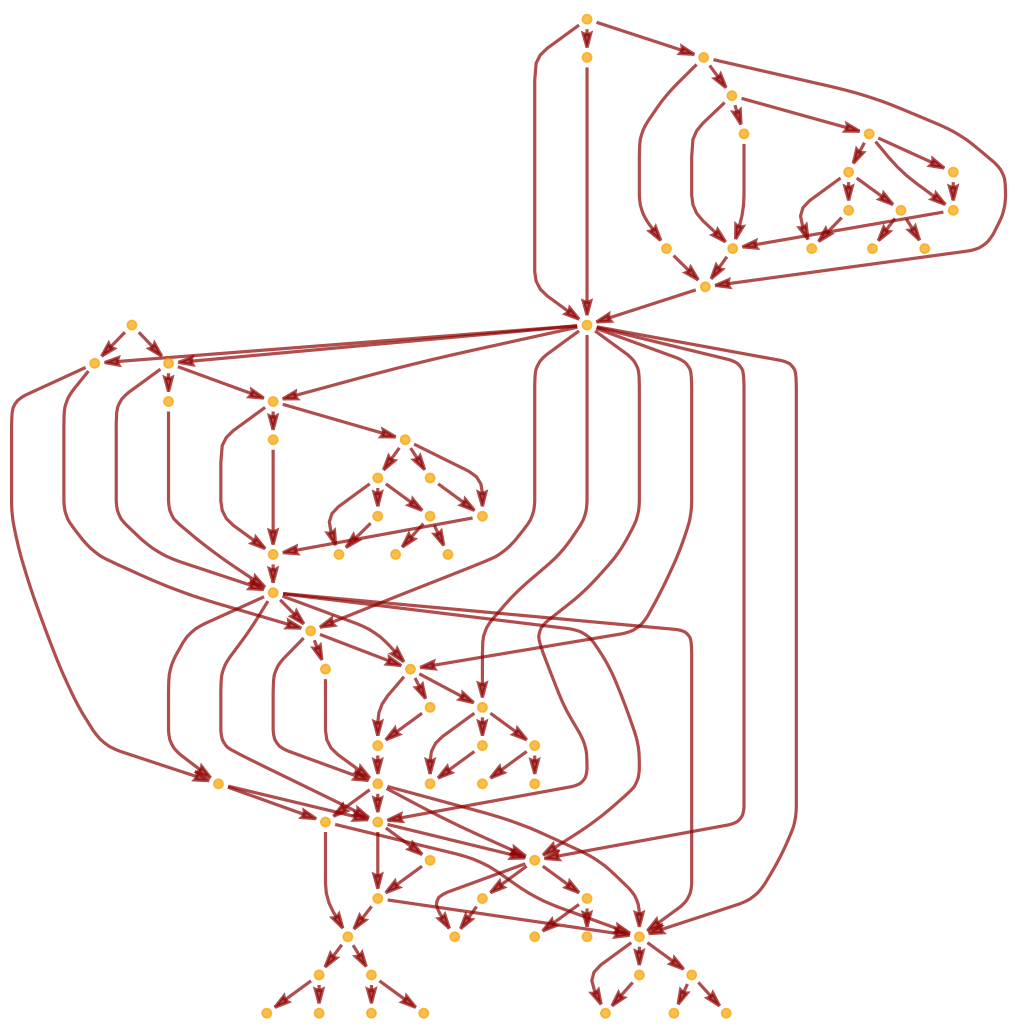

Informed by our Physics Project, we can think of each edge in the causal graph as representing a causal relationship between two events which follow each other in time. In the causal graph for an ordinary "single-way" Turing machine these two events may also be separated in space, so that in effect the causal graph defines how space and time are knitted together. In the (multiway) causal graph for a multiway Turing machine, events can be separated not only in space, but also in branchial space (or, in other words, they can occur on different branches in the evolution)—so the multiway causal graph can be thought of as defining how space, branchial space and time are knitted together.

We discussed above how states in a multiway graph can be thought of as being laid out in "multispace" which includes both ordinary space and branchial space coordinates. One can do more or less the same for events in a multiway causal graph---suggesting a "multispace rendering of a multiway causal graph":

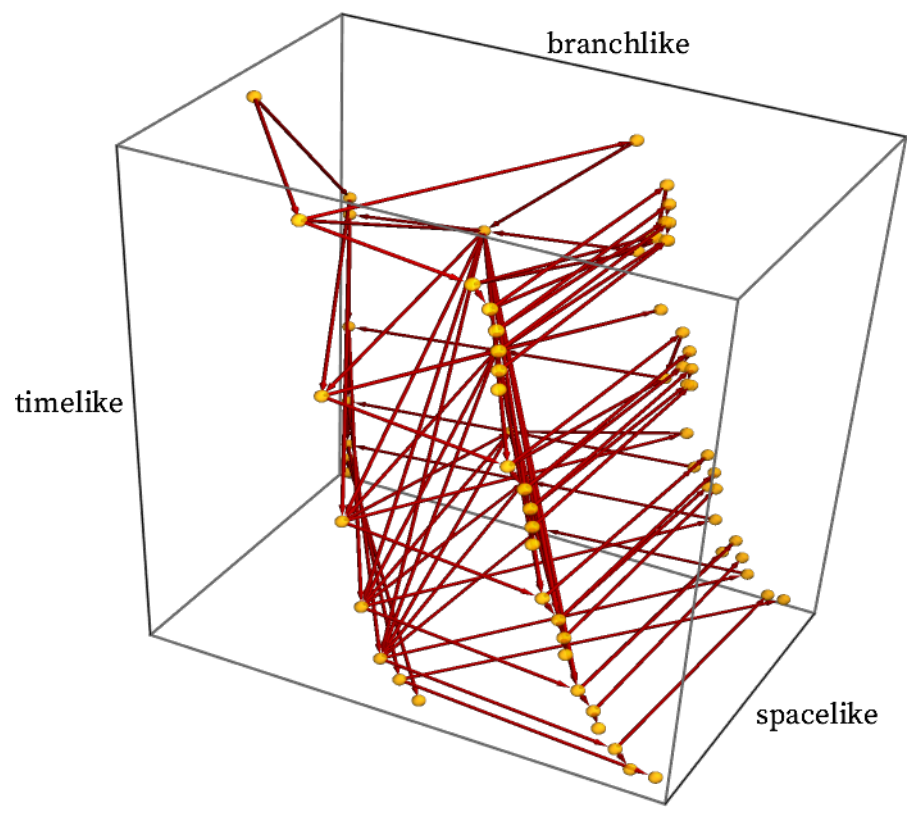

Different multiway Turing machines can have quite different multiway causal graphs. Here are some samples for various rules we considered above:

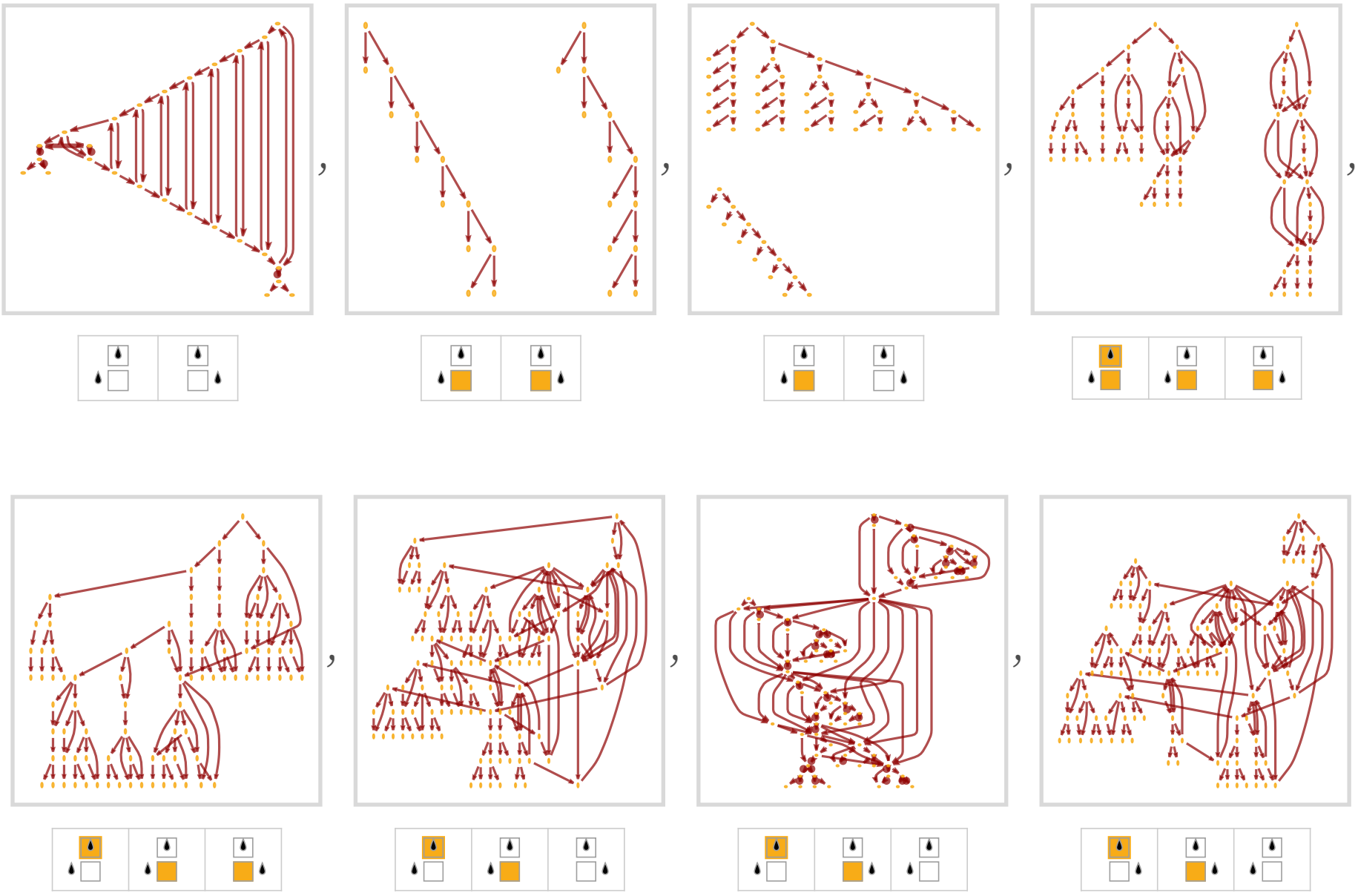

These multiway causal graphs in a sense capture all causal relationships within and between different possible paths followed in the multiway graph. If we pick a particular path in the multiway graph (corresponding to a particular sequence of choices about which case in the multiway Turing machine rule to apply at each step) then this will yield a "deterministic" evolution. And this evolution will have a corresponding causal graph---that must always be a subgraph of the full multiway causal graph.

## Causal Invariance

Whenever there is more than one possible successor for a given state in the evolution of a multiway Turing machine, this will lead to multiple branches in the multiway graph. Different possible "paths of history" (with different choices about which case in the rule to apply at each step) then correspond to following different branches. And in a case like this once one has picked a branch one is "committed": one can never subsequently reach any other branches—and paths that diverge never converge again.

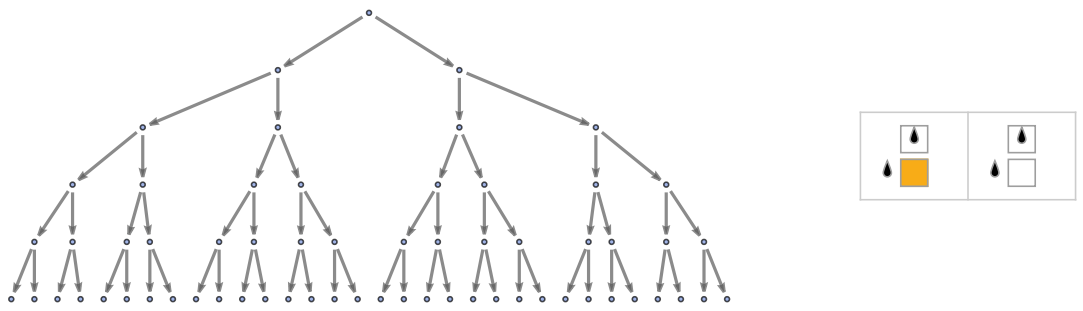

But in a case like this it's a different story because here—at least if one goes far enough in generating of the multiway graph—every pair of paths that diverge must eventually converge again. In other words, if one takes a "wrong turn", one can always recover. Or, put another way, whatever sequence of rules one applies, it is always possible to reach eventual consistency in the results one gets.

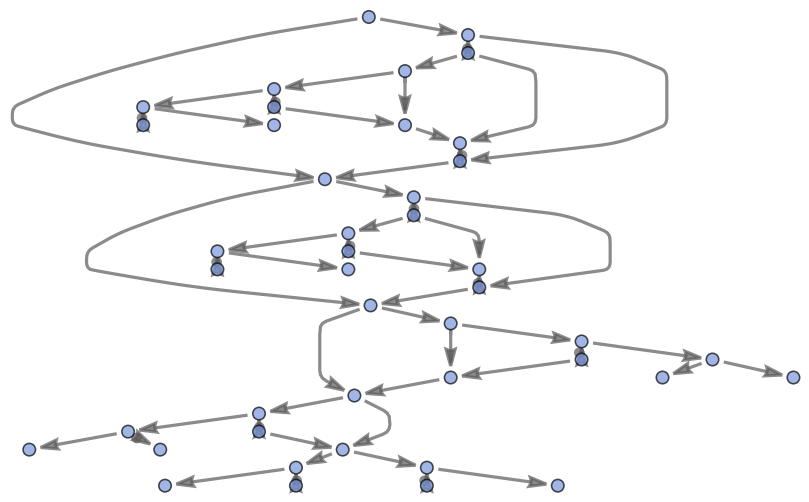

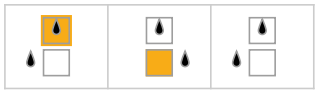

This property is closely related to a property that's very important in our Physics Project: causal invariance. When causal invariance is present, it implies that the causal graphs generated by following any possible path must always be the same. In other words, even though the multiway system in a sense allows many possible histories, the network of causal relationships obtained in each case will always be the same—so that with respect to causal relationships there is essentially just one "objective reality" about how the system behaves.

(By the way, the multiway causal graph contains more information than the individual causal graphs, because it also describes how all the causal graphs—from all possible histories—"knit together" across space, time and branchial space.

There's a subtlety which we will not explore in detail here. Whether one considers branches in the multiway graph to have converged depends on how one defines equivalence of states. In the multiway graphs we have drawn, we have done this just by looking at whether the instantaneous states of Turing machines are the same. And in doing this, the merging of branches is related to a property often called confluence. But to ensure that we get full causal invariance we can instead consider states to be equivalent if in addition to having the same tape configurations, the causal graphs that lead to them are isomorphic (so that in a sense we're considering a "causal multiway graph").

So what multiway Turing machines are causal invariant (or at least confluent)? If a multiway Turing machine has rules that make it deterministic (without halting), and thus effectively "single way", then it is trivially causal invariant. If a multiway Turing machine has multiple halting states it is inevitably not causal invariant—because if it "falls into" any one of the halting states, then it can never "get out" and reach another. On the other hand, if there is just a single halting state, and all possible histories lead to it, then a multiway Turing machine will at least be confluent (and in terms of its computation can be thought of as always "reaching a final unique answer", or "normal form"). Finally, in the "rulial limit" where the multiway Turing machine can use any possible rule, causal invariance is inevitable.

In general, causal invariance is not particularly rare among multiway Turing machines. For s = 1, k = 1 rules, it is inevitable. For s = 1, k = 2 rules here are some examples of multiway graphs and multiway causal graphs for rules that appear to be at least confluent

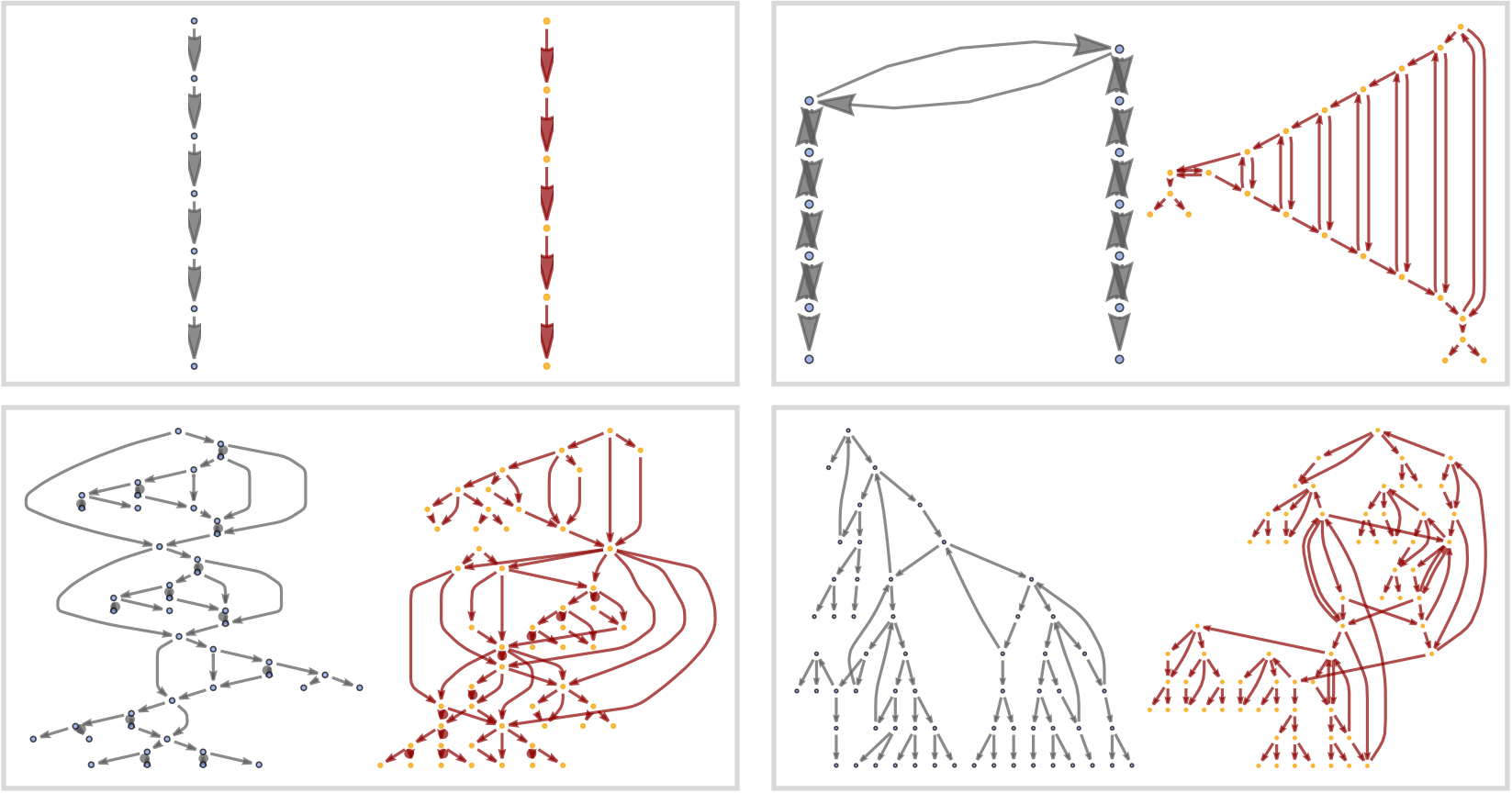

and ones that do not:

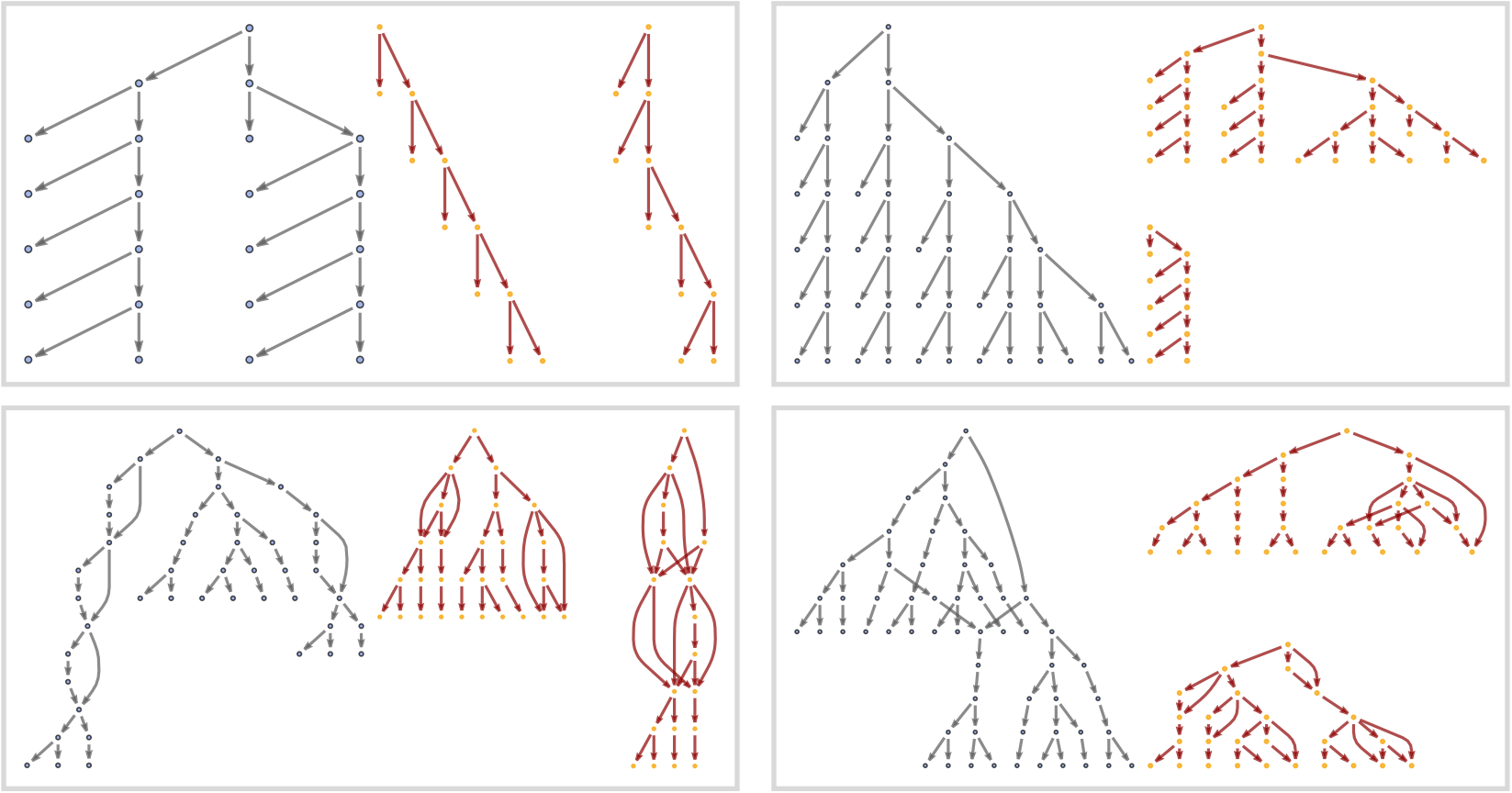

(Note that in general it can be undecidable if a given multiway Turing machine is causal invariant—or confluent—or not. For even if two branches eventually merge, there is no a priori upper bound on how many steps this will take—though in practice for simple rules it typically seems to resolve quite quickly.)

## Finite Tapes

So far we've always assumed that the tapes in our Turing machines are unbounded. But just as we did in our earlier project of studying the rulial space of Turing machines, we can also consider the case of Turing machines with bounded—say cyclic—tapes with a finite number of "cells" n.

Such a Turing machine has a total of $n\ s\ k^n$ possible complete states—and for a given n we can construct a complete state transition graph. For an ordinary deterministic Turing machine, these graphs always have a single outgoing edge from each state node (though possible several incoming edges). So, for example, with the rule

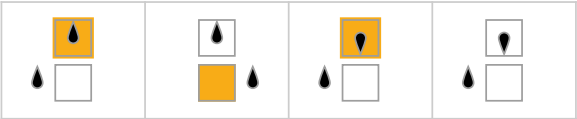

the state transition graph for a length-3 tape is:

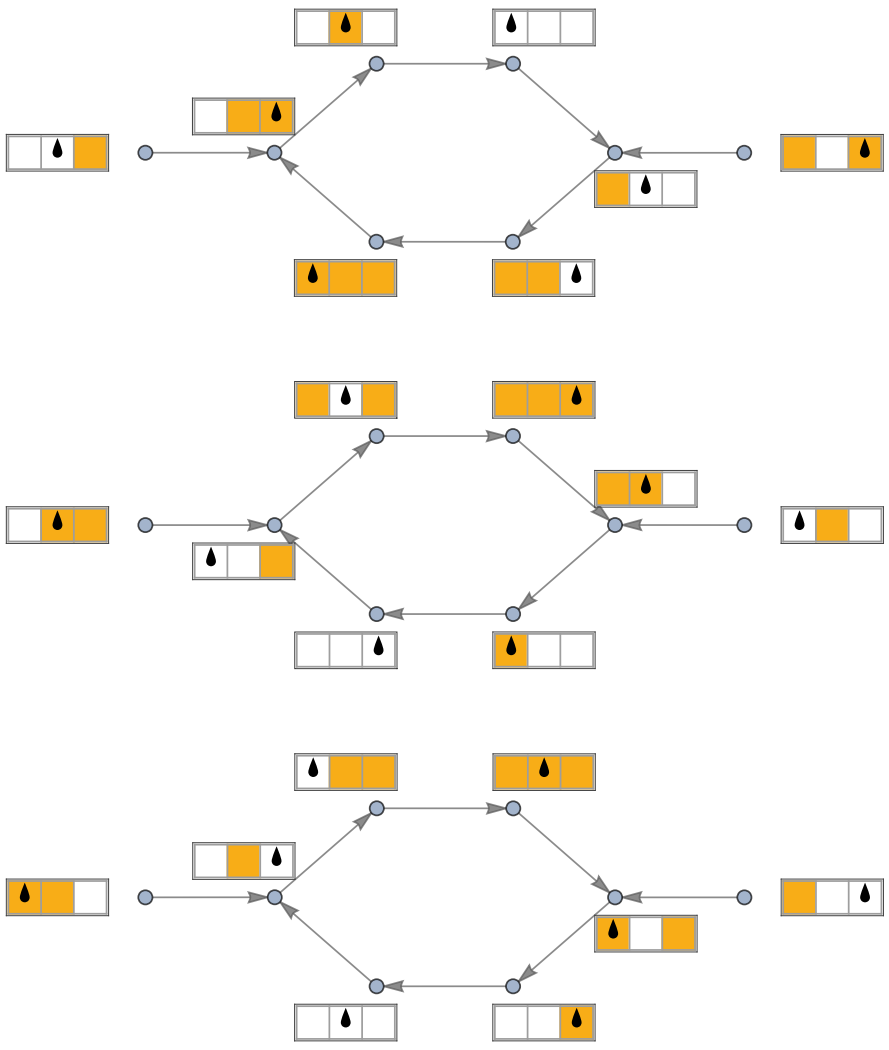

Here are some other s = 2, k = 2 examples (all for n = 3):

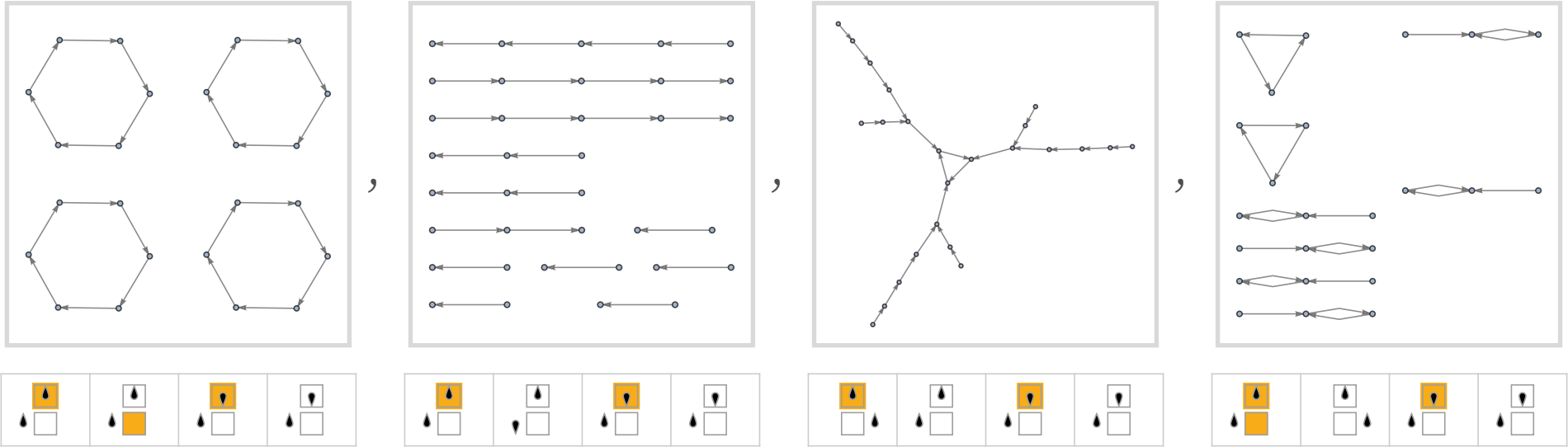

In all cases, what we see are certain cycles of states that are visited repeatedly, fed by transients containing other states.

For multiway Turing machines such as

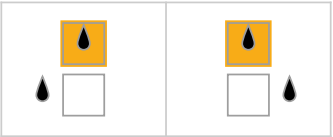

the structure of the state transition graph can be different, with nodes having both more or less than 1 outgoing edge:

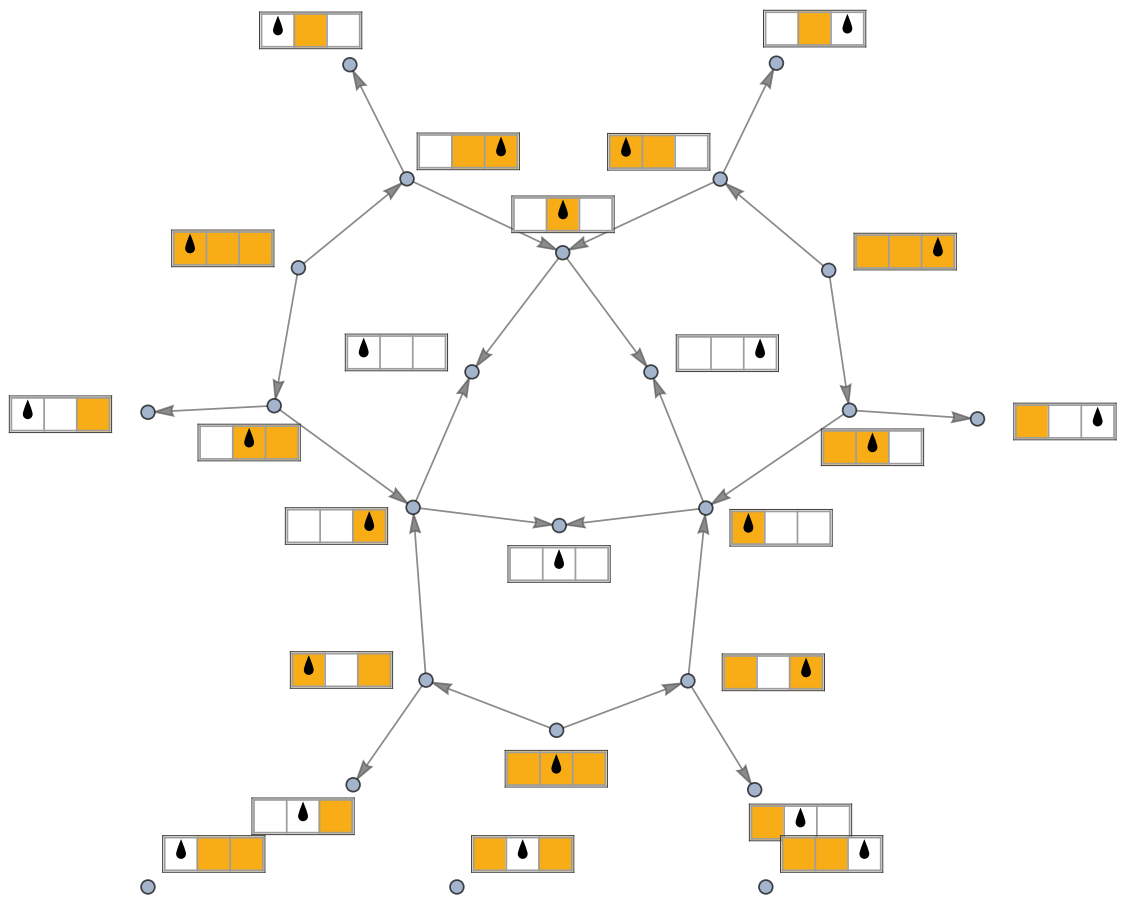

Starting from the node corresponding to a particular state, the subgraph reached from that state is the multiway graph corresponding to the evolution from that state. Halting occurs when there is outdegree 0 at a node—and the 3 nodes at the bottom of the image correspond to states where the Turing machine has immediately halted.

Here are results for some other multiway Turing machines, here with tape size n = 5, indicating halting states in red:

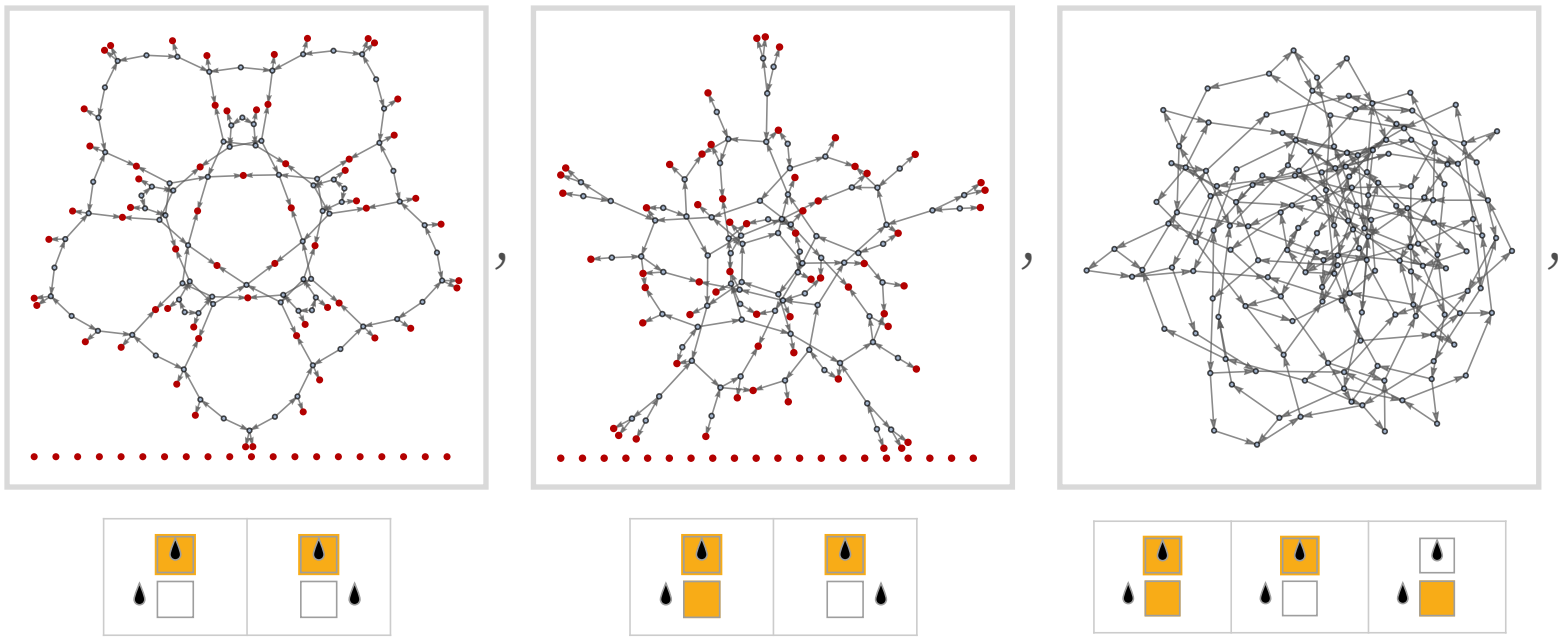

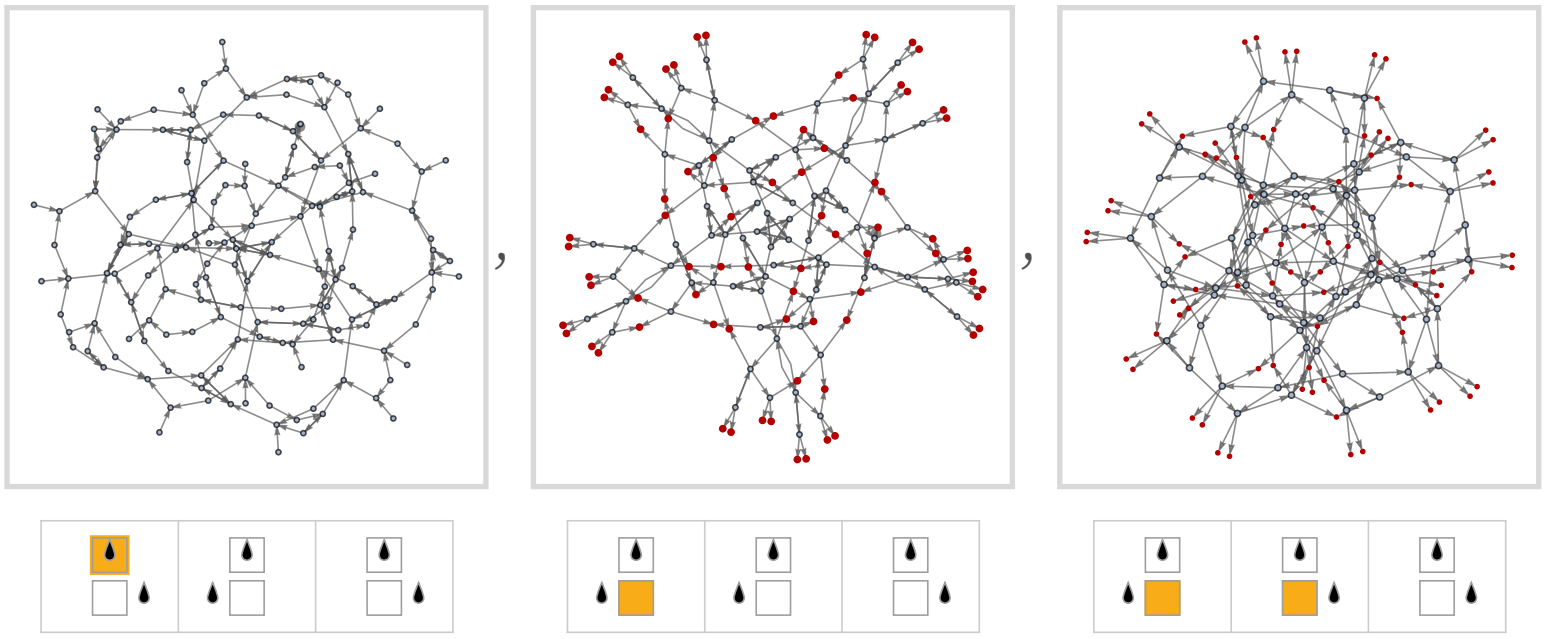

What happens when there are more cases p in the multiway Turing machine rule? If all possible cases are allowed, so that $p = 2\,s^2\,k^2$, then we get the full rulial multiway graph. For size n = 3 this is

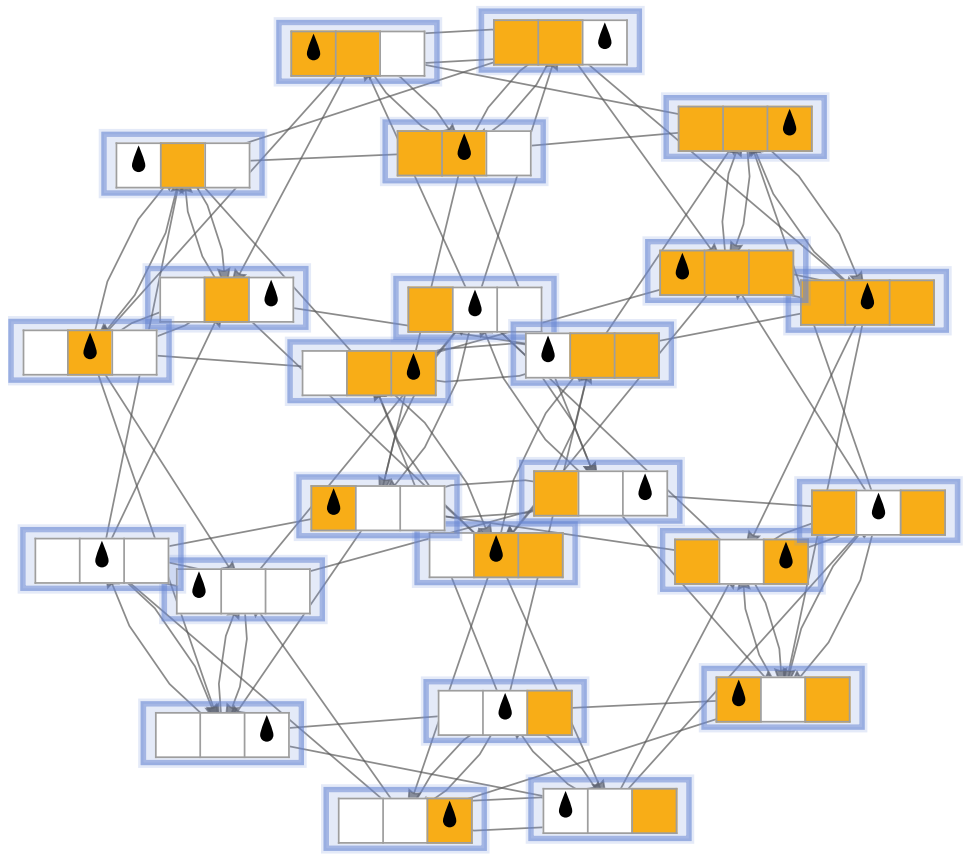

while for n = 5 it is:

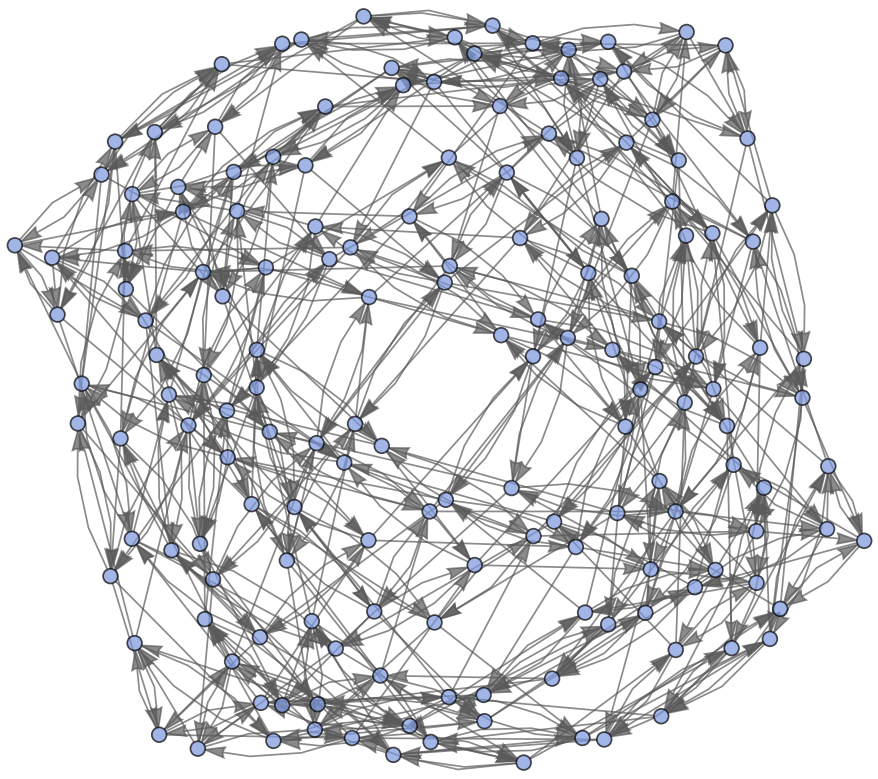

This “rulial-limit” multiway graph has the property that it is in a sense completely uniform: the graph is vertex transitive, so that the neighborhood of every node is the same. (The graph corresponds to the Cayley graph of a finite “Turing machine group”, here $TM^n_{1,2} = \mathbb{Z}_n \ltimes (\mathbb{Z}_2)^n$.)

But while this “rulial-limit” graph in a sense involves maximal nondeterminism and maximal branching, it also shows maximal merging, and in fact starting from any node, paths that branch must always be able to merge again---as in this example for the n = 3 case (here shown with two different graph renderings):

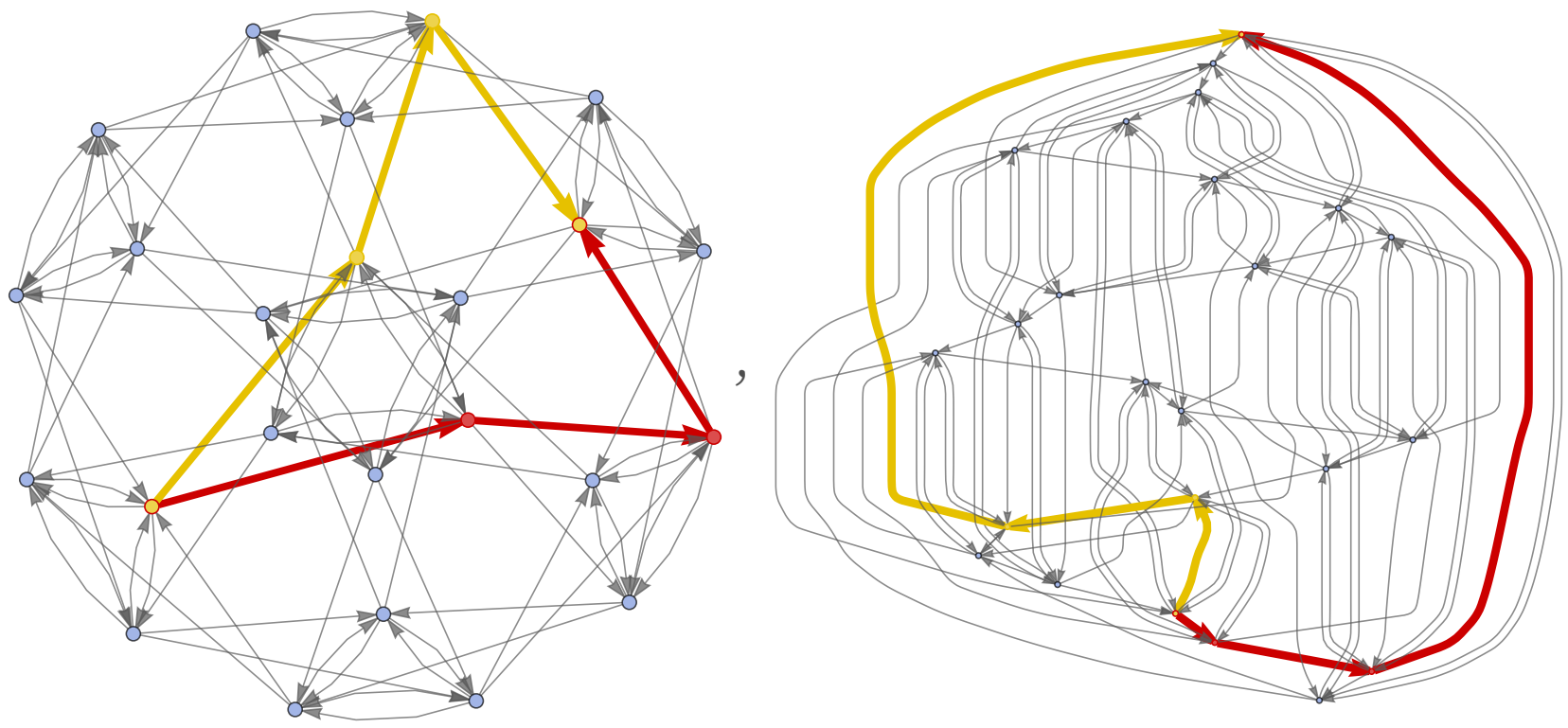

So what this means is that—just as in the infinite-tape case discussed above—multiway Turing machines must always be causal invariant in the rulial limit.

So what about "below the rulial limit"? When is there causal invariance (or the related property of confluence)? For s = 1, k = 1 it is inevitable:

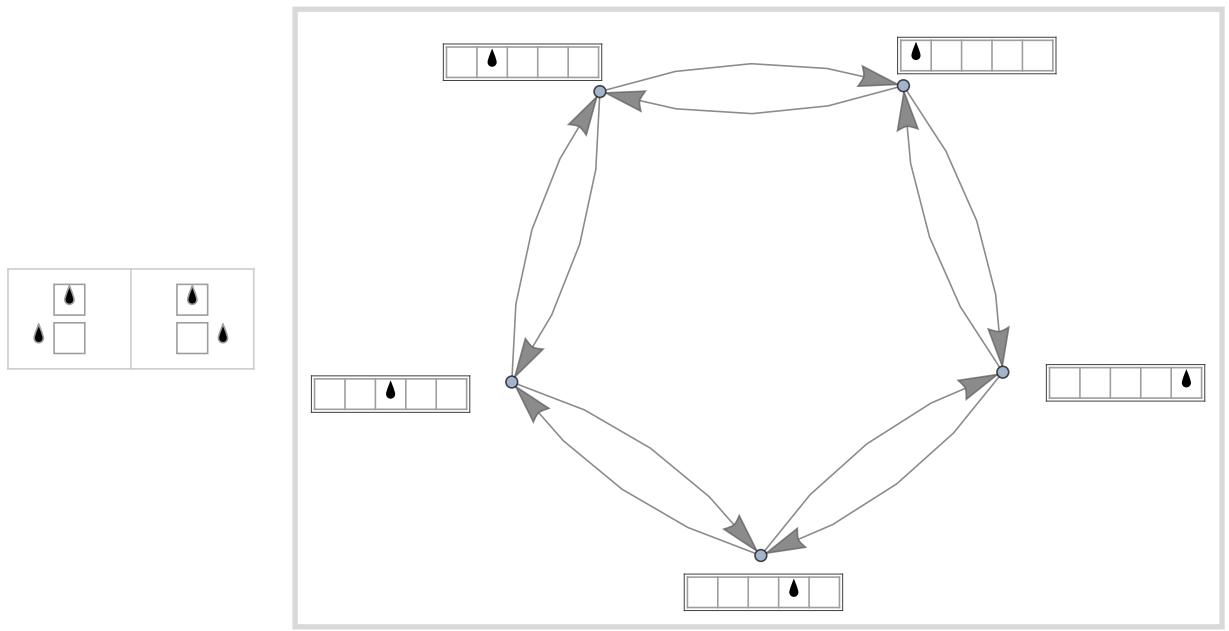

For s = 1, k = 2 it is somewhat rarer; here is an example where paths can branch and not merge:

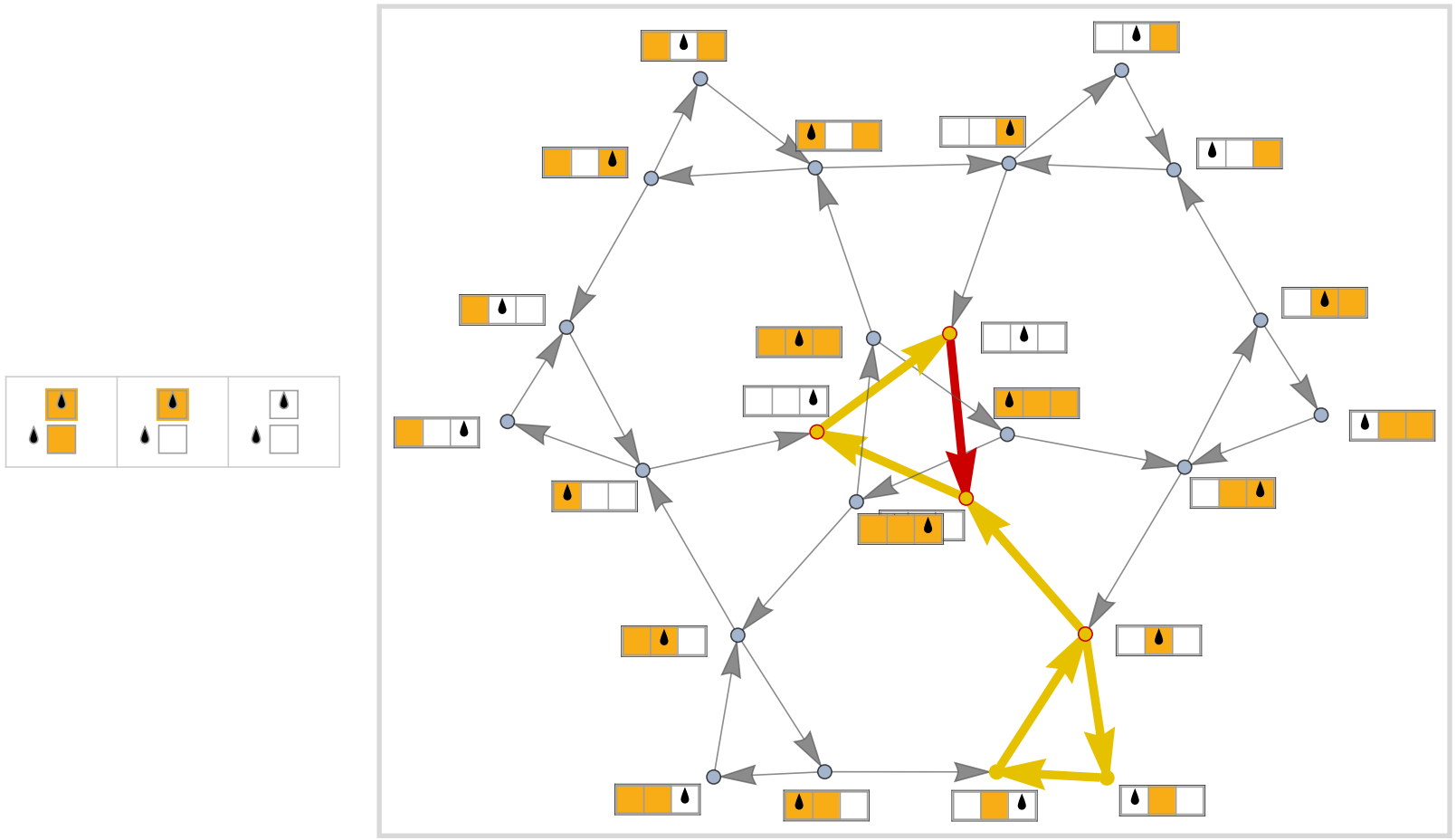

When we discussed causal invariance (and confluence) in the case of infinite tapes, we did so in the context of starting from a particular configuration (such as a blank tape), and then seeing whether all paths from this node in the multiway graph eventually merge. We can do something similar in the case of finite tapes, say asking about paths starting from the node corresponding to a blank tape. Doing this for tapes of sizes between 2 and 4, we get the following results for the number of s = 1, k = 2 rules (excluding purely deterministic ones) that exhibit confluence as a function of the number of cases p in the rule:

| | *n*=2 | | | *n*=3 | | | *n*=4 | | |
|---|---|---|---|---|---|---|---|---|---|
| *p* | × | ✓ | | × | ✓ | | × | ✓ | |
| 2 | 4 | 8 | 67% | 5 | 7 | 58% | 5 | 7 | 58% |
| 3 | 4 | 52 | 93% | 14 | 42 | 75% | 14 | 42 | 75% |
| 4 | 1 | 69 | 99% | 14 | 56 | 80% | 14 | 56 | 80% |
| 5 | 0 | 56 | 100% | 6 | 50 | 89% | 6 | 50 | 89% |
| 6 | 0 | 28 | 100% | 1 | 27 | 96% | 1 | 27 | 96% |
| 7 | 0 | 8 | 100% | 0 | 8 | 100% | 0 | 8 | 100% |
| 8 | 0 | 1 | 100% | 0 | 1 | 100% | 0 | 1 | 100% |

As expected, it is more common to see confluence with shorter tapes, since there is then less opportunity for branching in the multiway graph. (By the way, even for arbitrarily large n cyclic tapes make more machines confluent than infinite tapes, because they allow merging of branches associated with the head "going all the way around". If instead of cyclic boundary conditions, one uses boundary conditions that "reflect the head", fewer machines are confluent.)

With finite tapes, it is possible to consider starting not just from a particular initial condition, but from all possible initial conditions---leading to slightly fewer rules that show "full confluence":

| | *n*=2 | | | *n*=3 | | | *n*=4 | | |
|---|---|---|---|---|---|---|---|---|---|
| *p* | × | ✓ | | × | ✓ | | × | ✓ | |
| 2 | 8 | 4 | 33% | 12 | 0 | 0% | 12 | 0 | 0% |
| 3 | 16 | 40 | 71% | 32 | 24 | 43% | 36 | 20 | 36% |
| 4 | 14 | 56 | 80% | 28 | 42 | 60% | 30 | 40 | 57% |
| 5 | 6 | 50 | 89% | 12 | 44 | 79% | 12 | 44 | 79% |
| 6 | 1 | 27 | 96% | 2 | 26 | 93% | 2 | 26 | 93% |
| 7 | 0 | 8 | 100% | 0 | 8 | 100% | 0 | 8 | 100% |
| 8 | 0 | 1 | 100% | 0 | 1 | 100% | 0 | 1 | 100% |

Here are examples of multiway graphs for rules that show full confluence

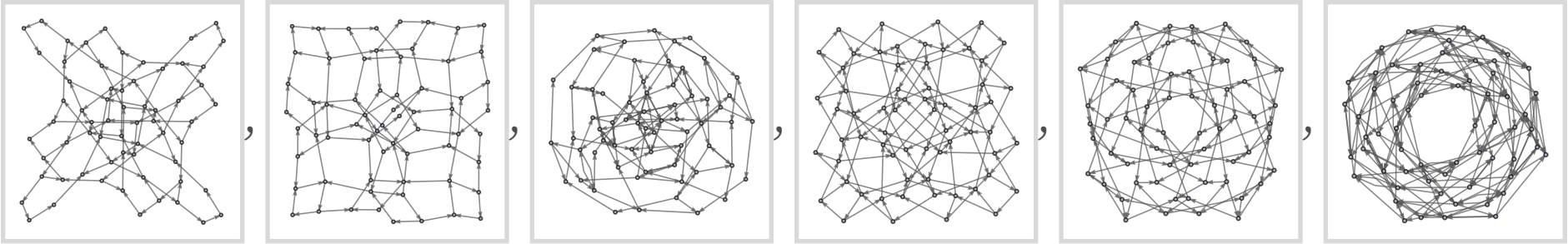

and ones that do not:

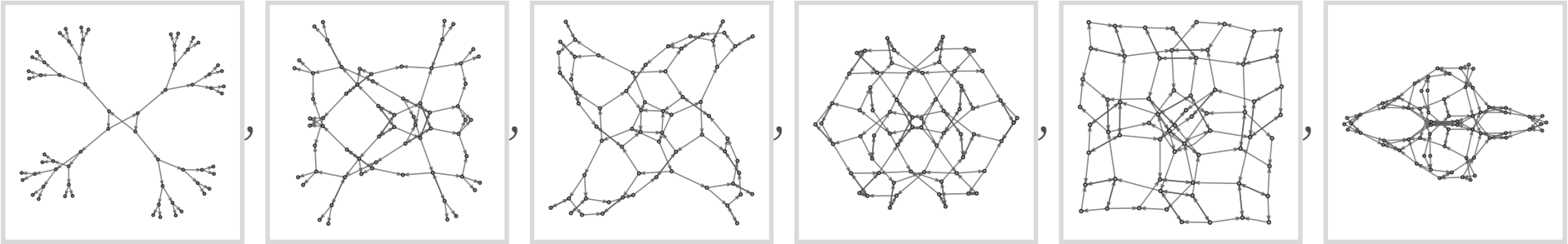

Typically non-confluent cases look more "tree like" while confluent ones look more "cyclic". But sometimes the overall forms can be very similar, with the only significant difference being the directionality of edges. (Note that if a multiway graph is Hamiltonian, then it inevitably corresponds to a system that exhibits confluence.)

## Notes

What I call multiway Turing machines are definitionally the same as what are usually called nondeterministic Turing machines (NDTMs) in the theory of computation literature. I use "multiway" rather than "nondeterministic", however, to indicate that my interest is in the whole multiway graph of possible evolutions, rather than in whether a particular outcome can "nondeterministically" be reached.

The idea of nondeterminism seems to have diffused quite gradually into the theory of computation, with the specific concept of nondeterministic Turing machines emerging around 1970, and promptly being used to formulate the class of NP computations. The original Turing machines from 1936 were purely deterministic. However, although they were not yet well understood, combinators introduced in 1920 already involved what we now think of as nondeterministic reductions, as did lambda calculus.

Many mathematical problems are of the form "Does there exist a ___ that does ___?", and the concept of looking "nondeterministically" for a particular solution has arisen many times. In a specifically computational context it seems to have been emphasized when formal grammars became established in 1956, and it was common to ask whether a particular string was in a given formal language, in the sense that some parse tree (or equivalent) could be found for it. Nondeterministic finite automata seem to have first been explicitly discussed in 1959, and various other forms of nondeterministic language descriptions appeared in the course of the 1960s.

In the early 1980s, quantum generalizations of Turing machines began to be considered (and I in fact studied them a little at that time). The obvious basic setup was structurally the same as for nondeterministic Turing machines (and now for multiway Turing machines), except that in the quantum case different "nondeterministic states" were identified as quantum states, assigned quantum amplitudes, and viewed as being combined in superpositions. (There was also significant complexity in imagining a physical implementation, with an actual Hamiltonian, etc.)

Turing machines are most often used as theoretical constructs, rather than being investigated at the level of specific simple rules. But while some work on specific ordinary Turing machines has been done (for example by me), almost nothing seems to have been done on specific nondeterministic (i.e. multiway) Turing machines.

The fact that two head states are sufficient for universality in ordinary Turing machines was established by Claude Shannon in 1956 (and we now know that s = 2, k = 3 is sufficient), but I know of no similar results for nondeterministic Turing machines.

Note that in my formulation of multiway Turing machines halting occurs just because there are no rules that apply. Following Alan Turing, most treatments of Turing machines introduce an explicit special "halt" state---but in my formulation this is not needed.

I have considered only Turing machines with single heads. If multiple heads are introduced in ordinary Turing machines (or mobile automata) there typically have to be special rules added to define how they should interact. But with my formulation of multiway Turing machines, there is no need to specify this; heads that would "collide in space" just end up at different places in branchial space.

A basic multiway generalization of the TuringMachine function in Wolfram Language is (note that the specification of configurations is slightly different from what is used in A New Kind of Science):

```
MWTMEvolveList[rule_, inits:{{{_, _}, _}...}, t_Integer] :=
  NestList[Union[#, Catenate[Map[MWTMStep[rule, #] &, #]]] &, inits, t]

MWTMStep[rule_List, {{s_, i_}, a_}] /; 1 ≤ i ≤ Length[a] :=
  Apply[{{#1, i + #3}, ReplacePart[a, i -> #2]} &, ReplaceList[{s, a[[i]]}, rule], {1}]
```

Running this for 2 steps from a blank tape gives:

```
MWTMEvolveList[{{1, 1} → {1, 0, −1}, {1, 0} → {1, 0, −1}, {1, 0} → {1, 1, 1}}, {{{1, 3}, {0, 0, 0, 0, 0}}}, 2]

{{{{1, 3}, {0, 0, 0, 0, 0}}}, {{{1, 2}, {0, 0, 0, 0, 0}}, {{1, 3}, {0, 0, 0, 0, 0}}, {{1, 4}, {0, 0, 1, 0, 0}}},
  {{{1, 1}, {0, 0, 0, 0, 0}}, {{1, 2}, {0, 0, 0, 0, 0}}, {{1, 3}, {0, 0, 0, 0, 0}},
    {{1, 3}, {0, 0, 1, 0, 0}}, {{1, 3}, {0, 1, 0, 0, 0}}, {{1, 4}, {0, 0, 1, 0, 0}}, {{1, 5}, {0, 0, 1, 1, 0}}}}
```

The number of distinct states increases as:

```
Length/@%
```

```
{1, 3, 7}
```

This creates a multiway graph for the evolution:

```
MWTMStep[rule_List, h_[{s_, n_}, a_]] :=
  With[{nn = Mod[n, Length[a], 1]},
    h@@@Apply[{{#1, Mod[nn+#3, Length[a], 1]}, ReplacePart[a, nn → #2]} &,
        ReplaceList[{s, a[[nn]]}, rule], {1}]]
```

```
NestGraph[MWTMStep[{{1, 1} → {1, 0, −1}, {1, 0} → {1, 0, −1}, {1, 0} → {1, 1, 1}}, #] &,
  Graph[{h[{1, 3}, {0, 0, 0, 0, 0}]}, {}], 4, VertexLabels → None]
```

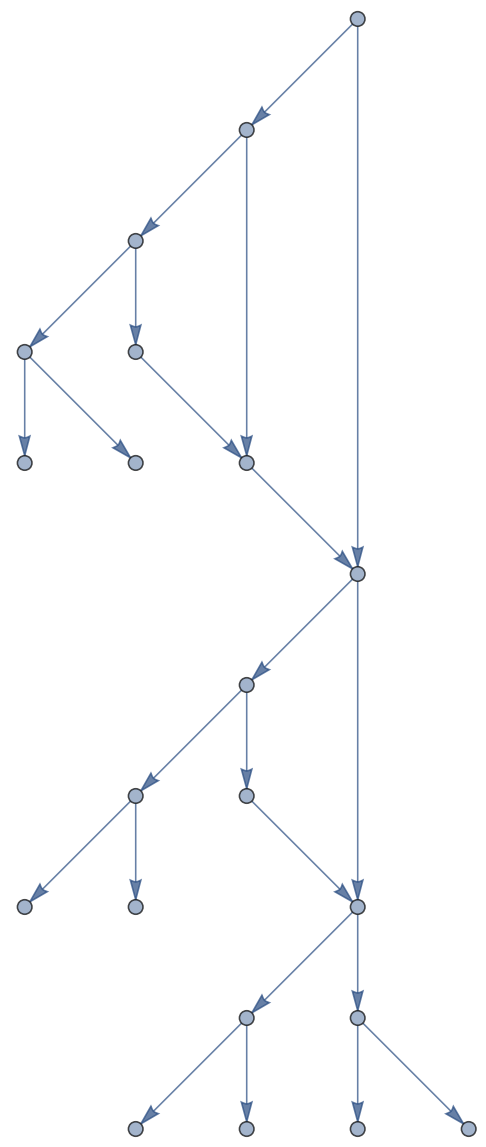

A more complete multiway Turing machine function is in the Wolfram Function Repository.

# References

*Links to references are included within the body of this document.*